\documentclass[aps,prd,11pt, tightenlines, twoside, secnumarabic, superscriptaddress, showpacs, preprintnumbers, nofootinbib, notitlepage, onecolumn, fleqn]{revtex4-1}

\pdfoutput=1
\usepackage[english]{babel}
\usepackage{amsmath,amssymb,amsfonts, bm,bbm,slashed}
\usepackage{graphicx}
\usepackage[sort&compress]{natbib}
\usepackage[dvipsnames]{xcolor}
\usepackage[normalem]{ulem}
\usepackage{hyperref}
\usepackage{cleveref}
\definecolor{red}{rgb}{1.0, 0, 0}
\usepackage{hyperref}
\usepackage{enumerate}
\usepackage{epsfig, subfigure}
\usepackage{setspace}
\usepackage{booktabs, tabularx}
\usepackage{units}
\usepackage{multirow}

\hypersetup{
    pdfnewwindow=true,   
    colorlinks=true,     
    linkcolor=blue,      
    citecolor=blue,      
    filecolor=blue,      
    urlcolor=blue        
}

\allowdisplaybreaks

\bibpunct{[}{]}{,}{n}{}{,}

\renewcommand{\vec}[1]{{\mathbf{#1}}}

\newcommand{\keV}{\,\text{keV}}

\newcommand{\GeV}{\,\text{GeV}}
\newcommand{\TeV}{\,\text{TeV}}

\newcommand{\mttwo}{m_{\text{T}2}}

\newcommand{\sigeff}{\ensuremath{\sigma_{\textrm{eff}}} }

\newcommand{\geff}{\ensuremath{g_{\textrm{eff}}} }

\begin{document}

\title{Leptonic WIMP Coannihilation and the Current Dark Matter Search Strategy}

\author{Michael J.\ Baker}
\email{baker@physik.uzh.ch}
\affiliation{Physik-Institut, Universit\"at Z\"urich, 8057 Z\"urich, Switzerland}
             
\author{Andrea Thamm}
\email{andrea.thamm@cern.ch}
\affiliation{Theoretical Physics Department, CERN, 1211 Geneva, Switzerland}
             
\date{\today}
\preprint{CERN-TH-2018-139}
\preprint{ZU-TH 22/18}


\begin{abstract}

We discuss the extent to which models of Weakly Interacting Massive Particle
(WIMP) Dark Matter (DM) at and above the electroweak scale can be probed conclusively in
future high energy and astroparticle physics experiments.  
We consider simplified models with bino-like dark matter and slepton-like
coannihilation partners, and find that perturbative models yield the observed relic abundance
up to at least $10\TeV$. We emphasise that coannihilation can either increase or decrease the
dark matter relic abundance.
We compute the sensitivity of direct detection experiments to 
DM-nucleus scattering, 
consider indirect detection bounds 
and estimate the sensitivity 
of future proton colliders to slepton pair production.
We find that current and future experiments will 
be able to probe the Dirac DM models up to at least $10\TeV$. 
However, current and future searches will not be sensitive to 
models of Majorana dark 
matter for masses above $2$ or $4\TeV$, for one or ten coannihilation partners respectively, 
leaving around $70\%$ of the 
parameter space unconstrained.
This demonstrates the need for new experimental 
ideas to access models of coannihilating Majorana dark matter.

\end{abstract}

\maketitle


\section{Introduction}

Understanding the nature of Dark Matter (DM) is one of the most pressing questions in particle physics.  
Its existence is well established by a wide range of astrophysical observations and its 
energy density is measured to 2\% accuracy~\cite{Patrignani:2016xqp}.
A thermally produced WIMP (Weakly Interacting Massive Particle) has long been the dominant
paradigm.  
In this picture, dark matter is assumed to have non-negligible interactions with 
Standard Model (SM) particles.  In the early universe, the temperature was very high so the 
standard model particles and dark matter populated a thermal bath.
As the temperature cooled below the mass of the dark matter particle, it self-annihilated 
more often than it was produced, and so its abundance dropped (it became Boltzmann suppressed).
As the universe expanded, the annihilation became inefficient, and 
the dark matter particles could no longer annihilate -- the dark matter froze-out -- 
leaving behind a relic abundance.
This picture successfully predicts the observed relic abundance of dark matter if 
there is a weak-scale interaction cross-section with SM particles.
This success, as well as other hints that beyond the standard model physics, such as 
supersymmetry (SUSY) or new strong dynamics, may be found slightly above the weak scale,
have led to a strong theoretical and experimental exploration of the thermal WIMP.

The canonical WIMP is the lightest neutralino 
of the minimal supersymmetric standard model (MSSM). 
It is both a well motivated dark matter candidate in its own right and, as an
admixture of neutral binos, winos and higgsinos, a powerful parameterisation of
a wide range of WIMP models.  
As such, there has been a large effort to probe its parameter space.
Although direct and indirect detection
experiments are currently probing this parameter space, 
no signal has yet been seen.  
The LHC is also probing the motivated parameter space, but is yet to find 
signals of a WIMP or any other new physics particles.  
Although there are many experiments planned for the future,
the clear question to answer is: `will they probe the whole neutralino parameter space?'  As such, 
it is important to identify viable scenarios in the sub-GeV and multi-TeV mass region, and to consider 
whether the suite of proposed experiments will successfully probe the entire region.
It has been shown~\cite{Bramante:2014tba, Bramante:2015una} 
that future experiments will be able to probe the neutralino 
relic surface (where the parameters of the theory are restricted to produce 
the observed relic abundance via thermal freeze-out) up to masses of 
$4\TeV$, if the sfermions are decoupled.  
However, once coannihilation with sfermions is taken 
into account, a larger and more challenging parameter space becomes accessible.  
It is precisely this scenario we consider in the current work.

Coannihilation~\cite{Griest:1990kh} has been studied for a long time. 
It occurs during thermal freeze-out when there are other dark sector particles, $\phi$, 
similar in mass to the dark matter particle, $\chi$.  
Freeze-out occurs at temperatures where the abundances of 
 $\chi$ and $\phi$ are significantly Boltzmann suppressed.  In this situation the relic abundance of 
 $\chi$ may be reduced if it can
 effectively annihilate via $\phi$, or its relic abundance may be increased if $\phi$ cannot 
 effectively annihilate~\cite{Edsjo:1997bg, Profumo:2006bx, Edsjo:2003us}.  
 This effect can dramatically change the relic abundance and 
 consequently has an important impact on the relic surface.
There has recently been considerable interest in
the range of possible coannihilation models~\cite{Baker:2015qna}, 
their role in producing
viable sub-GeV~\cite{Cheng:2018vaj, DAgnolo:2018wcn}  and multi-TeV scale~\cite{Garny:2014waa,Abdughani:2018bhj, Ellis:2018jyl, ElHedri:2017nny, ElHedri:2018atj} 
dark matter candidates and in
coannihilating models at the LHC and future colliders~\cite{Aboubrahim:2017aen, Duan:2018rls, Dutta:2017nqv}.
In this work we use a simplified model framework to explore the impact of 
coannihilation on multi-TeV dark matter.  We consider 
a bino motivated (gauge-singlet Dirac or Majorana fermion) dark matter candidate accompanied by 
$n$ dark-sector scalars with unit hypercharge.    
In the MSSM, a pure bino with no other nearby states cannot efficiently annihilate, 
resulting in overclosure of the universe.  However, when sfermions are included, the 
observed relic abundance can be recovered for a relatively wide range of masses. 
We consider the three possible Yukawa couplings with SM electrons, muons and taus 
individually.
This minimal setup lets us study the impact of coannihilation in isolation, and vary the 
degree of coannihilation by changing the mass difference (between dark matter and the new scalars) and the number of coannihilating 
partners.  We first find, in \cref{sec:model}, the relic surface for a range of models, demonstrating that they provide 
a viable multi-TeV dark matter candidate.  We then consider the reach of a range of 
direct detection experiments (\cref{sec:directdetection}), indirect detection experiments (\cref{sec:indirectdetection}), and proton colliders (\cref{sec:collider}).  
We will see that there is a large region of viable parameter space for Majorana dark matter 
which future experiments will be unable to probe, motivating the need for 
new experimental ideas.

The experimental landscape in dark matter physics consists of 
colliders, direct detection experiments and indirect detection experiments.
The LHC is currently running at $13\TeV$ and has delivered approximately $100\,$fb$^{-1}$ of 
integrated luminosity to the ATLAS and CMS experiments.  This has allowed the experiments 
to place significant bounds on simplified dark matter models mostly via mono-$\gamma, Z, W, h, t$ and mono-jet searches. ATLAS and CMS also search directly for the mediators in di-jet or di-lepton plus missing energy searches~\cite{Aaboud:2018jiw}. For an exhaustive list of possible coannihilating DM searches at the LHC see, e.g,.~\cite{Baker:2015qna}. A higher energy collider will be required to efficiently produce multi-TeV particles.  Currently under discussion 
are a $27\TeV$ high energy upgrade to the LHC, dubbed HE-LHC, which would deliver approximately 
$15\,$ab$^{-1}$ of integrated luminosity~\cite{HELHC} and a $100\TeV$ collider, either in Europe or in China, which 
would deliver approximately $20\,$ab$^{-1}$~\cite{Hinchliffe:2015qma}.  
Future lepton colliders, such as the ILC \cite{Fujii:2017vwa}, CLIC \cite{CLIC:2016zwp} and the FCC-ee \cite{FCCee}, are also under consideration. CLIC is designed to reach the highest centre of mass energy among these machines with $\sqrt{s}=3\TeV$.  However, this is still not high enough to produce the multi-TeV particles we discuss. 
It should be noted that although a collider 
may produce dark sector particles, it cannot  
determine that any particle is the cosmologically stable dark matter
since it can only test particle stability on detector scales.

Direct detection experiments consist of a body of shielded target material. 
A dark matter particle in our galaxy may interact with the target and deposit energy, which 
may then be detected as light, heat or ionisation.  These experiments are most sensitive to the 
dark matter mass range $10\GeV$ -- $1\TeV$ (although there are substantial efforts to extend the sensitivity to lower masses), and the current leaders are 
LUX~\cite{Akerib:2016vxi}, PandaX-II~\cite{Cui:2017nnn} and XENON1T~\cite{Aprile:2018dbl}, which all use xenon as their target.  In what follows we take XENON1T as the illustrative example.
In the future, the most ambitious is the planned DARWIN experiment, which aims to 
have a sensitivity 100 times better than these experiments with an exposure of 
500 $\text{ton}\cdot\text{years}$~\cite{Aalbers:2016jon}.

The final class of experiment is indirect detection of dark matter.  
Although dark matter stopped annihilating when it froze-out, due to its low number density, 
gravity interactions have now caused dark matter to cluster, in haloes which encompass galaxies and galaxy clusters.
Indirect detection experiments look for annihilation of dark matter where its abundance is expected to be largest.  
Since thermal dark matter particles could annihilate into any standard model particles, there are a range of strategies 
looking for photons, neutrinos and a range of anti-matter produced in 
the galactic centre or in dwarf galaxies.  In this work we find the best limits from searches for an 
excess in continuum photons (as opposed to mono-energetic photons).  We find that the strongest 
constraints are placed on our model by the Fermi-LAT -- MAGIC collaboration~\cite{Ahnen:2016qkx} 
and HESS~\cite{HESS:2015cda}.
The Fermi-Large Area Telescope (Fermi-LAT) is the principle scientific instrument on the Fermi Gamma Ray Space Telescope spacecraft and is a high-energy gamma-ray telescope covering the energy range from about $20\,$MeV to more than $300\,$GeV.
MAGIC (Major Atmospheric Gamma Imaging Cherenkov Telescopes) is a system of two ground-based atmospheric Cherenkov telescopes.
HESS (High Energy Stereoscopic System) is an array of four ground-based atmospheric Cherenkov telescopes 
which measure cosmic photons in the energy range from 10s of GeV to 10s of TeV.  
In the future, the most ambitious planned experiment which improves on these bounds is CTA~\cite{Carr:2015hta}.
CTA (Cherenkov Telescope Array) is the next generation ground-based array, which will operate 
in a similar energy band with several tens of telescopes.

\section{The Models and their Relic Surfaces}
\label{sec:model}

\begin{table}
  \centering
  \begin{minipage}{13cm}
    \begin{ruledtabular}
\begin{tabular}{cccccc}
Field & Spin & $su(3)\times su(2)_L \times u(1)_Y$ & $\mathbb{Z}_2$ & Copies & DOF\\
\hline
$\chi$ & 1/2 & (1,1,0) & -1 & 1 & 4\\
$\phi_i$ & 0 & (1,1,-1) & -1 & $n$ & $2n$
\end{tabular}
    \end{ruledtabular}
  \end{minipage}
  \caption{The new particles we introduce in \cref{sec:model} with their respective charges, 
    the number of copies we consider and the number of degrees of freedom.}
  \label{tab:particles}
\end{table}

In this work we focus on several related simplified models.  As shown in \cref{tab:particles}, we introduce dark matter as a Majorana or Dirac fermion, $\chi$, and $n$ copies of an uncoloured scalar coannihilation partner, $\phi_i$, with unit hypercharge. The dark matter particle and each coannihilation partner couple to a standard model right-handed charged lepton. In addition to kinetic and mass terms, the Lagrangian only has one new interaction term (ignoring the scalar quartic, which plays no role in our phenomenology)
\begin{align}
\mathcal{L} \supset \overline{\chi} (i \slashed \partial - m_\chi) \chi + \frac{1}{2} |D_\mu \phi_i|^2 - \frac{1}{2} m_\phi^2 \phi_i^2 + (y_\chi \phi_i \overline{\chi} \ell_R + h.c. ) \,,
\end{align}
where $D_\mu = \partial_\mu - i g' Y B_\mu$ and the coupling is taken to be universal, i.e., $y_\chi$ is the same for all $\phi_i$. We consider the cases $\ell_R = e_R, \mu_R$ and $\tau_R$, and assume that all $\phi_i$ have the same mass, $m_{\phi_i} = m_\phi$, and that $m_\chi < m_\phi$.
For illustration, we focus on $n \in \{1, 3, 10\}$, which will allow us to show the impact of 
one, several and many coannihilation partners. In a supersymmetric context, the DM particle $\chi$ would correspond to a bino  and the scalar $\phi_i$ can be identified with a right-handed slepton. Note that, in SUSY, the number of degrees of freedom of one right-handed slepton corresponds to $n=1$, all right-handed sleptons corresponds to $n=3$, while all right-and left-handed sleptons correspond to $n=9$. Here, we follow a simplified model approach in order to isolate the effect of coannihilation from the added complications due to considering several flavours at once or non-trivial $su(2)$ quantum numbers.
Depending on the single lepton flavour involved, we will refer to our models as electron, muon and tau type. 

\begin{figure}
  \centering
  \includegraphics[width=\textwidth]{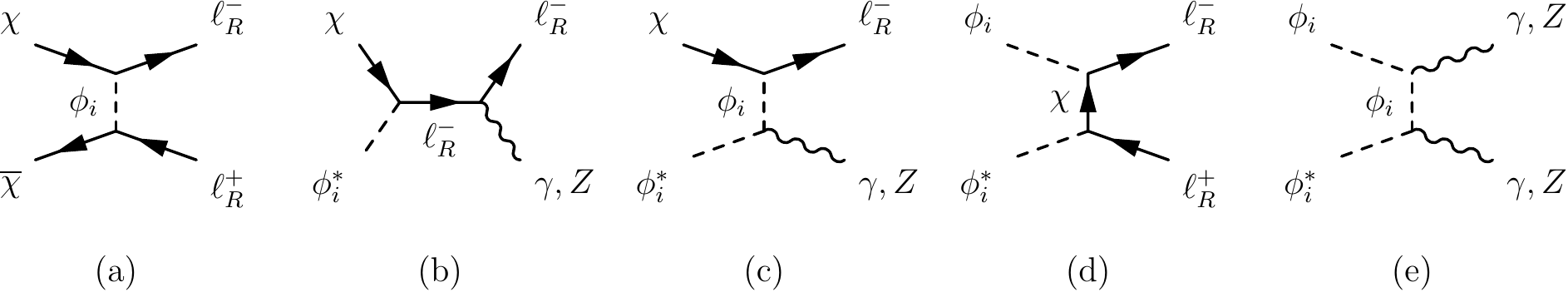}
  \caption{Some representative annihilation and coannihilation diagrams.  In (a) we 
  show the only annihilation process, in (b) and (c) we show typical $\chi$ -- $\phi_i$ coannihilation processes, 
  in (d) and (e) we show typical annihilation 
  process of the coannihilating partner, $\phi_i$.  Not shown are the processes 
  $\phi_i \phi_i^* \rightarrow \gamma, Z \rightarrow \bar{\psi} \psi$ and 
  $\chi \phi_i^* \rightarrow \tau_R^- \rightarrow \tau_L^- h$ (where $\psi$ is a generic 
  standard model particle).
  }
  \label{fig:coannihilation-diagrams}
\end{figure}

We are interested in the slice of parameter space where the models produce the observed relic abundance of $\chi$ via thermal freeze-out. 
In the following we will denote generic standard model bath particles as $\psi$, 
whether they are fermions or bosons.
In the coannihilation regime, where 
\begin{align}
\Delta \equiv \frac{m_{\phi}-m_\chi}{m_\chi} \lesssim 0.2\,,
\end{align}
the Boltzmann equation for the abundance of $\chi$ 
becomes a coupled set of differential equations which also track the abundance of $\phi_i$.
These can be combined~\cite{Griest:1990kh} into a single differential equation, 
the same as the usual Boltzmann equation for a single species, 
if the $\chi \bar{\chi} \rightarrow \bar{\psi} \psi$ annihilation 
cross-section is replaced by
\begin{align} \label{eq:sigeff}
 \sigeff &= \sum_{ij}^N \frac{g_i g_j}{\geff^2} \sigma_{ij}
(1+\Delta_{i})^\frac{3}{2}(1+\Delta_{j})^\frac{3}{2}
e^{-x(\Delta_i+\Delta_j)} \, ,
\end{align}
where $i, j$ index the DM particle and its coannihilation partners $\{\chi, \phi_1, \phi_2,\ldots, \phi_N \}$, 
$g_i$ is the number of degrees of freedom of particle $i$,
$\Delta_i = (m_i - m_\chi)/m_\chi$, the cross-section is
$\sigma_{ij} = \sigma(i j \rightarrow \bar{\psi} \psi)$ 
and $x = m_\chi / T$. The effective number of degrees of freedom is given by 
\begin{align}
 \geff &= \sum_{i=1}^N g_i(1+\Delta_i)^\frac{3}{2} e^{-x\Delta_i} \, .
\end{align}
Note that $ \geff$ is always greater than $g_\chi$. For Majorana and Dirac DM $g_\chi = 2$ and $4$, respectively. The degrees of freedom of a complex scalar $\phi_i$ are $g_\phi = 2$. 
Some representative coannihilation diagrams which contribute to \sigeff in our model
are shown in \cref{fig:coannihilation-diagrams}.

We can understand some features of these equations on physical grounds.  
The abundances of both $\chi$ and $\phi$ are 
similarly Boltzmann suppressed during freeze-out.  This means that the rate 
of dark sector annihilation $\chi \bar{\chi} \rightarrow \bar{\psi} \psi$, $\chi \phi \rightarrow \bar{\psi} \psi$ 
and $\phi \phi^* \rightarrow \bar{\psi} \psi$, where two rare particles are in the initial state, becomes exponentially smaller than the conversion processes $\chi \psi \rightarrow \phi \psi$, which requires only one rare particle and one bath particle. 
Thus, even if $\chi$ annihilation has a very small cross-section,
$\chi$ can be efficiently depleted by first converting into $\phi$ and then annihilating.  
That is, if $ \sigma_{\chi \phi}\gg \sigma_{\chi \bar{\chi}}$ and/or 
$\sigma_{\phi \phi} \gg \sigma_{\chi \bar{\chi}}$, and both $\Delta$ and $g_\text{eff}$ are not too large, then 
\begin{align}
\sigma_\text{eff} \gg  \sigma_{\chi \bar{\chi}}\,,
\end{align} 
and,
since $\Omega_\chi h^2 \sim 1/ \langle \sigma v \rangle$, coannihilation reduces the relic abundance.
This is the usual mechanism used to deplete bino dark matter.  As $\Delta$ becomes larger, $\phi$ is 
Boltzmann suppressed to a larger degree, and there is not enough thermal energy 
to efficiently achieve $\chi \psi \rightarrow \phi \psi$, so the mechanism becomes 
ineffective.

We note that although coannihilation is usually thought to increase the effective 
cross-section (as above), it can also reduce the 
effective cross-section, increasing the dark matter relic abundance (as noted in, 
e.g.,~\cite{Edsjo:1997bg, Profumo:2006bx, Edsjo:2003us}).  
If $\sigma_{\chi \bar{\chi}} \gg \sigma_{\chi \phi}, \sigma_{\phi \phi}$ but 
$\chi \psi \rightarrow \phi \psi$ is still efficient, then the terms 
in the sum with $i\neq \chi$ and $j \neq \chi$ become negligible and we are left with
\begin{align} 
 \sigeff &\approx  \frac{g_\chi^2}{\geff^2} \sigma_{\chi \bar{\chi}} \ll  \sigma_{\chi \bar{\chi}}\, ,
\end{align}
i.e., the cross-section has effectively been reduced by $g_\chi^2/ g_\text{eff}^2$.  
We can understand this situation by imagining a temperature above 
the temperature at which $\chi$ freezes out, but
below that at 
which $\phi$ would have frozen-out 
if $\chi$ were not present.  
The energy density which resides in $\phi$ cannot go into the thermal 
bath via $\phi \phi \rightarrow \bar{\psi} \psi$, since the cross-section is too small, 
but it can go via $\phi \psi \rightarrow  \chi \psi$, since this rate is not 
doubly Boltzmann suppressed.  This can be thought to `top-up' the abundance 
of $\chi$ during freeze-out, resulting in a higher final $\chi$ abundance.
However, coannihilation becomes ineffective as $\Delta$ becomes larger than around 20\% as 
$\phi$ can fully annihilate before $\chi$ has frozen-out, 
eliminating the `top-up'.


\begin{figure}
  \begin{center}
  \begin{tabular}{cc}
    \includegraphics[width=0.42\columnwidth]{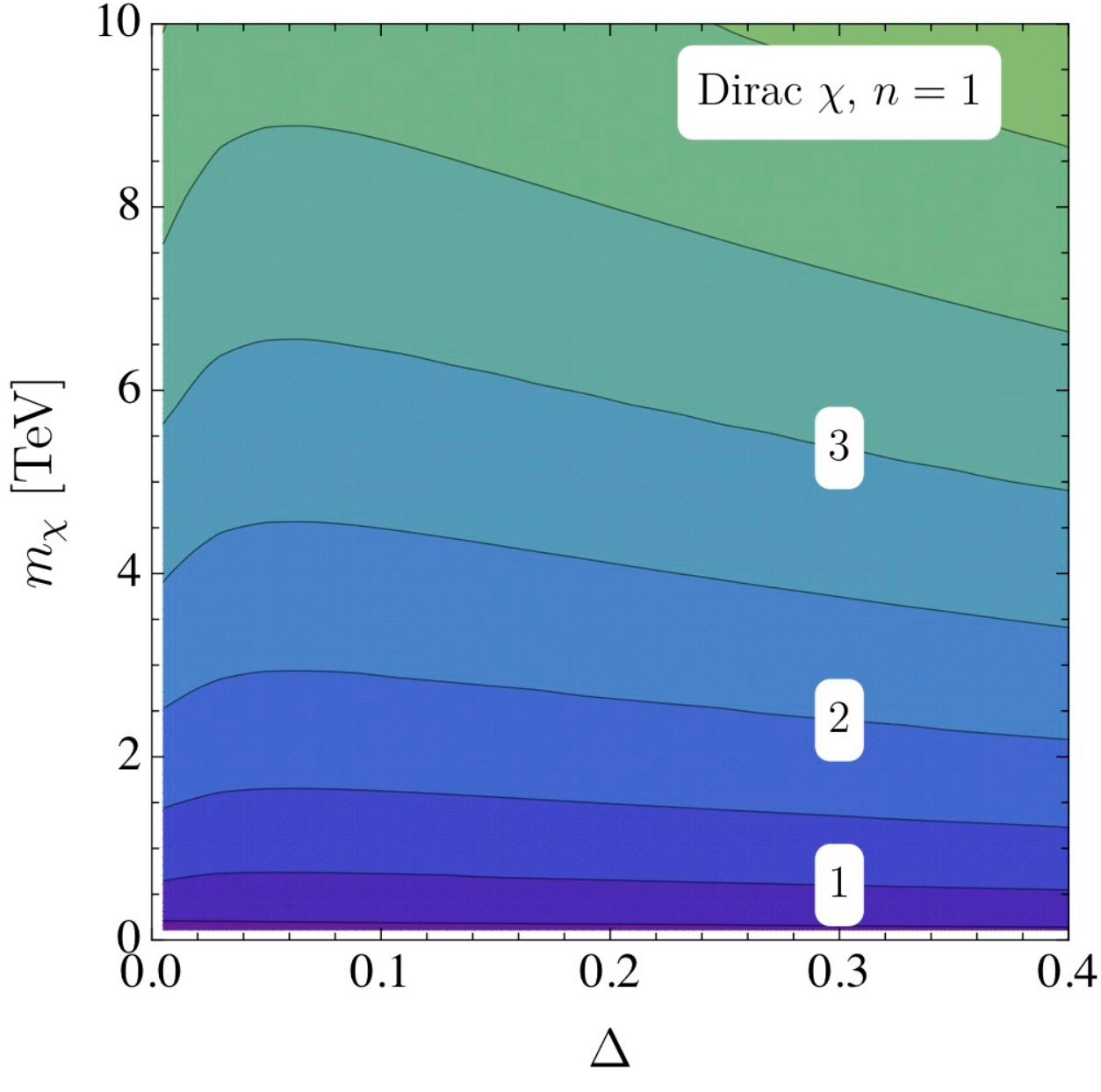}&
    \includegraphics[width=0.42\columnwidth]{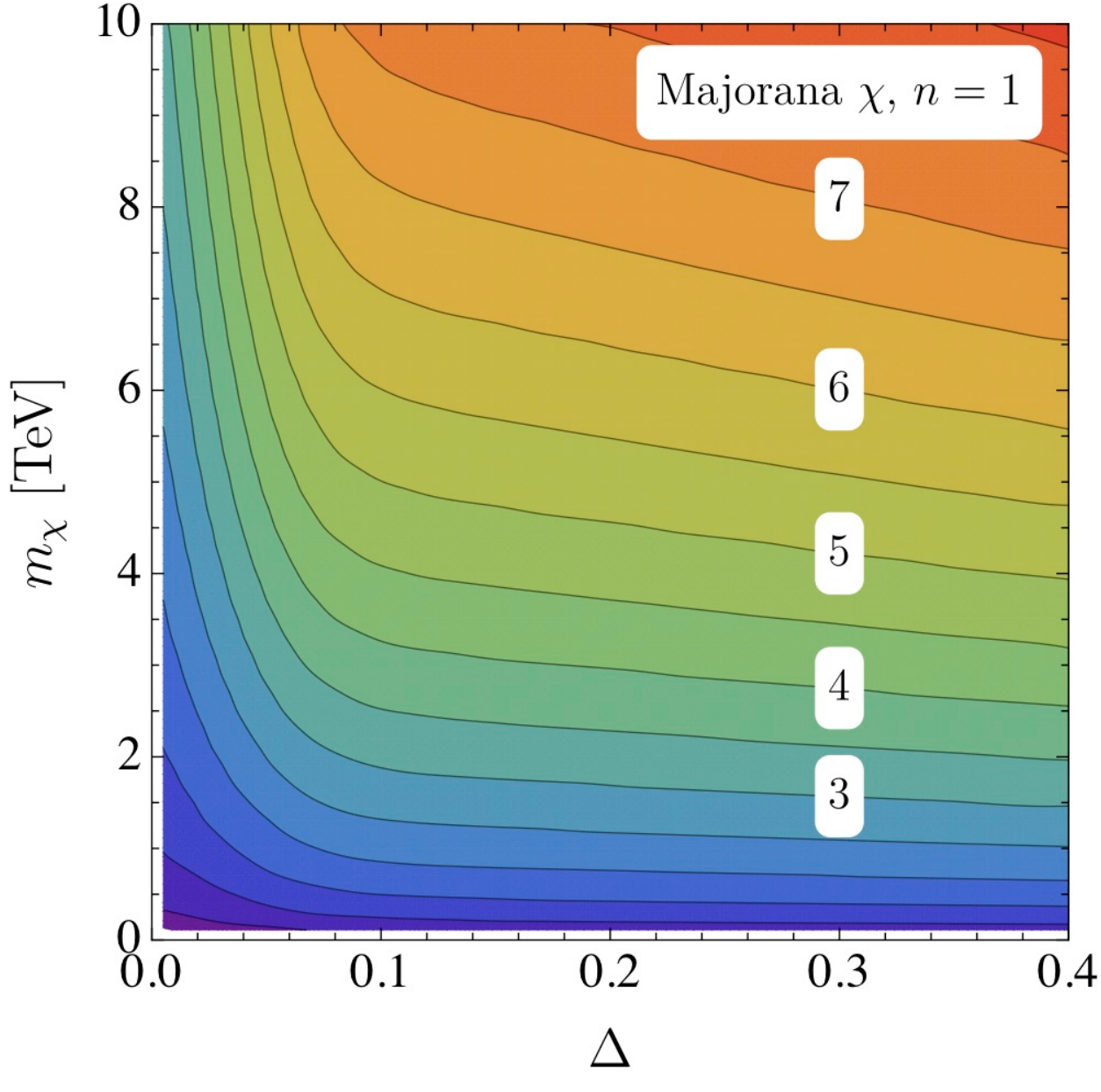}\\
    \includegraphics[width=0.42\columnwidth]{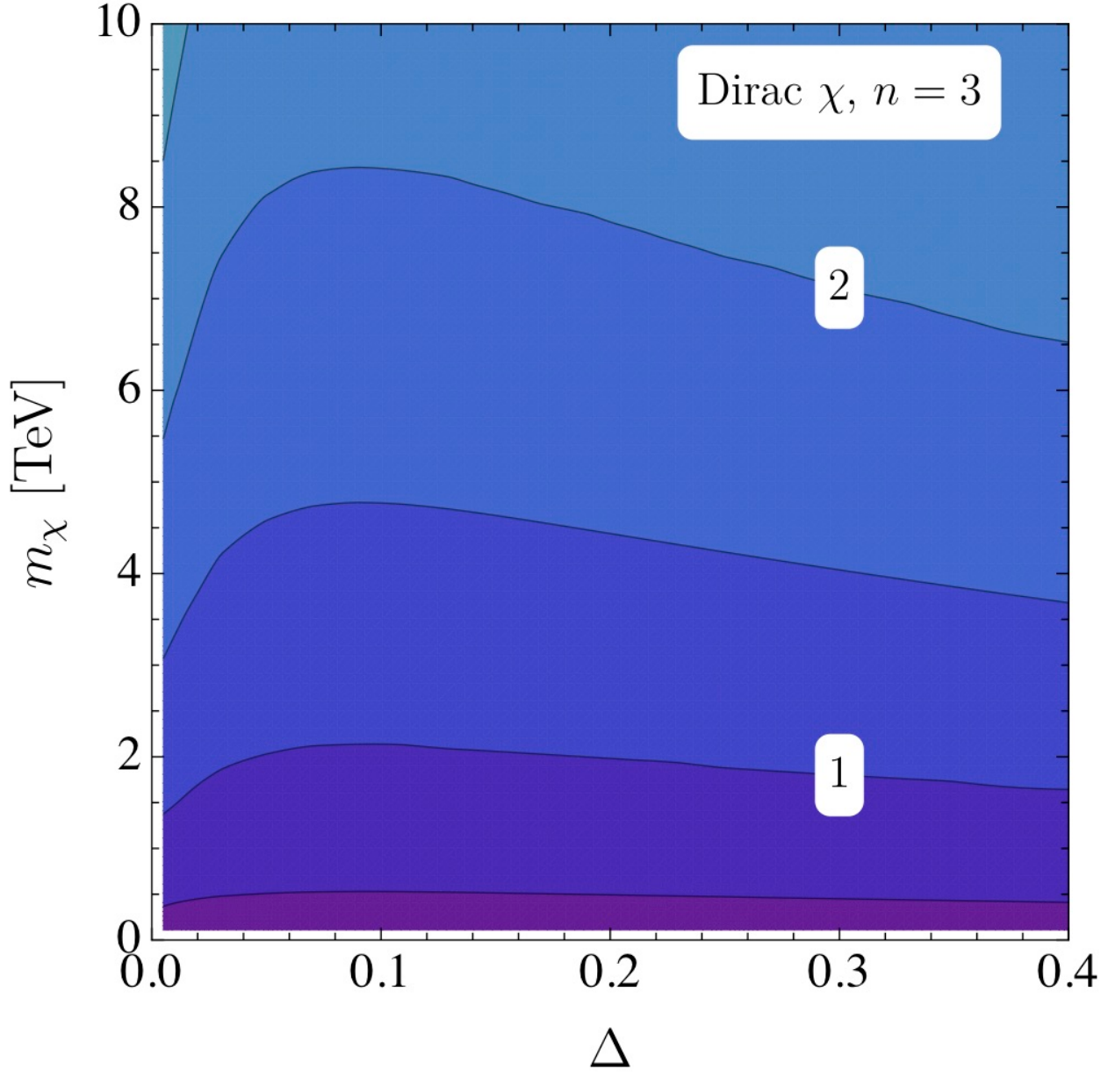}&
    \includegraphics[width=0.42\columnwidth]{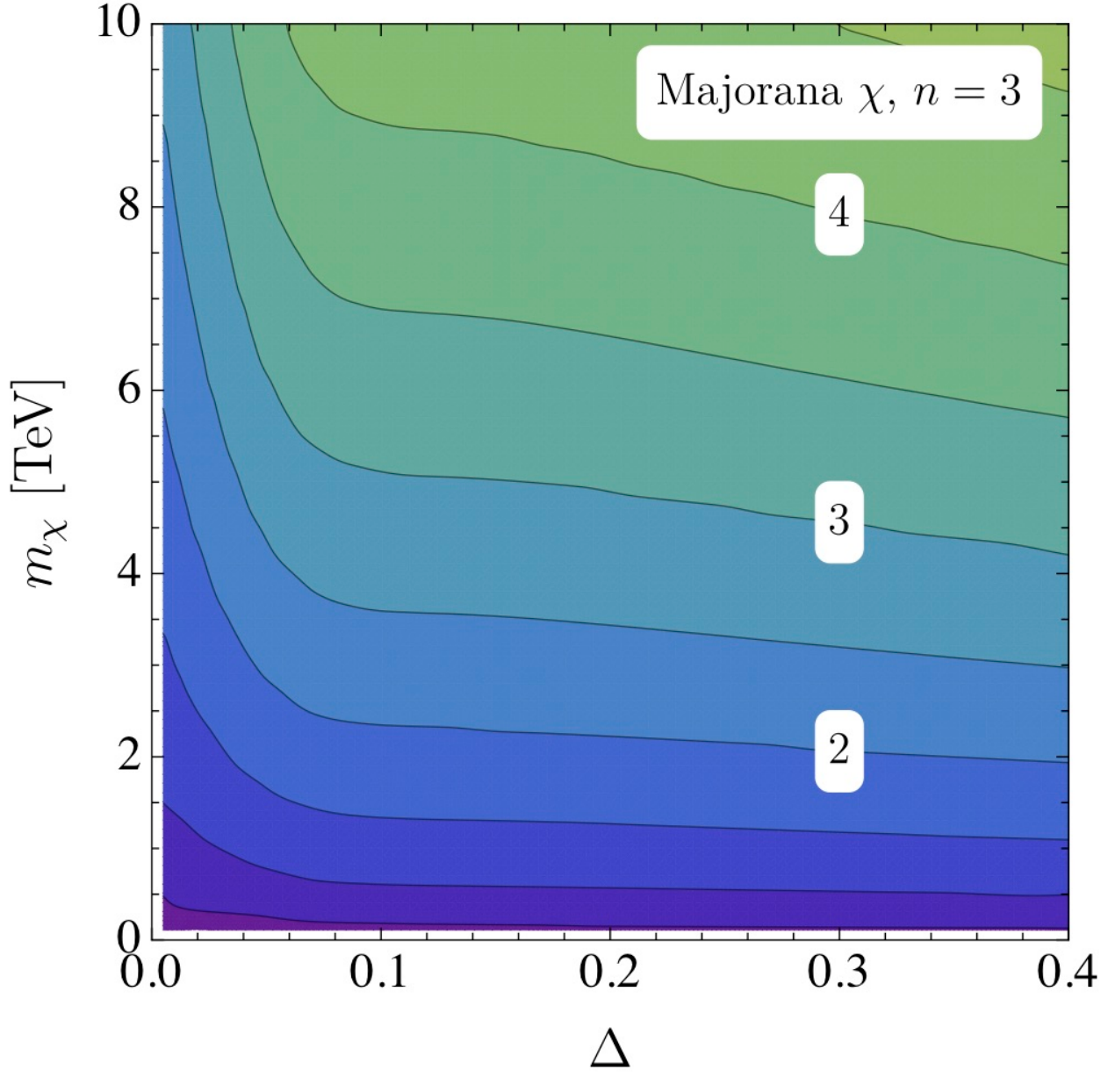}\\
    \includegraphics[width=0.42\columnwidth]{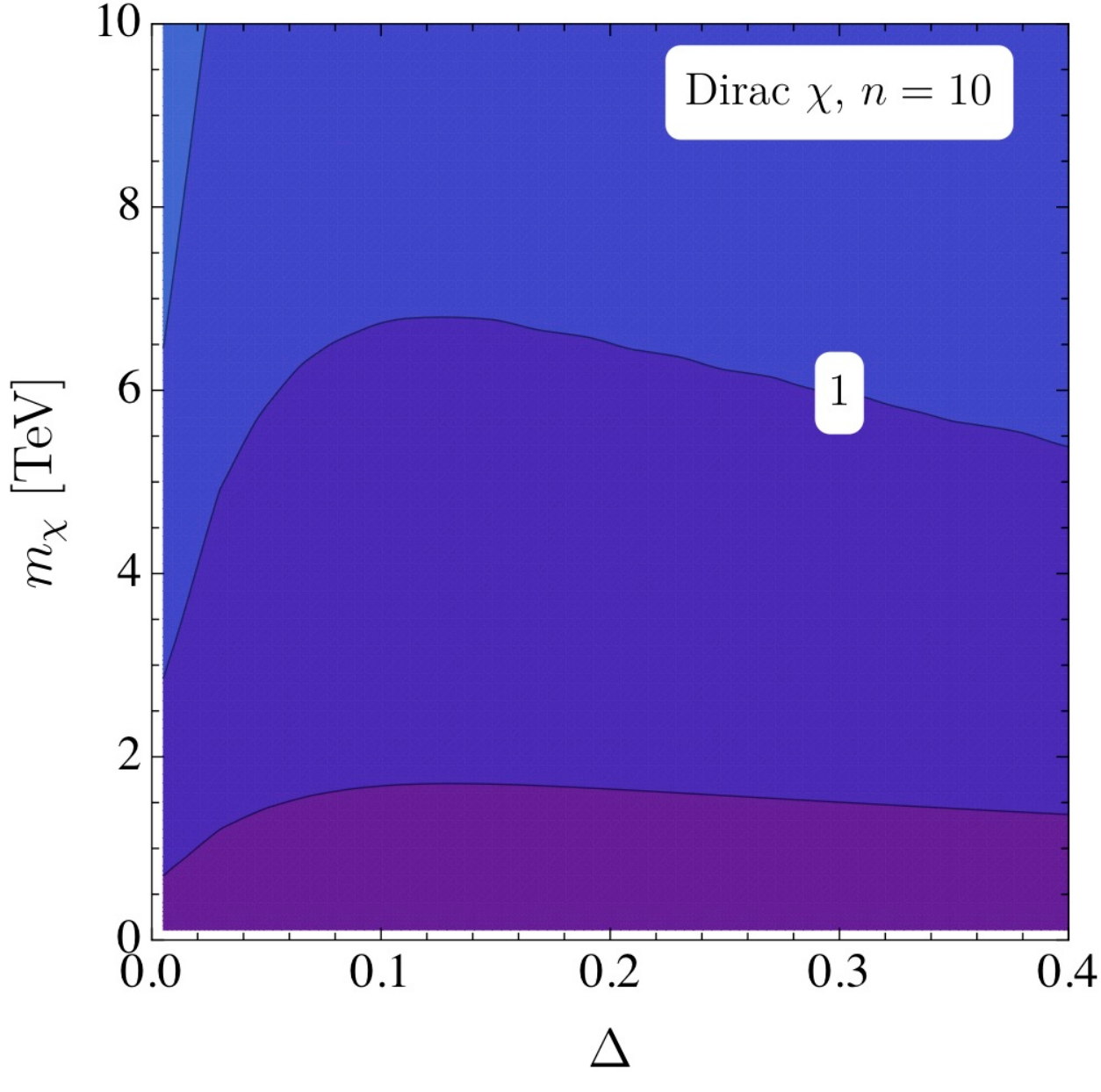}&
    \includegraphics[width=0.42\columnwidth]{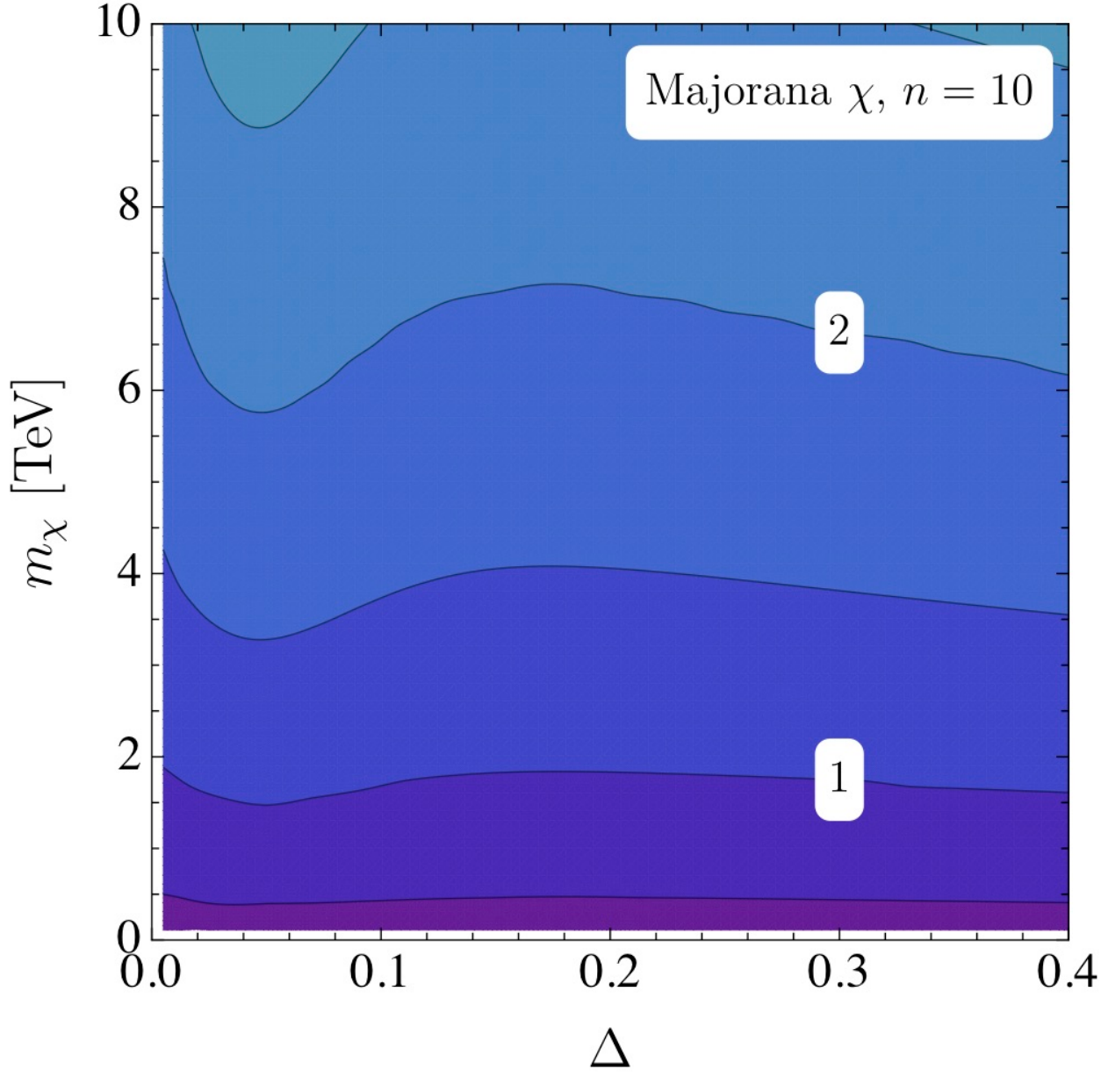}
 \end{tabular}
  \end{center}
  \caption{The value of $y_\chi$ required to give the observed relic abundance for 
  Dirac dark matter (left) and Majorana dark matter (right), for $n=1$ (top), 
  $n=3$ (middle) and $n=10$ (bottom) coannihilation partners.
  Since we are always in a regime where $m_\chi \gg m_\ell$, the relic surfaces are independent of 
  which lepton interacts with $\chi$.}
  \label{fig:model-1-relic-surface}
\end{figure}


To calculate the relic abundances in our models, we use model files written with
\texttt{SARAH\,v4.12.1} \cite{Staub:2008uz} and calculate the relic abundance using \texttt{micrOMEGAs\,v4.3.5}~\cite{Belanger:2014vza}, which implements coannihilation.  
We then interpolate the results of a 3-dimensional scan (in $m_\chi$, $\Delta$ and $y_\chi$) 
to determine the value of the coupling $y_\chi$ which will result in the observed relic abundance~\cite{Ade:2015xua} of
\begin{align}
\Omega_\chi h^2 = 0.1186 \pm 0.0020 \, .
\end{align}
The results for $n=1,3, 10$ are shown in \cref{fig:model-1-relic-surface}.  
We consider the parameter space  
$100\GeV < m_\chi < 10\TeV$ and $0.01 < \Delta < 0.4$ for all 18 cases (Majorana and Dirac 
$\chi$, each with $n \in \{ 1, 3, 10\}$, each for the electron, muon and tau type models). 
We focus on this region of parameter space since 
a range of strategies are being pursued for coannihilating dark matter around 
$100\GeV$~\cite{Gori:2013ala, Schwaller:2013baa, Evans:2016zau, Khoze:2017ixx, Mahbubani:2017gjh}
and for sub-GeV dark matter~\cite{Hochberg:2017wce,Knapen:2017xzo,Knapen:2017ekk}, 
while upcoming proton-proton colliders have little prospect of probing 
particles heavier than $10\TeV$.  The $\Delta < 0.01$ region is extremely fine-tuned, 
requiring significant theoretical motivation, while the region $\Delta > 0.4$ will not 
exhibit significant coannihilation.
The relic surfaces for all 18 models can be presented in six plots.  
Since we are always in the limit $m_\chi \gg m_\ell$, the impact of the lepton 
mass on the relic surface is negligible, so the relic surface is independent of the 
lepton flavour of the model.
We see that the observed relic abundance can be reached for perturbative 
couplings ($y_\chi < 4 \pi / \sqrt{n} \approx \{13,\, 7.3,\, 4.0\}$ for $n=\{1,\,3,\,10\}$, respectively). 
Although the coupling remains $< 4 \pi / \sqrt{n}$, it is relatively large in much of 
our parameter space, which suggests that higher order corrections to our tree-level 
and one-loop calculations may not be insignificant.

In \cref{fig:model-1-relic-surface} (top-right) we see the relic surface for the Majorana 
models with $n=1$.  
As $m_\chi$ increases, the required coupling increases, as is expected since 
the annihilation cross-section scales as $\sigma_{\chi\bar{\chi}} \sim y_\chi^4 m_\chi^{-2}$.  
Since the relic abundance is roughly inversely proportional to $\sigma_{\chi\bar{\chi}}$, 
$y_\chi$ needs to increase as $m_\chi$ increases to keep $\sigma_{\chi\bar{\chi}}$  
approximately constant.  As $\Delta$ becomes smaller than 0.1, we begin to see the effect of 
coannihilation.  For Majorana $\chi$, $\sigma_{\chi\bar{\chi}}$ is velocity suppressed and so 
is significantly smaller than $\sigma_{\phi \phi}$.  As coannihilation becomes relevant, 
$\sigeff$ becomes larger, which would reduce the relic abundance, if $y_\chi$ did not reduce 
to compensate.  On the relic surface, we see the required reduction in $y_\chi$.  
Although coannihilation is not active above $\Delta\approx0.15$, we note that the scalar 
partners still allow $\chi$ to have the correct relic abundance (which would not be the case 
if they were completely decoupled).

In \cref{fig:model-1-relic-surface} (top-left) we see the relic surface for the Dirac models 
with $n=1$.  
Again we see $y_\chi$ increasing as $m_\chi$ increases.  However, as $\Delta$ reduces 
below 0.1, 
$y_\chi$ now increases.  This is both because $\sigma_{\chi\bar{\chi}}$ is not velocity suppressed 
for Dirac dark matter and because in most of the parameter space $y_\chi > 1$, which is greater than 
the electromagnetic coupling of $\phi$.  As such, the extra cross-sections we add into 
\cref{eq:sigeff} are small, and the dominant effect is to reduce $\sigeff$ 
due to the increase of $g_\text{eff}$.
In this situation, coannihilation increases the relic abundance.

In \cref{fig:model-1-relic-surface} (middle) and (bottom) we see the impact of 
increasing the number of coannihilation partners.  Extra 
partners change the required $y_\chi$ both at large and small 
$\Delta$.  
We see an effect at large $\Delta$, where the relic abundance is set by $\sigma_{\chi\bar{\chi}}$,
 since the only contribution to 
$\sigma_{\chi\bar{\chi}}$ is a $t$-channel diagram with a $\phi$ propagator.  As such, 
increasing the number of partners will increase this cross-section by $n^2$.  To maintain 
the observed relic abundance, the coupling $y_\chi$ has to decrease accordingly.
As $\Delta$ goes to zero, for Dirac $\chi$, more coannihilating partners simply 
give a stronger increase in $y_\chi$. 
For Majorana $\chi$, we see a balance between an increase 
in the effective cross-section due to the extra coannihilation processes and a 
decrease in $\sigeff$ due to an increasing $g_\text{eff}$, 
which is especially pronounced in the $n=10$ case.  
For Majorana dark matter, $\sigma_{\chi\bar{\chi}}$ is velocity suppressed  
while $\sigma_{\phi\phi}$ grows with $n^2$.  
The $\chi$ contribution to $g_\text{eff}$, however, 
is not suppressed in any way, 
so $g_\text{eff}^2$ is not simply proportional to $n^2$.
At $\Delta < 0.05$, the dominant effect comes from an increase in 
$\sigma_\text{eff}$ due to $\sigma_{\phi\phi}$.  
At $0.01 < \Delta < 0.2$,  we see that an increased $g_\text{eff}$, which reduces $\sigma_\text{eff}$, 
is the dominant effect.

\section{Direct detection}
\label{sec:directdetection}


\begin{figure}
  \centering
  \includegraphics[width=0.4\textwidth]{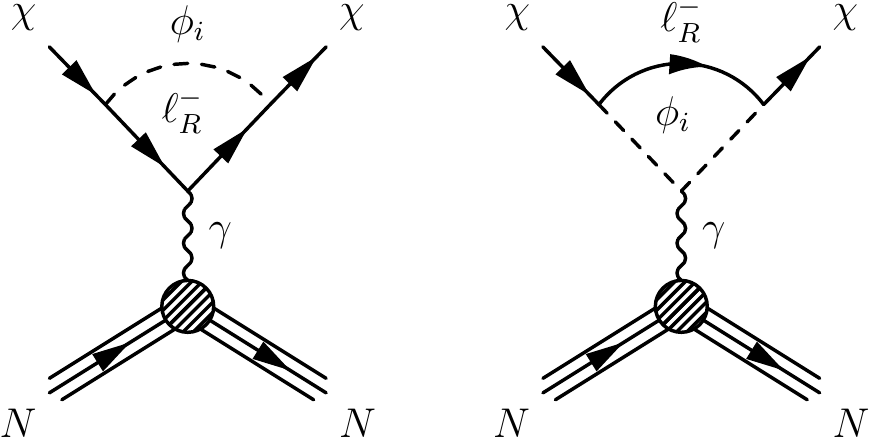}
  \caption{The dominant contribution to direct detection for Dirac DM in our model. For Majorana DM two more diagrams with crossed $\chi$ legs are present.
  }
  \label{fig:DD-diagram}
\end{figure}


Direct detection experiments place important constraints on our models.
In these experiments, dark matter in the neighbourhood of the earth may pass through the 
detector and interact with the nucleus of one of the target atoms.  
The energy deposited causes the emission of light, charge (electrons) and heat.  Direct detection 
experiments are typically sensitive to two of these three signals, and use them to place a limit on the rate of interactions seen in the target material. They then translate these into a bound on the DM-nucleon interaction cross-section, assuming a contact interaction.  However, as we will see, the models we are discussing do not have a contact interaction but instead dipole and anapole interactions.  As such, we will consider the expected interaction rate between our DM models and the target nuclei and compare them to the rates that can be probed in experiments.  The DM-nucleus scattering rate per unit target mass is given by
\begin{align} \label{eq:dRdEr}
\frac{dR}{dE_r} = \frac{\rho_0}{m_\chi m_N} \int_{v_{min}}^{\infty} v f_{MW}(\vec{v} + \vec{v_e}) \, \frac{d\sigma_{\chi N}}{dE_r} \,d^3 v \,,
\end{align}
where $\rho_0 \sim 0.3\GeV/\text{cm}^3$ is the local DM density, $m_N$ is the target nucleus mass, $v_{min} = \sqrt{E_r/2} (m_\chi + m_N)/(m_\chi m_N)$ is the minimal DM velocity required to give a recoil energy $E_r$ and $f_{MW}(\vec{v} + \vec{v_e})$ describes the DM velocity distribution in the rest frame of the detector.  The particle physics interactions are contained within $\sigma_{\chi N}$, which we now discuss.

Although the dark matter in our models is uncharged, it can interact with a nucleon, and hence the nucleus, at one-loop level. In our model, the dominant contribution to direct detection comes from the one-loop diagrams shown in~\cref{fig:DD-diagram}. Note that for Majorana DM there are two additional diagrams with crossed $\chi$ legs. 
The loop diagrams can be mapped onto effective DM-photon interactions, where 
the most general effective Lagrangian for our interaction is given by 
\begin{align}
\mathcal{L}_\text{eff} = \frac{d_M}{2} \bar{\chi} \sigma^{\mu\nu} \chi F_{\mu\nu} + \frac{d_E}{2} \bar{\chi} \sigma^{\mu\nu} \gamma^5 \chi F_{\mu\nu} + \mathcal{A} \bar{\chi} \gamma^{\mu} \gamma^5 \chi \partial^\nu F_{\mu\nu} \,,
\end{align}
where $d_M$ and $d_E$ are the magnetic and electric dipole moments and $\mathcal{A}$ denotes the anapole moment. We see that the dipole operator appears at dimension five whereas the anapole operator is dimension six.  For Majorana DM the magnetic and electric dipole moments are identically zero (which ultimately leads to a dramatically reduced rate for the Majorana models). The loop diagrams and their contribution to the dark matter-nucleon scattering cross-section for one coannihilation partner in the $n=1$ case have been computed in~\cite{Kopp:2014tsa, Agrawal:2011ze, Bai:2014osa, Ibarra:2015fqa}. Here we briefly summarise the results relevant for our work.  

%
%
\begin{figure}
  \begin{center}
  \begin{tabular}{cc}
    \includegraphics[width=0.5\columnwidth]{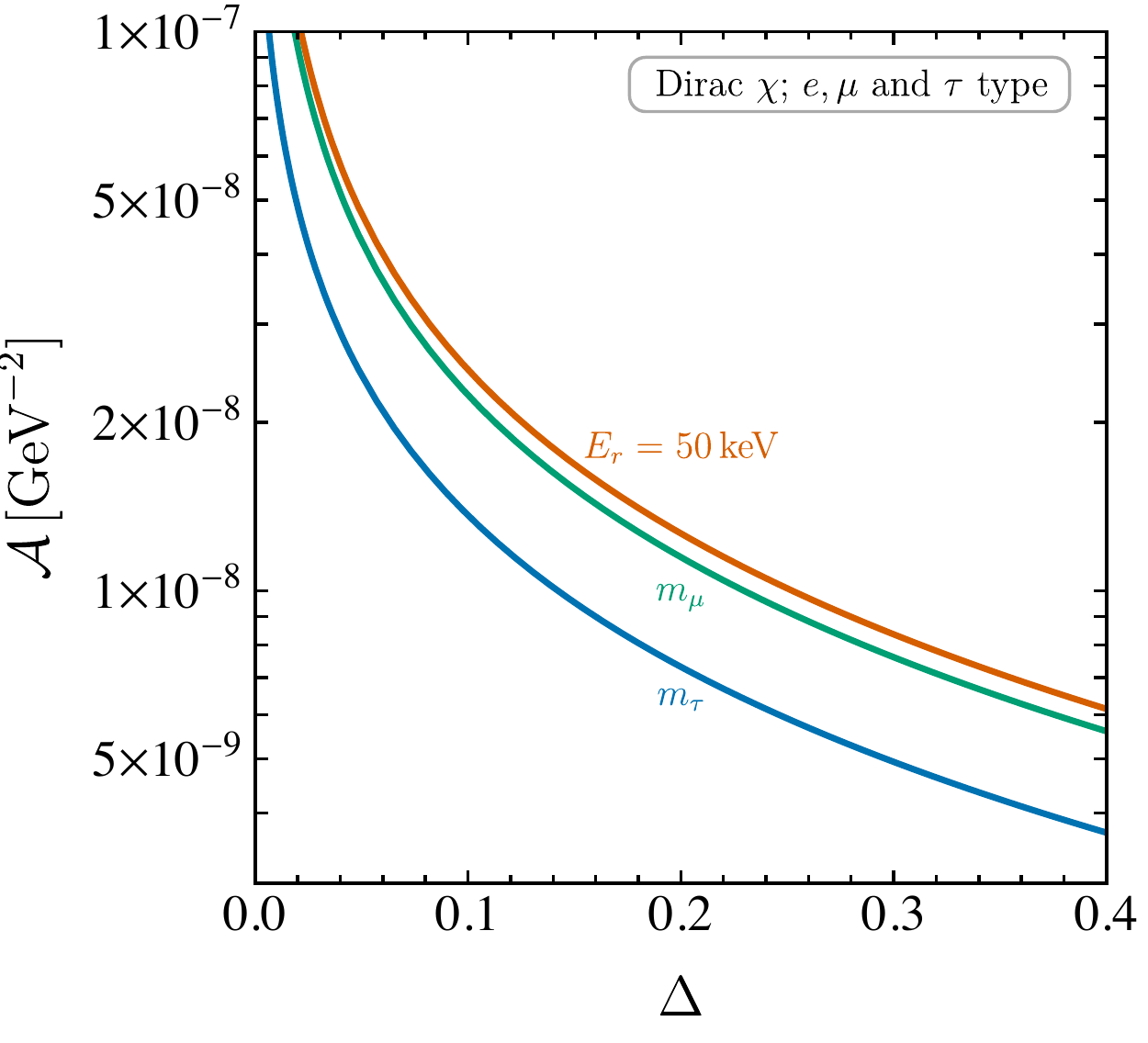}&
    \includegraphics[width=0.5\columnwidth]{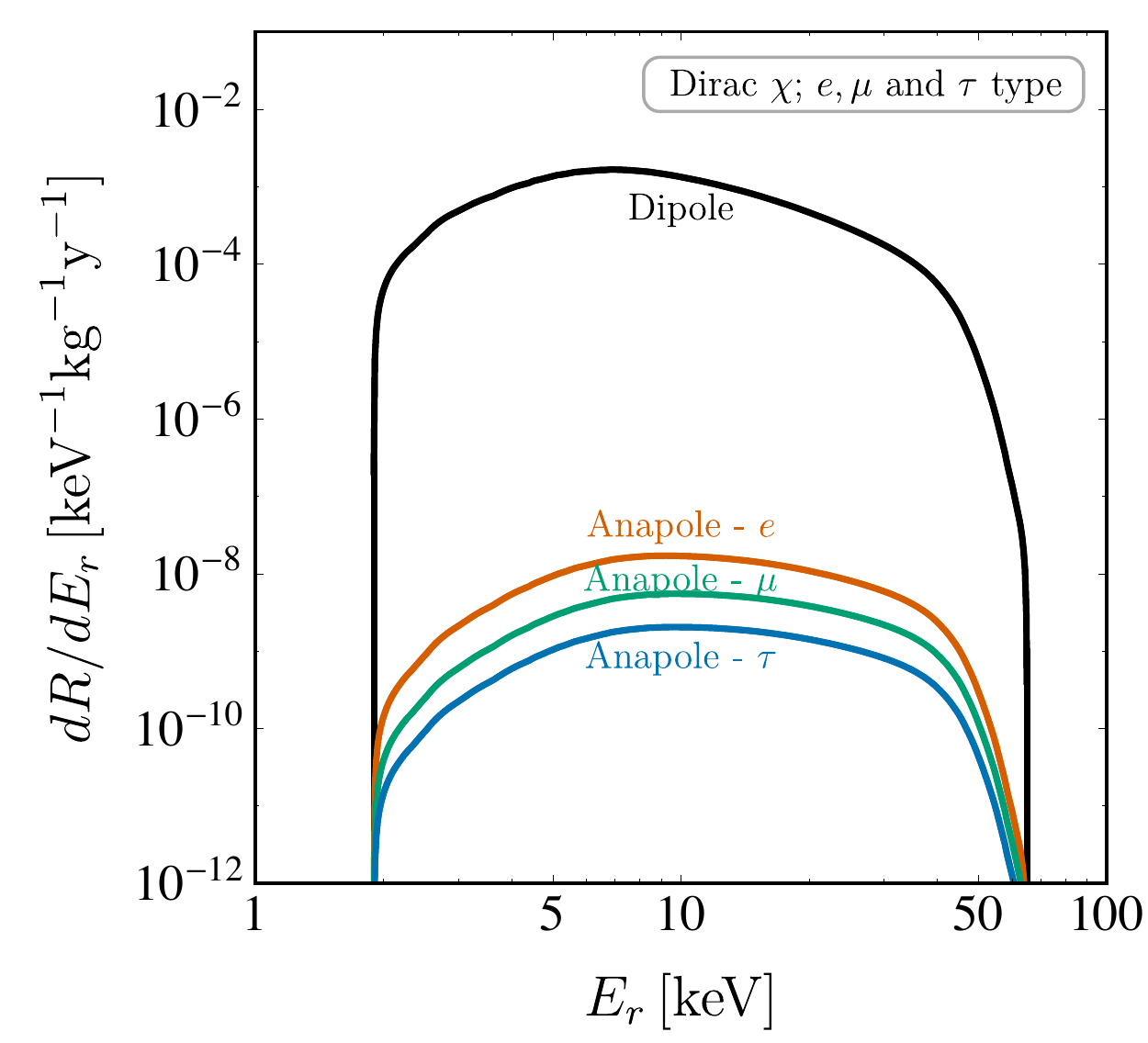}\\
  \end{tabular}
  \end{center}
  \caption{The anapole moment (left) for electrons (orange), muons (green) and taus (blue) running in the loop. For the electron case, we take a representative $E_r = 50\,$keV. The anapole moment is larger for smaller lepton masses. The differential rate (right) as a function of the recoil energy for a dipole moment (black) and anapole moment for electrons (orange), muons (green) and taus (blue), including the nuclear form factors and XENON1T efficiency (see text for details). We assume $m_\chi = 1\,$TeV and $\Delta=0.1$ and restrict $y_\chi$ to lie on the relic surface.}
  \label{fig:Anapole}
\end{figure}
%
%

For Majorana dark matter, the one-loop contribution to the anapole moment for the tau and muon models is given by
\begin{align} \label{eq:Anapole}
\mathcal{A}^{\mu,\tau}_\text{Maj} = - \frac{e \, n \, y_\chi^2}{96 \pi^2 m_\chi^2} \left[ \frac{3}{2} \log\frac{\mu}{\epsilon} - \frac{1 + 3\mu - 2\epsilon}{\sqrt{(\mu - 1 - \epsilon)^2 - 4\epsilon}} \, \text{tanh}^{-1} \left( \frac{\sqrt{(\mu - 1 - \epsilon)^2 - 4\epsilon}}{\mu - 1 + \epsilon}  \right) \right], |q^2| \ll m_\ell^2\,,
\end{align}
for momentum transfers $q$ much smaller than the lepton mass, $|q^2| \ll m_\ell^2$, and where $\mu =  (1+\Delta)^2$ and $\epsilon = m_\ell^2/m_\chi^2$. The factor $n$ accounts for the sum of diagrams when $n$ coannihilation partners are present. Since the momentum transfer is typically larger than the electron mass, we take the limit $|q^2| \gg m_\ell^2$ for the anapole moment of the electron models,
\begin{align} \label{eq:Anapole-electron}
\mathcal{A}^e_\text{Maj} = - \displaystyle \frac{e \, n \, y_\chi^2}{32 \pi^2 m_\chi^2} \left[ \frac{-10 + 12\log \left(\frac{\sqrt{|q^2|}}{m_\chi} \right) - (3+ 9\mu) \log(\mu - 1) - (3-9\mu) \log \mu)}{9(\mu - 1)} \right], |q^2| \gg m_\ell^2 \,.
\end{align}
In \cref{fig:Anapole} (left) we show the anapole moment as a function of $\Delta$ for electrons, muons and taus. For the electron case the anapole moment depends on the momentum transfer, which is given by $q = \sqrt{2 E_r m_N}$, and we show the moment for the exemplary value $E_r = 50\keV$. Assuming $m_\ell \approx 0$, the anapole moment for all models has a $\log$ divergence as $\Delta \rightarrow 0$, reflecting the fact that all particles in the loop can be on-shell simultaneously (when $m_\phi \approx m_\chi$).
The anapole moment has a further divergence which is regularised by either the lepton mass or the momentum transfer, which explains why it is larger for smaller lepton masses.  
Finally, the anapole moment tends to zero as $\Delta \rightarrow \infty$, since the coannihilation partners decouple in this limit.

For Dirac DM, the anapole moment is half as large as the moment for Majorana DM, \cref{eq:Anapole,eq:Anapole-electron},
\begin{align} \label{eq:Anapole-Dirac}
\mathcal{A}^{\ell}_\text{Dir} = \frac{1}{2}\mathcal{A}^{\ell}_\text{Maj}\,,
\end{align}
for $\ell \in \{ e, \mu, \tau\}$.  Additionally, it generates a dipole moment given by
\begin{align}
d_M = \displaystyle \frac{e \, n \, y_\chi^2}{32 \pi^2 m_\chi} \Bigg[- &1 + \frac{1}{2} (\epsilon - \mu) \log\frac{\epsilon}{\mu}  \\
- &\frac{(\mu - 1) (\mu - 2\epsilon)-\epsilon(3-\epsilon)}{\sqrt{(\mu - 1)^2 - 2\epsilon(\mu+1)+ \epsilon^2}} \, \text{tanh}^{-1} \left( \frac{\sqrt{(\mu - 1)^2 - 2\epsilon (\mu+1)+\epsilon^2}}{\mu - 1 + \epsilon}  \right)\Bigg]\, .
\end{align}
Since we are only interested in the limit where $m_\ell \ll m_\chi$, the above expression simplifies to
\begin{align}
d_M = \displaystyle \frac{e \, n \, y_\chi^2}{32 \pi^2 m_\chi} \left(\mu \log \frac{\mu}{\mu - 1} - 1 \right)\, ,
\end{align}
which is independent of the lepton mass. 

The anapole and dipole moments contribute to the differential DM-nucleus cross-section,
\begin{align} \label{eq:dsigmadErAna}
\frac{d\sigma_{\chi N}^{\text{Ana}}}{dE_r} &= 4 \alpha \, Z^2 \, \mathcal{A}^2  \, F_Z(E_r)^2 \left[2m_N - \left(1+\frac{m_N}{m_\chi} \right)^2 \frac{E_r}{v^2} \right] + 4 d_A^2 \mathcal{A}^2 \, F_s(E_r)^2 \left(\frac{J+1}{3J} \right) \frac{2 E_r m_N^2}{\pi v^2}\,, \\ 
\frac{d\sigma_{\chi N}^{\text{Dip}}}{dE_r} &= \frac{\alpha \, Z^2 \, d_M^2 }{2 m_N E_r} F_Z(E_r)^2 \left[2 m_N -\left(1+2\frac{m_N}{m_\chi} \right) \frac{E_r}{v^2} \right] + d_A^2 d_M^2 \, F_s(E_r)^2 \left(\frac{J+1}{3J} \right) \frac{m_N}{\pi v^2} \,, \label{eq:dsigmadErDip}
\end{align}
where the first term in both expressions corresponds to the spin-independent part (where the DM scatters on the nuclear charge $Z$), while the second term parameterises the spin-dependent interaction (where the DM scatters on the nuclear magnetic moment, $d_A$). Here, $\alpha$ is the fine-structure constant, $J$ and $m_N$ are the spin and the mass of the nucleus, respectively, $v$ is the velocity of the incoming DM particle and $E_r$ the recoil energy. Note that \cref{eq:dsigmadErAna} and \cref{eq:dsigmadErDip} are written for one nucleon isotope only and the spin-dependent and spin-independent parts have to be summed separately over the relevant isotopes. Each summand is weighted by the isotope abundance. In the present paper we focus on Xenon as the target material, with $Z=54$ and sum over the isotopes given in~\cite{Banks:2010eh, 8ad88931fb144c15b303a77d05f9ab6e}. 
For the nuclear charge and spin form factors we use~\cite{Banks:2010eh}
\begin{align}
F_Z(E_r) &= 3 \exp^{-q^2 s^2/2} \frac{ \sin(qr) - qr \cos(qr)}{(qr)^3}\,, \\ F_s(E_r) & = 
\begin{cases}\displaystyle{ \frac{\sin(qR_s)}{qR_s}}\,,& \text{for}\,\, qR_s < 2.55 \; \text{or} \; qR_s > 4.5\,,\\[14pt]
\displaystyle 0.217\,,&\text{otherwise}\,,
\end{cases}
\end{align}
where $q = \sqrt{2 E_r m_N}$, $s=1\,$fm, $r=\sqrt{R^2 - 5 s^2}$, $R=1.2 A^{1/3}\,$fm, $R_s= A^{1/3}\,$fm and $A$ is the nuclear mass number.


\begin{figure}
  \begin{center}
  \begin{tabular}{cc}
    \includegraphics[width=0.45\columnwidth]{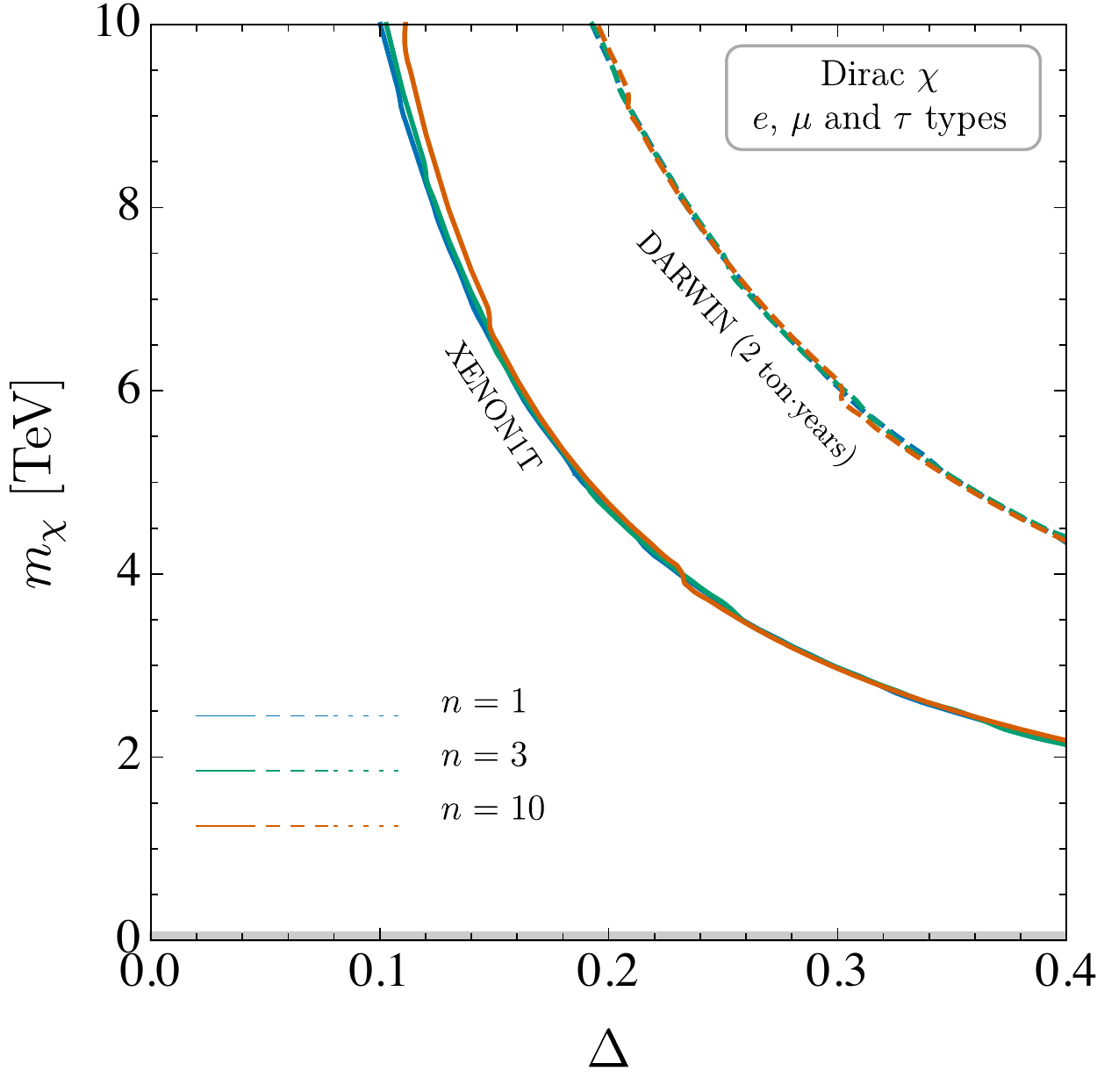}&
    \includegraphics[width=0.45\columnwidth]{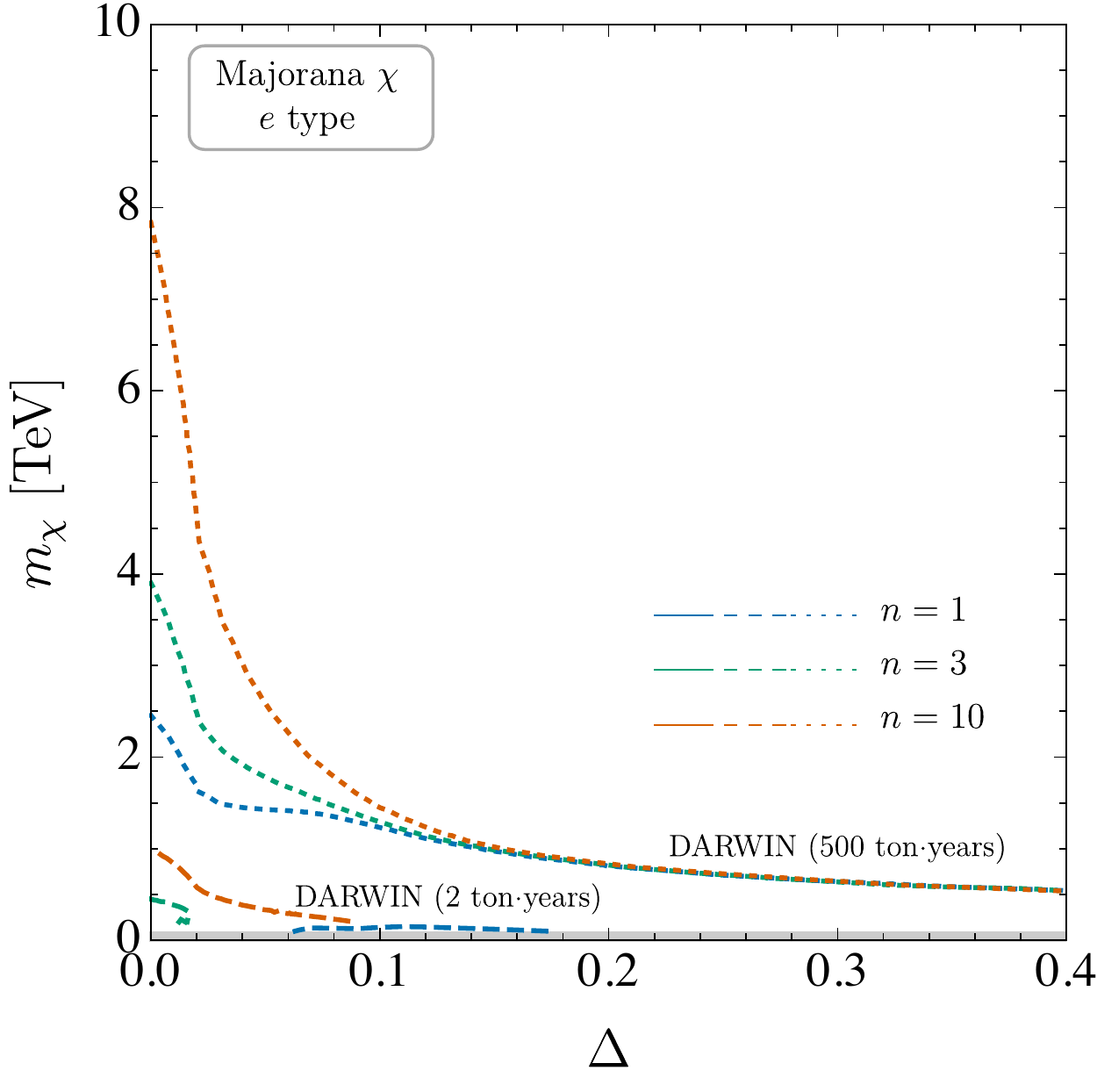}\\
    \includegraphics[width=0.45\columnwidth]{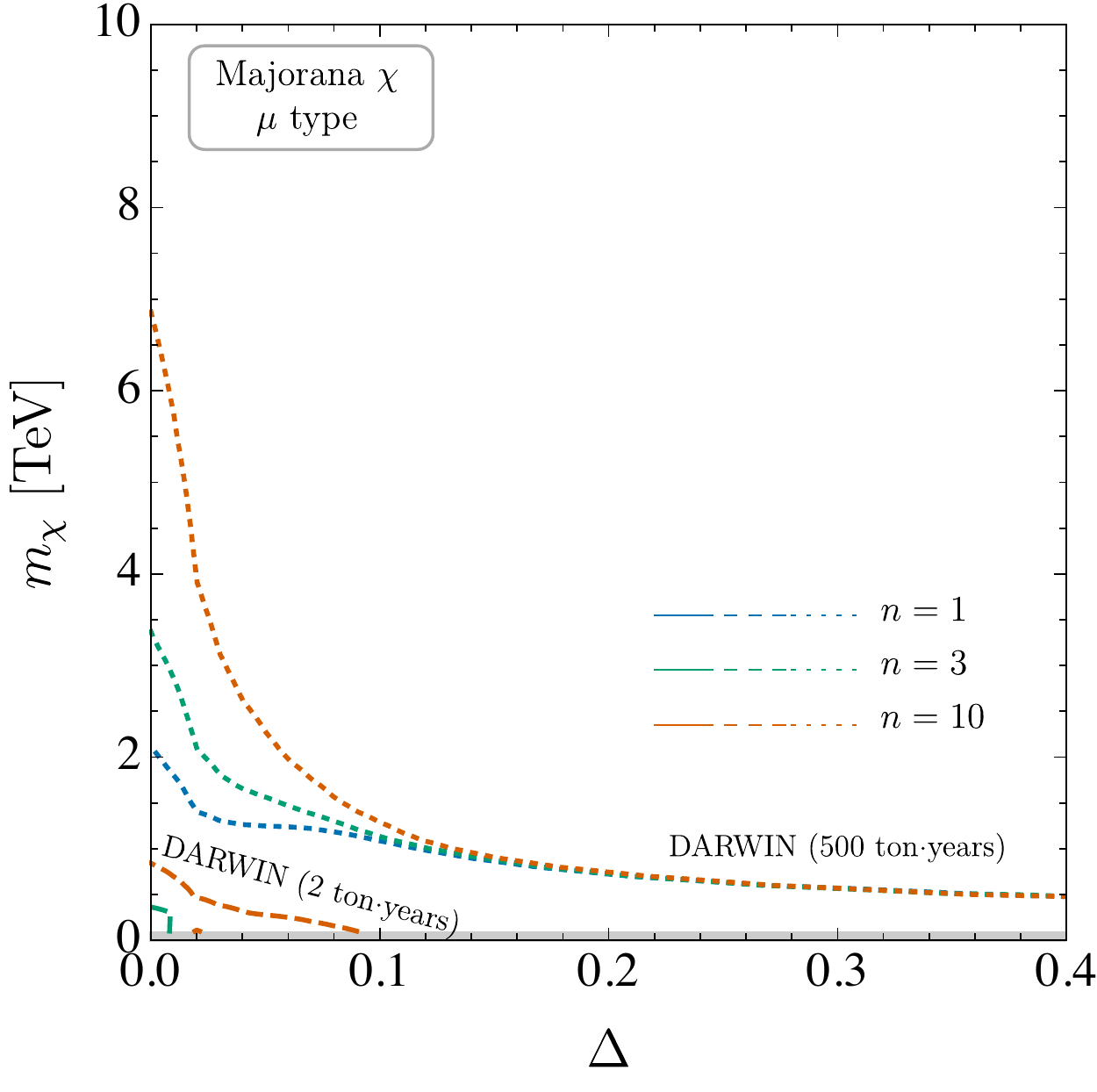}&
    \includegraphics[width=0.45\columnwidth]{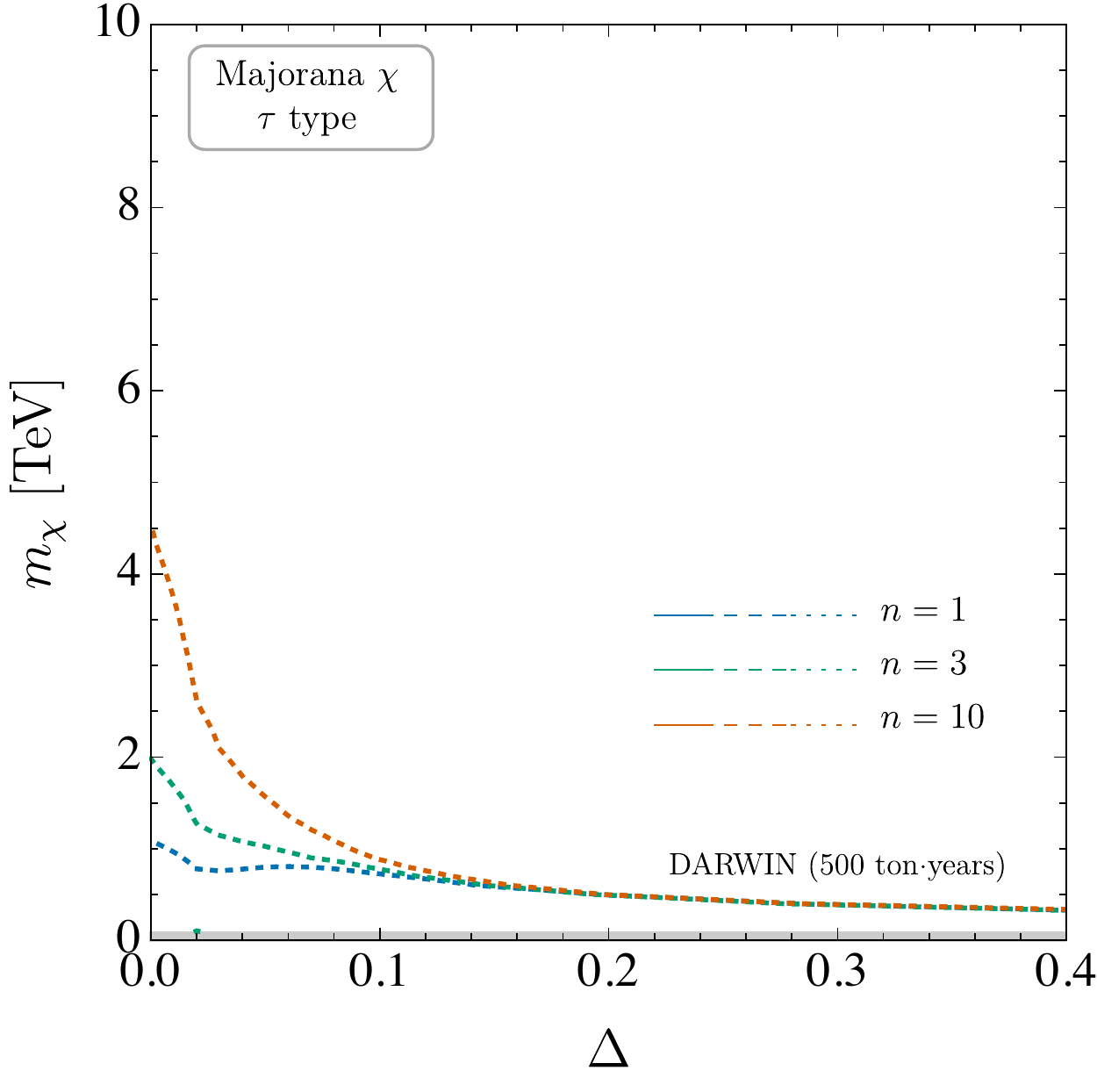}
 \end{tabular}
  \end{center}
  \caption{Current and future direct detection 90 \% C.L bounds on our models, where 
  solid lines are bounds from existing XENON1T data, while dashed and dotted 
  lines gives prospective DARWIN bounds from $2\, \text{ton}\cdot\text{years}$ and $500\, \text{ton}\cdot\text{years}$ exposure, respectively, and
  blue, green and red lines refer to 1, 3 and 10 coannihilation partners, respectively:
 (top-left) constraints for Dirac DM, where the constraints are independent 
  of the lepton flavour;
  (top-right) Majorana DM, electron-type;
  (bottom-left) Majorana DM, $\mu$-type;
  (bottom-right) Majorana DM, $\tau$-type. The excluded regions are below and to the left of the curves.  We do not include the 
  region $m_\chi < 0.1\TeV$ (grey) in our analysis.}
  \label{fig:model-1-sigma-DD}
\end{figure}


Turning now to the astrophysical quantities in \cref{eq:dRdEr}, we assume the standard halo model (an isotropic and isothermal sphere) for the DM distribution, which leads to a Maxwell-Boltzmann velocity distribution in the galactic frame smoothly truncated at the galaxy escape velocity, $v_{\text{esc}} = 550\,$km~\cite{McCabe:2010zh}, given by
\begin{align}
f_{MW}(\vec{v}) = 
\begin{cases}
\displaystyle{ \frac{1}{N} \left[\exp\left(\frac{\vec{v}^2}{v_0^2} \right) -\exp\left(\frac{v_{\text{esc}}^2}{v_0^2}\right) \right]} \,,& v < v_{\text{esc}}\,,\\[14pt]
\displaystyle{0} \,,&  v > v_{\text{esc}}\,,
\end{cases}
\end{align}
where $N$ is a normalisation constant and $v_0 = 220\,$km/s~\cite{McCabe:2010zh} describes the velocity of the sun about the centre of the galaxy. Since the velocity distribution is given in the rest frame of the Milky Way, we use a Galilean transformation to move to the rest frame of the detector, $f_{MW}(\vec{v} + \vec{v_e})$, where $\vec{v_e}$ is the velocity of the Earth relative to the galactic centre. 
For simplicity, we neglect the velocity of the Earth with respect to the sun and take $v_\text{e} = 220\,$km/s. Note that all terms in~\cref{eq:dsigmadErAna} and \cref{eq:dsigmadErDip} are either independent of the velocity or proportional to $1/v^2$~\cite{Kahlhoefer:2016eds}. The integral for velocity dependent terms can be solved analytically (see for instance~\cite{Savage:2006qr,McCabe:2010zh}), while the integral with constant terms can be solved analytically, once numerical values for $v_\text{esc}$, $v_0$ and $v_\text{e}$ have been provided.

We now have all the ingredients necessary to calculate $dR/dE_r$,~\cref{eq:dRdEr}, and we show an example spectrum in~\cref{fig:Anapole} (right).  The differential rate for the dipole moment dominates over the rate for the anapole moment by $\sim5$ orders of magnitude, reflecting the fact that the dipole operator is dimension five whereas the anapole operator is dimension six. The figure assumes $m_\chi = 1\,$TeV and $\Delta=0.1$, while $y_\chi$ is restricted to lie on the relic surface. We include the effect of the form factors and XENON1T efficiency (see below).

We are now ready to compute bounds from existing XENON1T data and derive projections for future experiments, such as DARWIN, using the statistical procedure outlined in~\cite{Fairbairn:2008gz}. 
The current XENON1T exposure is $1\,\text{ton}\cdot\text{years}$~\cite{Aprile:2018dbl}, while DARWIN aims at exposures of $2\,$ and $500\, \text{ton}\cdot\text{years}$~\cite{Aalbers:2016jon}. We take into account the efficiency of nuclear recoil event detection in XENON1T given in~\cite{Aprile:2018dbl} and assume the same efficiency profile for future experiments. 

Figure \ref{fig:model-1-sigma-DD} shows current XENON1T (solid lines) and projected DARWIN (dashed and dotted lines) bounds for Dirac (top-left) and Majorana (top-right and bottom) DM as a function of $\Delta$ and $m_\chi$. The line colours correspond to one, three and ten coannihilation partners. As shown in \cref{fig:Anapole} (right), the dipole contribution dominates over the anapole contribution which can thus be neglected for Dirac DM. Since the dependence of the dipole on the lepton mass is negligible, the bounds 
for Dirac DM are the same for all leptons. 
Current XENON1T results (solid lines) exclude Dirac DM masses at around $2\TeV$ for large values of $\Delta$, but up to $10\,$TeV in the coannihilation region, $\Delta \lesssim 0.1$. 
XENON1T bounds are slightly more stringent for models with more coannihilation partners. An exposure of $2\,\text{ton}\cdot\text{years}$ at DARWIN (dashed line) can exclude masses up to $4\,$TeV for large $\Delta$ and $10\,$TeV for $\Delta \lesssim 0.2$. With the nominal exposure of $500\,\text{ton}\cdot\text{years}$, this region of parameter space for Dirac DM will be probed completely. 

Majorana DM contributes only to the anapole moment, which depends on the lepton mass. The top-right, bottom-left and 
bottom-right panels show the bounds for Majorana DM coannihilating with an $e, \mu$ and $\tau$-type partner, respectively. The bounds are significantly weaker than for Dirac DM, and XENON1T does not currently constrain any of the parameter space.
An exposure of $2\,\text{ton}\cdot\text{years}$ can access masses up to $1\,$TeV and $\Delta < 0.1$ for the model with ten electron-type coannihilation partners, but is barely sensitive to the models with fewer partners or models that are muon or tau type. 
The bounds are strongest for electrons, since the anapole moment is larger for smaller lepton masses. 
An exposure of $500\,\text{ton}\cdot\text{years}$ can exclude Majorana DM masses $\sim 0.5\TeV$ for all $\Delta$, and 
up to a maximum of $8\TeV$ in the coannihilation region $\Delta < 0.1$.

Note that in addition to generating a magnetic dipole moment for $\chi$, our models also contribute to the 
lepton magnetic dipole moments.  Using the results derived for $n=1$ in~\cite{Fukushima:2014yia,Kopp:2014tsa}, we find that the current experimental and theoretical uncertainties are three to four orders of magnitude too large to set bounds on our parameter space for all electron and muon models, and more than six orders of magnitude too large for the tau models.

\section{Indirect detection}
\label{sec:indirectdetection}

We now consider the bounds from indirect detection.  In regions of high dark matter 
density such as the galactic centre or dwarf galaxies, $\chi$ and $\bar{\chi}$ can annihilate and form pairs of high-energy 
opposite-sign leptons at tree-level, which may decay.  In the process, photons, positrons and anti-protons may be produced, along with other SM particles.  At one-loop level, pairs of mono-energetic photons 
may also be produced.
Current and future 
experiments search for excesses of these particles above the expected 
astrophysical backgrounds, which may be interpreted as a signal of dark matter.  
We first focus on continuum photon searches and consider the constraints placed on our model by Fermi-LAT, HESS and the region of parameter space that will be probed by CTA.  
These will turn out to be the most important constraints on our models.

Indirect search strategies aim to maximise the potential signal and minimise the background.  
The two main targets commonly considered are the Galactic Centre (GC) and dwarf spheroidal galaxies 
(dSphs).  
Dark matter density is highest at the GC, and the signal from the GC is expected to be several 
orders of magnitude larger than that from dwarf galaxies.  However, the GC suffers from 
both large background sources of gamma rays and significant uncertainty in the local DM density.  
For this reason, the GC is considered a likely target to provide the first measurement of 
a DM signal, but subsequent measurement of a signal in dwarf galaxies will usually be needed 
to make the claim that the signal is unambiguously due to DM.  
The DM density in the Milky Way is well measured away from the centre, but is poorly known 
in the inner $\sim 2\,\text{kpc}$.  DM models are either cusped, e.g., Einasto profile and 
Navarro-Frenk-White (NFW) profile, or cored, e.g., Burkett profile.  
Although $N$-body simulations suggest a cuspy profile, interactions with baryonic matter 
could lead to a cored profile (where the dark matter density is constant below $\sim 2\,\text{kpc}$).
HESS have developed strategies 
for both situations~\cite{Abdallah:2016ygi, Rinchiuso:2017pcx, HESS:2015cda}, 
but their sensitivity is much higher 
if the distribution is cusped.  
In our GC analysis we 
assume the profile is cusped.  If the distribution is 
in fact cored, neither CTA nor HESS can place any limits on the models we consider. 
We also determine weaker but more robust bounds obtained from observing dwarf galaxies. 

Annihilating dark matter can produce photons both via direct emission (primary) and by secondary 
production via Inverse Compton Scattering (ICS) of $\ell^\pm$ on the ambient photon background.
The differential flux of photons from direct emission, in a solid angle $\Delta \Omega$, is given by
\begin{align}
\frac{d \Phi_\gamma(E_\gamma, \Delta \Omega)}{d E_\gamma} =&\,
\frac{\langle \sigma v \rangle}{8 \pi m_\chi^2} J(\Delta \Omega) \frac{dN_\gamma(E_\gamma)}{d E_\gamma}
\end{align}
where $\langle \sigma v \rangle$ is the thermally averaged dark matter annihilation cross-section, 
$N_\gamma$ is the photon yield per annihilation
and the $J$-factor integrates the square of the dark matter density along the line of sight 
over the solid angle $\Delta \Omega$.
The spectrum of secondary photons produced via ICS can be calculated by convolving the 
$e^\pm$ injection spectrum with a halo function for the inverse Compton process~\cite{Lefranc:2015pza}.  
For electrons and muons, a non-negligible photon flux is generated from ICS.  However, the 
precise contribution depends strongly on the assumed halo function and 
on the spatial region considered in the analysis.
For the HESS limits, we use the results in~\cite{Abdallah:2016ygi} which ignore the contribution from ICS.  
For the prospective CTA limits we use the results in~\cite{Lefranc:2015pza} which include this contribution, 
strengthening these limits.

All of the dependence on the particle physics model is contained within the $\langle \sigma v \rangle$ 
term, the rest being dependent on astrophysical quantities.
Since DM in the Milky Way is travelling relatively slowly ($v \approx 10^{-3} c$), we 
can take the non-relativistic limit for the annihilation cross-section and assume that all 
dark matter particles are travelling with the same speed.
For the Dirac case of our model, 
taking the limit $m_\ell \ll m_\chi$,
the annihilation cross-section is
\begin{align}
\langle \sigma v \rangle \approx&\, \frac{y_\chi^4 n^2 m_\chi^2}{32 \pi (m_\chi^2 + m_\phi^2)^2}
\label{eq:cs-ann-dir}
\intertext{which agrees with~\cite{Carpenter:2016thc} for $n=1$.  For the Majorana case we have}
\langle \sigma v \rangle \approx&\, \frac{y_\chi^4 n^2 v^2 m_\chi^2 (m_\chi^4 + m_\phi^4)}{48 \pi (m_\chi^2 + m_\phi^2)^4}.
\label{eq:cs-ann-maj}
\end{align}
where $v$ is the relative velocity between the two annihilating dark matter particles, which agrees 
with~\cite{Bai:2013iqa, Garny:2015wea} for $n=1$.  
We see that in the Majorana case, the cross-section is velocity suppressed, as is well known.  
This suppression means that indirect detection is a poor probe of this case.

The strongest indirect detection constraints come from measurements of continuum photons from 
the galactic centre, assuming a cusped halo profile.
To calculate the current limits from HESS, we use the 95 \% C.L. upper limits 
 for $\chi\bar{\chi}\rightarrow\tau^+\tau^-$ and 
$\chi\bar{\chi}\rightarrow\mu^+\mu^-$ presented in~\cite{Abdallah:2016ygi}.\footnote{Fermi-LAT also 
provide limits at the GC~\cite{TheFermi-LAT:2017vmf}. Since they are weaker than HESS limits, we do not show them here.}  
To produce their limit, a difference measurement is preformed 
between a region near the GC and another region further away, and a 2D 
binned Poisson maximum likelihood analysis is used to distinguish signal from 
background using both spatial and spectral information.  As mentioned above,  
photons produced by ICS are ignored.
To estimate the 
sensitivities to the electron model, we rescale the muon limit using the integrated 
flux of prompt photons between $160\GeV$ (the threshold for HESS) and $m_\chi$, 
using the results presented in~\cite{Cirelli:2010xx}.  We use the muon limit since 
the prompt spectrum from muons is closest to that from electrons.
To calculate the prospective limit from CTA, we use the  95 \% C.L. sensitivity limits  on 
$\chi\bar{\chi}\rightarrow\tau^+\tau^-$, 
$\chi\bar{\chi}\rightarrow\mu^+\mu^-$ and 
$\chi\bar{\chi}\rightarrow e^+e^-$ presented in~\cite{Lefranc:2015pza}.  
These limits assume a cuspy Einasto profile and 500 hours of observation, and use
a likelihood ratio statistical test to derive the 95\% C.L. sensitivity limits.  
These limits include photons produced via ICS and 
ignore systematic uncertainties in the datasets and the galactic diffuse emission. 

As mentioned above, limits derived from observations of dwarf galaxies are more robust as they 
do not depend on assumptions about the dark matter halo in the centre of the Milky Way.
We consider 95 \% C.L. upper limits set by Fermi-LAT on $\chi\bar{\chi}\rightarrow\tau^+\tau^-$ and 
$\chi\bar{\chi}\rightarrow\mu^+\mu^-$~\cite{Ahnen:2016qkx}, and rescale the muon limit to electrons using the integrated flux of prompt photons.
Neither the HESS~\cite{Abramowski:2014tra} nor CTA~\cite{Carr:2015hta} constraints 
from dwarf galaxies are large enough to place any constraints on our models.


\begin{figure}
  \begin{center}
  \begin{tabular}{ccc}
    \includegraphics[width=0.325\columnwidth]{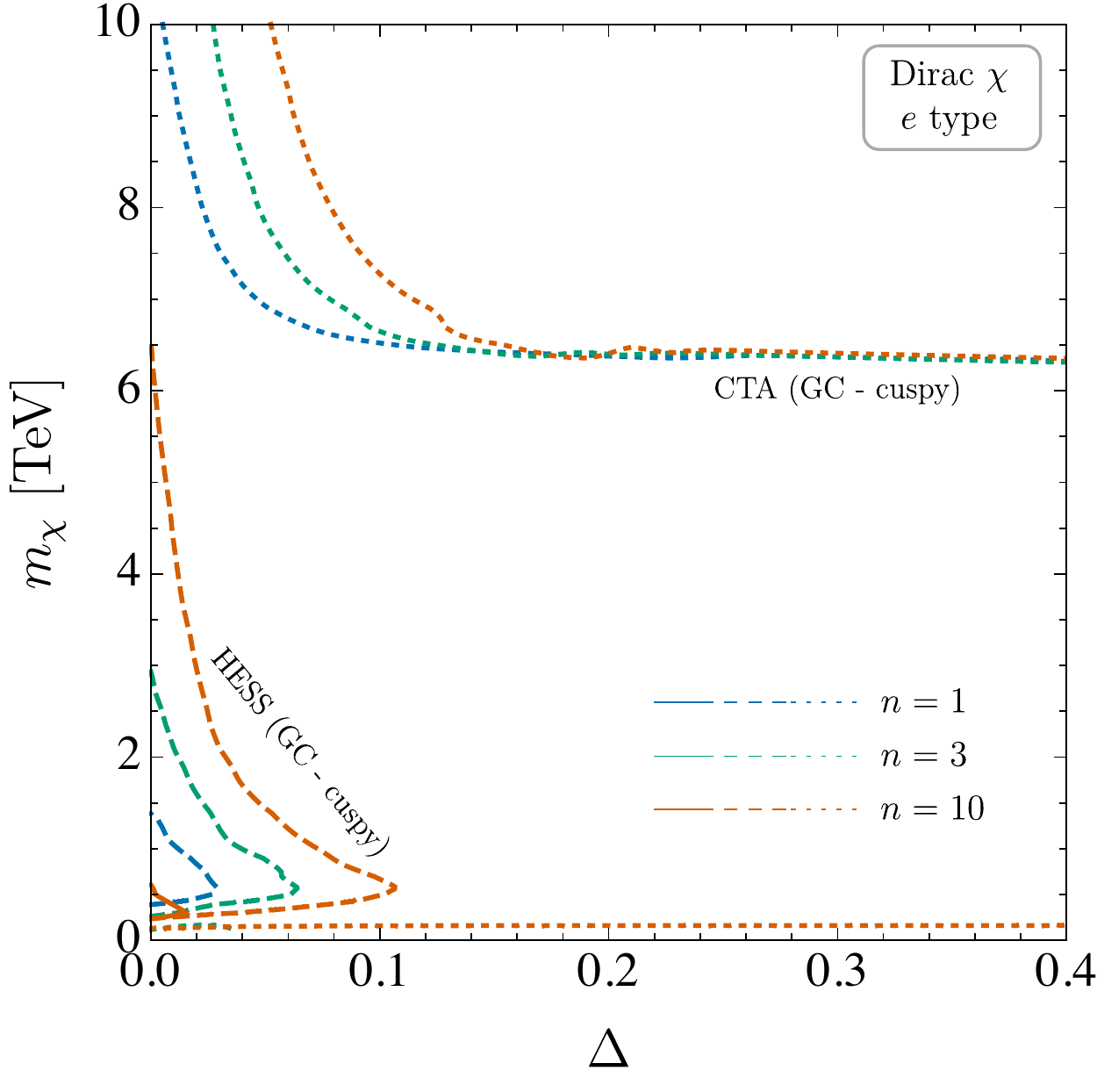}&
     \includegraphics[width=0.325\columnwidth]{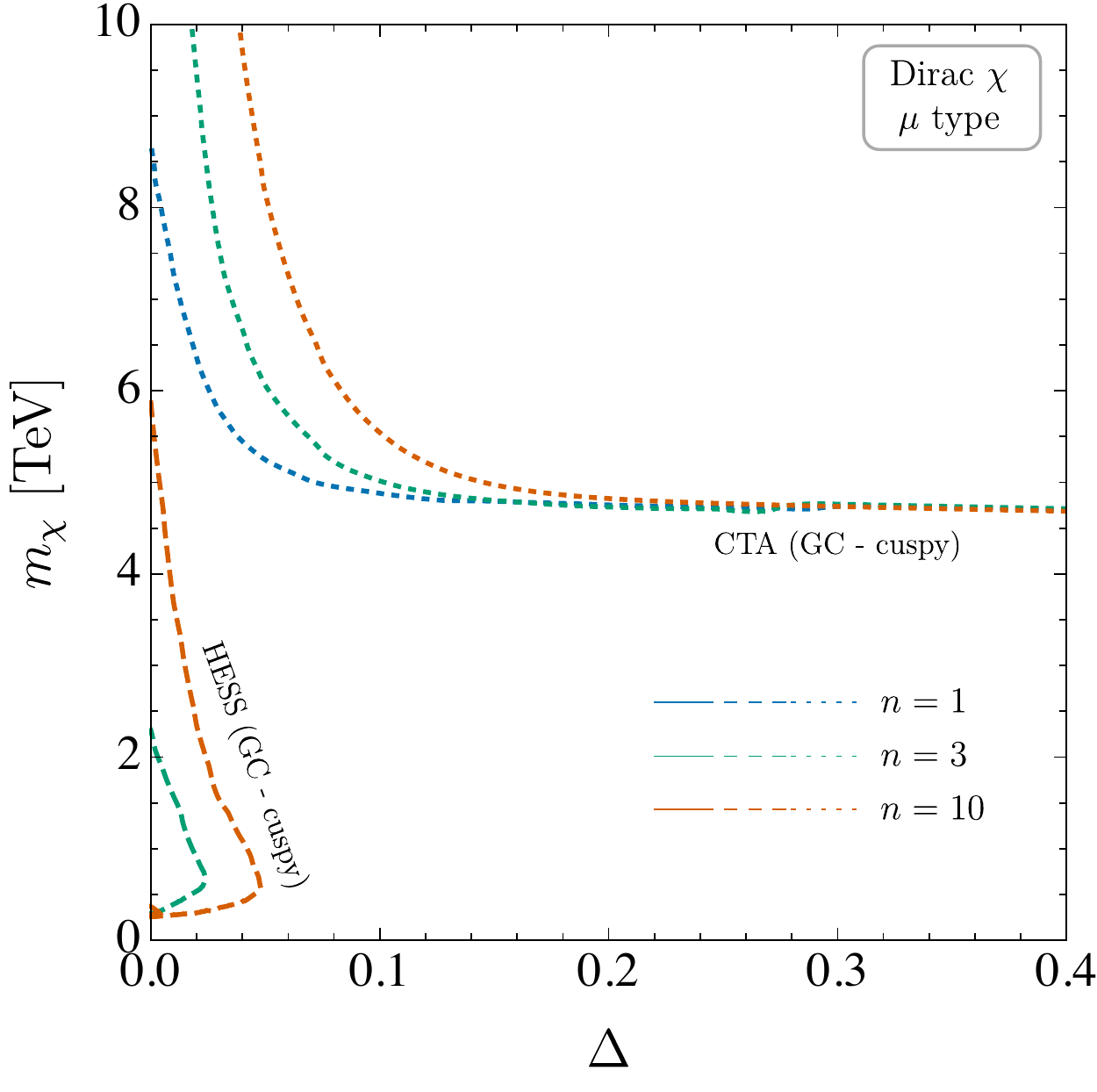}&
    \includegraphics[width=0.325\columnwidth]{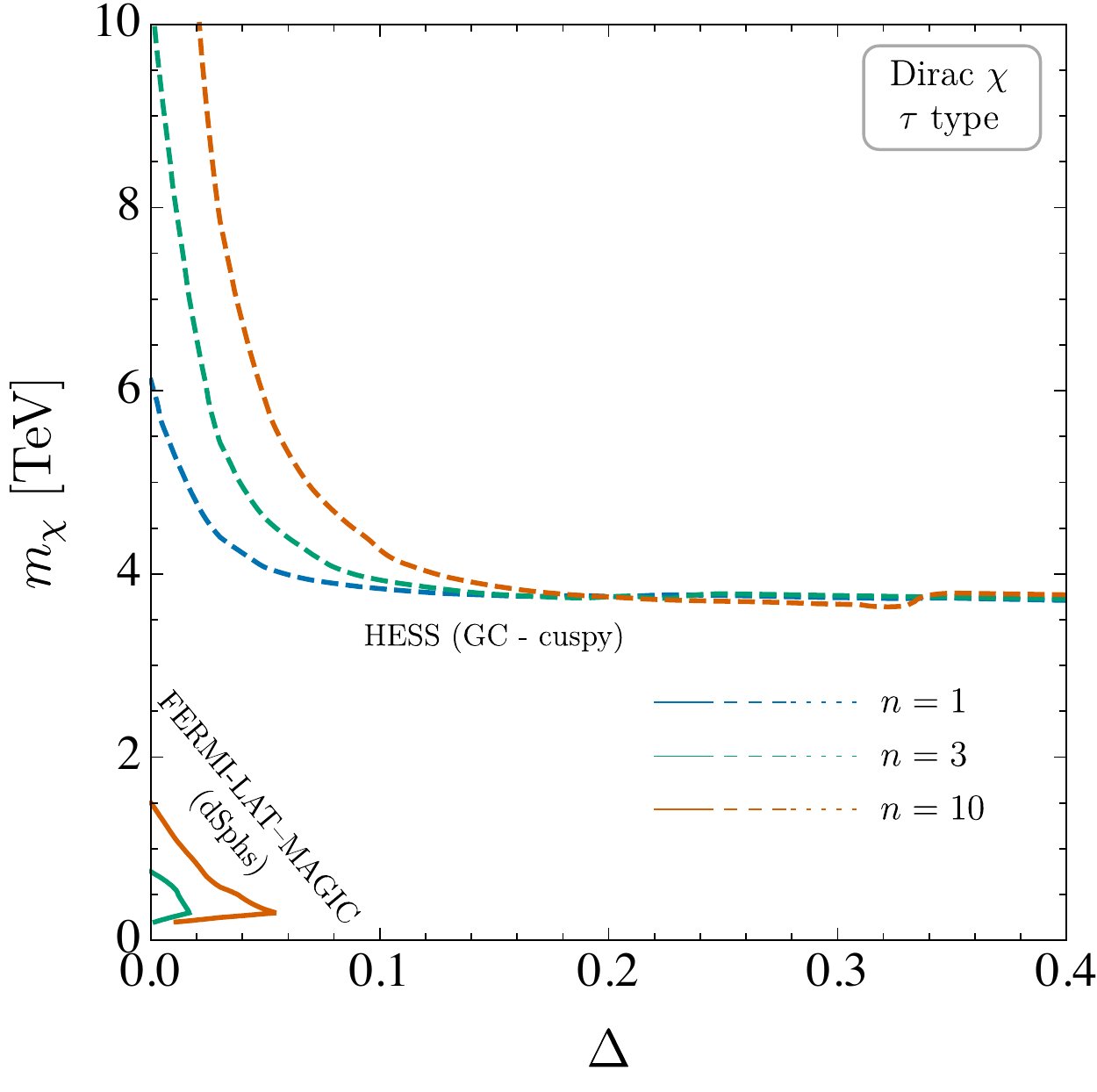}
 \end{tabular}
  \end{center}
  \caption{
Current indirect detection 95 \% C.L. bounds from Fermi-LAT -- MAGIC, HESS and CTA for
Dirac $\chi$ and $n$ right-handed electron (left), muon (middle) and tau (right) type coannihilation partners. 
The Fermi-LAT results are obtained from observations of dwarf galaxies and so 
do not depend on assumptions about the dark matter halo in the centre of the Milky Way.
The HESS and CTA bounds both rely on the assumption of a cusped (Einasto) dark matter halo.  
For HESS, photons produced via ICS are ignored, so the electron and muon bounds are conservative in this sense.  The HESS constraints on the electron model are estimated by scaling the muon bound by the ratio of prompt gamma rays.  For the CTA bounds, photons produced via ICS are taken into account.  The excluded regions are to the left of the curves.}
  \label{fig:ID-limits}
\end{figure}


In \cref{fig:ID-limits} we show current and prospective limits from HESS and CTA 
on our models.  Only the models with Dirac dark matter are shown as even in the most 
optimistic scenarios, CTA cannot probe any of the parameter space for Majorana dark matter.  
This is due to the velocity suppression in the annihilation cross-section, seen in \cref{eq:cs-ann-maj}.  
As mentioned above, if a cored dark matter profile is assumed, HESS and CTA place no limits on 
any of the parameter space of any of the models discussed in this paper.

In \cref{fig:ID-limits} (left) we see the constraints on the  electron type models for $n = 1,3,10$.  
HESS excludes the region of low dark matter mass and small $\Delta$.  
The CTA limits are significantly stronger and probe the electron type models up to $m_\chi \sim 7\TeV$ for large $\Delta$, and even higher for 
$\Delta \lesssim 0.1$.  
On dimensional grounds, the strongest constraints are always expected at low dark matter masses, 
since this is the characteristic scale of dark matter annihilation and $\langle \sigma v \rangle \sim m_\chi^{-2}$.  
At large $\Delta$, the diagram responsible for the indirect detection signal 
is the same as that responsible for dark matter freeze-out, so the variation in $n$ and $\Delta$ is cancelled by variation in $y_\chi$.  
However, for $\Delta \lesssim 0.1$, coannihilation diagrams contribute significantly to freeze-out  
causing $y_\chi$ to increase as $\Delta$ decreases, leading to strong constraints. The $n$ dependence also does not cancel in this region due to the interplay between numerator and the denominator in the coannihilation formula, \cref{eq:sigeff}.  This means that stronger constraints are seen with models with more coannihilation partners.  
In \cref{fig:ID-limits} (middle) we show the limits on the  muon type model.  We see that the results are broadly 
similar to the electron type models, but the region of parameter space which can be probed is slightly smaller.  
This is because muons produce fewer primary and ICS photons than electrons~\cite{Lefranc:2015pza}.  
In \cref{fig:ID-limits} (right) we show the limits on the tau type model.  Here the limits are significantly stronger.  
HESS can probe the parameter space up to $m_\chi \sim 4\TeV$ for any $\Delta$ while CTA probes the whole 
parameter space.  This is due to the large number of primary photons produced by taus. 
Figure~\ref{fig:ID-limits} also shows the limits from Fermi-LAT -- MAGIC observations of dwarf galaxies on our 
Dirac dark matter models. 
The bounds are weak and only constrain the 
$n = 10$ electron model ($m_\chi < 0.5\TeV$, $\Delta < 0.02$),
the $n = 3$ tau model ($m_\chi < 0.75\TeV$, $\Delta < 0.02$)
and the $n = 10$ tau model ($m_\chi < 1.5\TeV$, $\Delta < 0.06$).
However, these constraints do not depend on any assumptions 
about the DM halo profile or ICS and are therefore more 
robust than those from the galactic centre. 
Future CTA bounds for dwarf galaxies~\cite{Lefranc:2016dgx} are 
too weak to place any constraints on any of our models.
Furthermore, all the searches we consider are too weak to constrain 
any of the Majorana DM models. 
If CTA were strengthened by a factor of $\approx10$, it would begin to probe this 
parameter space~\cite{Garny:2015wea}.

The lepton pairs produced in the DM annihilation can also lead to primary or secondary positrons and anti-protons. Experimental limits on these final states~\cite{Accardo:2014lma,Aguilar:2014mma,Aguilar:2016kjl} extend only to DM masses of $\sim0.5\TeV$. Since we are primarily interested in heavier DM, we do not show these bounds in detail. Although AMS-02 may see an excess in the positron fraction~\cite{Accardo:2014lma}, there are many uncertainties in determining the background and a detailed analysis is beyond the scope of this paper. A preliminary study shows, however, that these positron bounds are very weak and only 
constrain the Dirac models with ten coannihilation partners at $m_\chi < 0.5\TeV$ and $\Delta < 0.01$.  
Furthermore, this model can only produce a signal larger than the astrophysical background, 
at any positron energy, for $m_\chi < 0.5\TeV$ or $\Delta < 0.01$.

Finally, we mention mono-energetic photons.  These may be produced via a one-loop diagram, with $\phi$ and $\ell$ running in the loop.  We find that the loop suppression is too large for Fermi-LAT, HESS or CTA observations~\cite{Lefranc:2016fgn} to provide any constraints.  Internal bremsstrahlung also produces a sharp feature in the gamma ray spectrum.  Although this process is not velocity suppressed, even for Majorana dark matter, the cross-section is too small for the experiments we consider to constrain the models, 
due to phase-space suppression.  For a detailed discussion of these gamma-ray constraints, see~\cite{Garny:2015wea}.

\section{Collider Constraints}
\label{sec:collider}


\begin{figure}
  \centering
  \raisebox{0.5\height}{\includegraphics[width=0.3\textwidth]{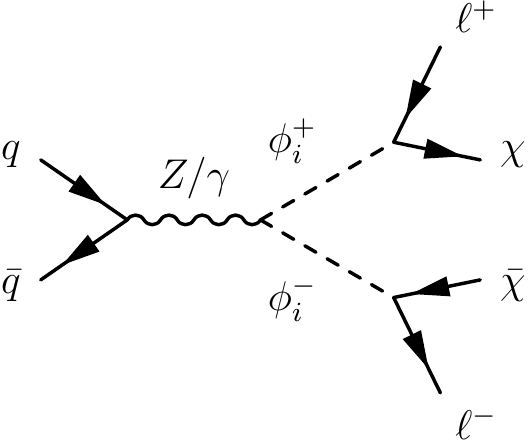}}
  \hspace{1.5cm}
   \includegraphics[width=0.5\columnwidth]{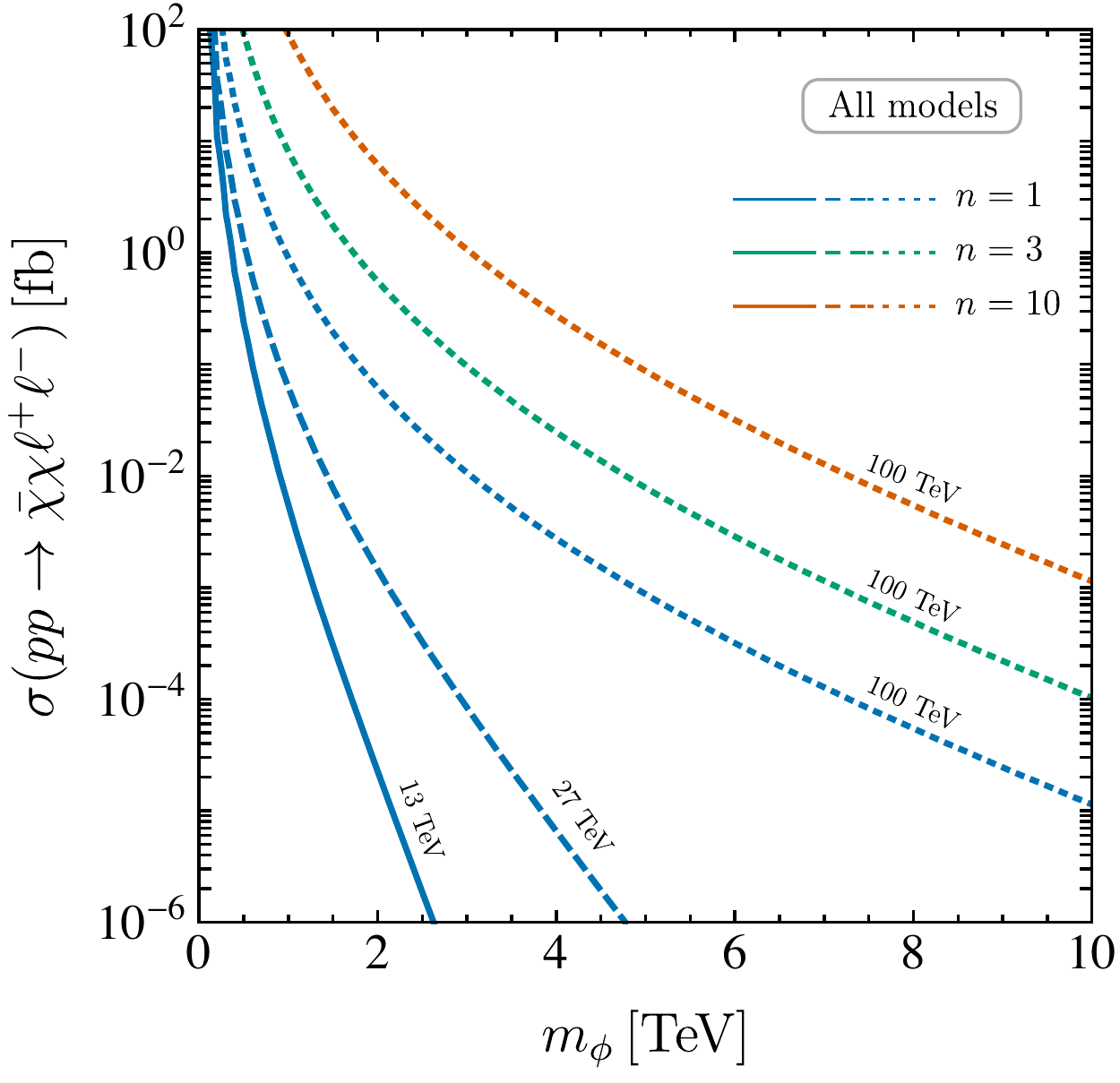}
  \caption{The leading order partonic process contributing to $p p \to \phi^+ \phi^- \to \bar \chi \chi \ell^+ \ell^-$ (left) and 
  its cross-section at $13\TeV$, $27\TeV$ and $100\TeV$ including a $K$-factor of 2 (right).
  }
  \label{fig:collider-diagrams}
\end{figure}


It is challenging to search for our dark matter models directly at a hadron collider, since the dark matter is a gauge singlet which only couples to leptons. The coannihilation partner, $\phi_i^\pm$, however is a charged scalar of similar mass.  It will be pair produced in the process $p p \to \phi_i^+ \phi_i^-$ with a subsequent decay of $\phi_i^\pm$ to a lepton, $\ell$, and $\chi$, depicted in \cref{fig:collider-diagrams} (left), where BR$(\phi_i^\pm \to \chi \ell^\pm)=1$. We focus on final states containing two opposite-sign same-flavour leptons and missing energy. As $\tau$ reconstruction at future colliders is particularly challenging to model, we do not provide collider limits for the $\tau$ models. However, we can assume that the collider reach on $\tau$ models will be somewhat worse than the limits on the models involving electrons and muons.  For the tables and figures in this section we focus on the muon type model, and provide the limits for the electron type models in \cref{sec:appColliderE}.

Since we are interested in multi-TeV dark matter, the LHC at $13\TeV$ only provides weak constraints.  
E.g.,~\cite{Aaboud:2018jiw} excludes our $n=1$ models only for 
$m_\chi < 0.3\TeV$.
We therefore present sensitivity projections for the HE-LHC with $\sqrt{s} = 27\TeV$ assuming an integrated luminosity of $15\,$ab$^{-1}$~\cite{HELHC} and for the FCC-hh with $\sqrt{s} =100\TeV$ and $20\,$ab$^{-1}$~\cite{Hinchliffe:2015qma}.
We estimate the sensitivity of future colliders to our models by adapting the analysis 
used in~\cite{Aad:2014vma} to search for slepton pair production with subsequent decay to neutralinos and leptons.

The signal $pp \to \phi^+ \phi^-$ is simulated using a custom \texttt{SARAH\,v4.12.1}~\cite{Staub:2008uz} model, we generate the signal and background parton level events using \texttt{MadGraph5\,v2.6.2}~\cite{Alwall:2014hca}, simulate the showering using \texttt{Pythia6\,v6.4.28}~\cite{Sjostrand:2006za} and perform the detector simulation with \texttt{Delphes\,v3.3.3}~\cite{deFavereau:2013fsa}.
For our $27\TeV$ simulations, we use the default Delphes card.  For the simulations at $100\TeV$ we use the FCC Delphes card implementing the configurations proposed by the FCC working group~\cite{FCCDelphes}. For the signal simulation, we adapt the card to treat the DM particle as missing energy. We use the LO partonic production cross-sections and multiply by a generous $K$-factor of 2, as we want to find the exclusion limits in the optimistic case, \cref{fig:collider-diagrams} (right). To validate our analysis, we reproduce the relevant backgrounds in~\cite{Aad:2014vma} and find good agreement when using
$\sigma_{WW} = 72$\,pb~\cite{Aad:2014vma, Aad:2016wpd, Gehrmann:2014fva, Heinemeyer:2013tqa},  
$\sigma_{WZ} = 26$\,pb~\cite{Aad:2014vma,Grazzini:2016swo},
$\sigma_{ZZ} = 9.0$\,pb~\cite{Aad:2014vma,Cascioli:2014yka},
$\sigma_{t \bar{t}} = 230$\,pb~\cite{Khachatryan:2016yzq} and
$\sigma_{Wt} = 23\,$pb~\cite{Gabaldon:2125723}.

The main SM backgrounds to our signal are $WW$,  $VV$, $WV$, $t \bar{t}$, $Wt$ and $V$+jets, where $V = Z, \gamma$. While only $WW$ and $VV$ are irreducible backgrounds, $WV$, $t \bar{t}$ and $Wt$ contribute if a lepton or one or two $b$-jets are missed. The $V$+jets background is important at low values of $\mttwo$, but is negligible above $\mttwo \approx 100\GeV$. In order to isolate the signal, we impose the following cuts.  Two opposite-sign same-flavour light leptons are required with $p_T > 35\GeV$ and $p_T > 20\GeV$ for leading and subleading leptons, respectively. We veto events with any other leptons, which reduces the $WV$ background. Removing events with $m_{\mu\mu} < 20\GeV$ and $|m_{\mu\mu} - m_Z | < 10\GeV$ significantly reduces backgrounds with a $Z$-boson in the final state. Finally we cut on the transverse mass~\cite{Lester:1999tx, Barr:2003rg}, $m_{\text{T}2} > 200\GeV$ , where we use
\begin{align}
m_{\text{T}2} =&\, \underset{\mathbf{q}_\text{T}}{\text{min}}\bigg [ \text{max}\Big [ m_\text{T} (\mathbf{p_\text{T}^{\mu^-}},\mathbf{q}_\text{T}),m_\text{T}(\mathbf{p}_\text{T}^{\mu^+},\mathbf{p}_\text{T}^\text{miss}-\mathbf{q}_\text{T})\Big] \bigg] \, ,
\end{align}
where $\mathbf{p}_\text{T}^{\mu^+}$ and $\mathbf{p}_\text{T}^{\mu^-}$ are the transverse momenta of the leptons, $\mathbf{q}_\text{T}$ is an arbitrary two-vector which represents the unknown transverse momenta 
of the dark matter particle associated with $\mu^-$, and
\begin{align}
m_\text{T}(\mathbf{p}_\text{T},\mathbf{q}_\text{T})=&\, \sqrt{2(p_\text{T} q_\text{T} - \mathbf{p}_\text{T} \cdot \mathbf{q}_\text{T})}.
\end{align}
For a process where two particles each decay to a lepton and missing energy, the $\mttwo$ distribution will have an end point at the mass of the heavier particle~\cite{Barr:2009jv}.  Although in~\cite{Aad:2014vma} a cut of $m_{\text{T}2} > 90\GeV$ is used, we increase this to $m_{\text{T}2} > 200\GeV$.  This has a small effect on our signal efficiency, as we are mostly interested in dark matter candidates with mass larger than $200\GeV$, while strongly reducing the background from $t \bar{t}$, $Wt$.  However, even with this large cut, we find a significant background from $WW$, $WV$ and $VV$, where at least one of the vector bosons is extremely off-shell.  
To include this effect in \texttt{MadGraph} we simulate $pp \rightarrow \ell^+ \ell^- \nu_\text{all} \bar{\nu_\text{all}}$  
and $pp \rightarrow \ell^+ \ell^- \ell_\text{all} \nu$, where $\nu_\text{all}$ is $\nu_e$, $\nu_\mu$ or $\nu_\tau$ 
and $\ell_\text{all}$ is any charged lepton.
We do not find a similar large contribution from off-shell particles in the $t \bar{t}$ and $Wt$ channels. Even though the cross-section of these gluon initiated channels grows faster than the di-boson processes as the collider energy is increased, they remain a subdominant background as the $t$ is narrower and 
as this background only passes the cuts if a jet is missed, reducing the $\mttwo$ endpoint. 
Finally, we checked that the contribution from jets faking muons is negligible.
In \cref{tab:cut-flow-1} we show the cross-sections at each stage in the analysis 
for the background and for an example signal, $m_\chi = 0.6 \TeV$, $\Delta = 0.34$ ($27\TeV$) and $m_\chi = 0.8 \TeV$, $\Delta = 0.2$ ($100\TeV$), both for the $n=10$ muon type model. 

\begin{table}
  \centering
  \begin{minipage}{16cm}
    \begin{ruledtabular}
\begin{tabular}{ccccccc}
 Channel& \multicolumn{2}{c}{$\mu^+ \mu^- \nu_\text{all} \bar{\nu_\text{all}}$} & 
 \multicolumn{2}{c}{$\mu^+ \mu^- l_\text{all} \nu$}  & 
 \multicolumn{2}{c}{Example Signal}\\
 Energy [TeV]& 27  & 100 & 27 & 100 & 27 & 100\\
\hline
No Cuts & 
2100 & 6900 & 560 & 1800 & 17 & 100\\
\begin{minipage}{7cm}$p_T^{\mu_1 (\mu_2)} > 35 (20)\GeV$ \& Lepton veto\end{minipage}& 
1100 & 620 & 120 & 160 & 12 & 14\\
Jet veto & 
690 & 530 & 45 & 61 & 3.3 & 9.4\\
\begin{minipage}{7cm}$m_{\mu\mu}>20\GeV$ \& $|m_{\mu\mu} - m_Z | > 10\GeV$\end{minipage}&  
470 & 370 & 6.6 & 13 & 3.3 & 8.9\\
$m_{\text{T}2} > 200\GeV$ & 
0.26 & 0.44 & 0.022 & 0.076 & 1.3 & 2.5
\end{tabular}
    \end{ruledtabular}
  \end{minipage}
  \caption{Cross-sections at each stage in fb.  The example signals are for the 
  muon type model with $n=10$ for the parameter points 
  $m_\chi = 0.6 \TeV$, $\Delta = 0.34$ ($27\TeV$) and $m_\chi = 0.8 \TeV$, $\Delta = 0.2$ ($100\TeV$).}
  \label{tab:cut-flow-1}
\end{table}

\begin{figure}
  \begin{center}
   \includegraphics[width=0.45\columnwidth]{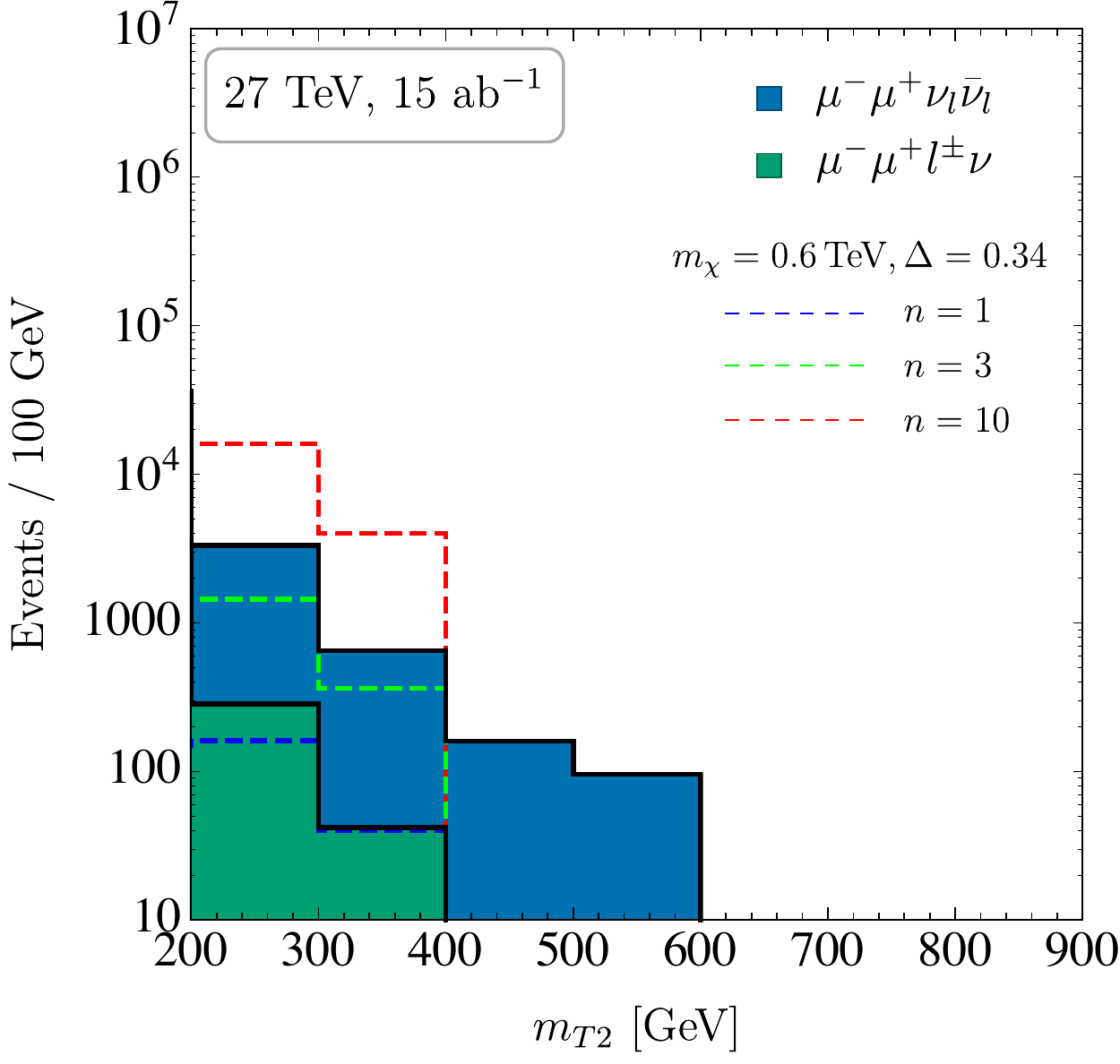}
   \includegraphics[width=0.45\columnwidth]{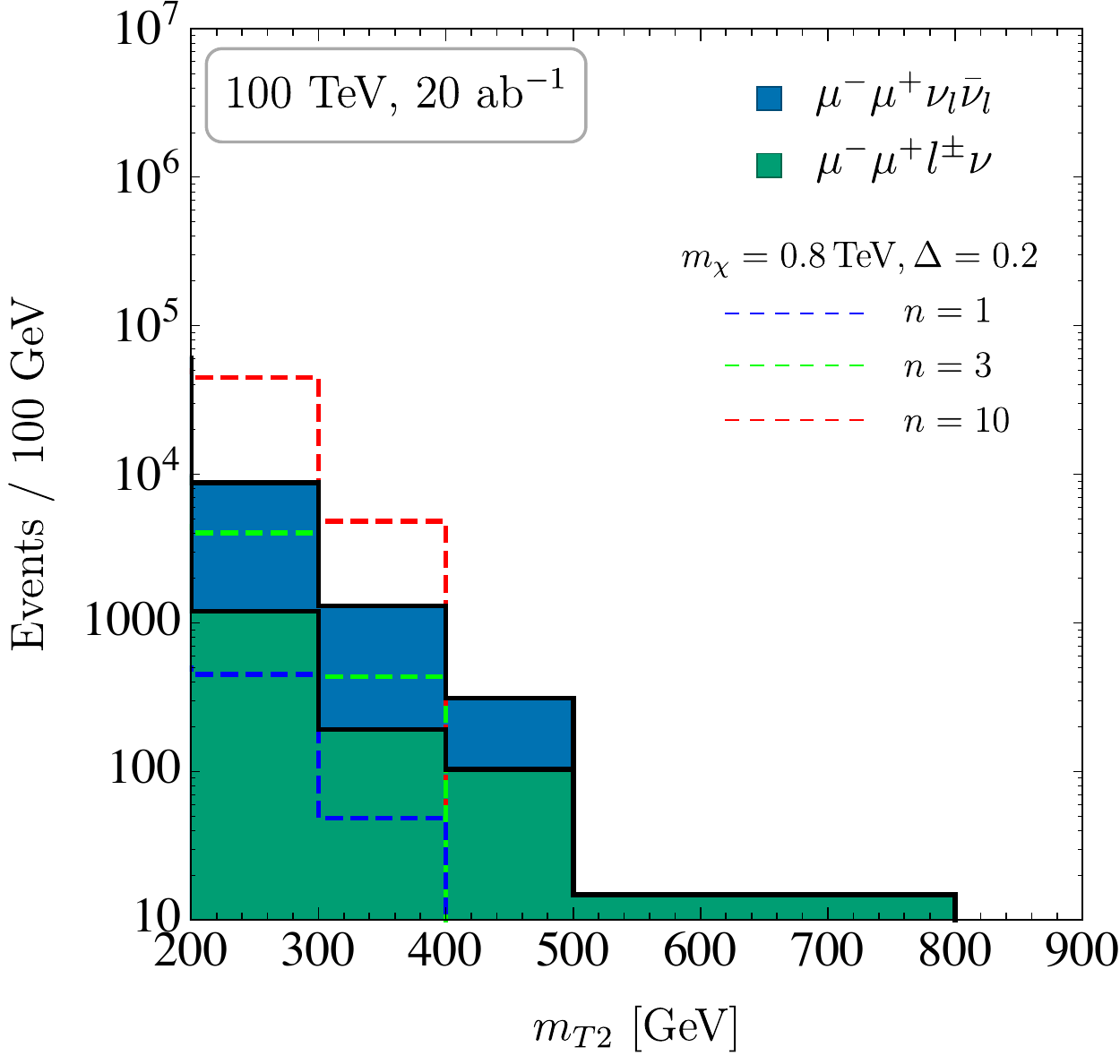}
  \end{center}
  \caption{The $\mttwo$ distribution for background events passing all cuts for the muon model, 
  and an example signal for $n=1,3,10$, at $27\TeV$ (left) and $100\TeV$ (right).  We do not use this information 
  in determining the reach, 
  but simply perform a cut-and-count analysis based on these events.}
  \label{fig:muon-mt2}
\end{figure}

In \cref{fig:muon-mt2} we show the differential distribution in $\mttwo$ for the events passing all cuts, for the background and example signal.  
We see that $\mu^+ \mu^- \nu_\text{all} \bar{\nu_\text{all}}$ is the dominant background, 
and $\mu^+ \mu^- \ell_\text{all} \nu$ is around an order of magnitude smaller.  This is due to 
both the smaller initial cross-section and the smaller efficiency.  We see that both the background and the example signal falls sharply from $\mttwo = 200 \GeV$ to $\mttwo \approx 500\GeV$.
However, the signal will continue to higher values of $\mttwo$ for other points in our parameter space.
We also see that at $27\TeV$, the $\mu^+ \mu^- \nu_\text{all} \bar{\nu_\text{all}}$ continues out to higher values 
of $\mttwo$, while at $100\TeV$ the situation is reversed.


To estimate the expected exclusion limit, we use a Poisson counting procedure 
for the signal and background events which pass all the cuts, 
based on a frequentist framework~\cite{Cowan:2010js, Patrignani:2016xqp}.  We use the likelihood ratio $\lambda(0) = L(0)/L(\hat{\mu})$, where
\begin{align}
L(\mu)&= \frac{(\mu s + b)^n}{n!}e^{-(\mu s+b)},
\end{align}
$n$ is the observed number of events, $s$ is the number of signal events, $b$ is the number of background events and $\hat{\mu} = (n-b)/s$.  In the large sample limit, the significance $Z_0$ is given by $Z_0 = \sqrt{-2\ln \lambda(0)}$.  Replacing $n$ with the expected value, $s+b$, we find that the significance is given by
\begin{align}
Z_0 &\,=\sqrt{2\left( (s+b) \ln \left( 1+\frac{s}{b}\right) -s\right)}.
\end{align}
In the limit $s\ll b$, this reduces simply to $s/\sqrt{b}$.  The significance is related to the $p$-value by $Z_0 = \Phi^{-1}(1-p)$, where $\Phi$ is the standard Gaussian cumulative distribution.  The 90\% confidence limit is given by $Z_0 \approx 1.64$.


\begin{figure}
  \begin{center}
    \begin{tabular}{cc}
    \includegraphics[width=0.42\columnwidth]{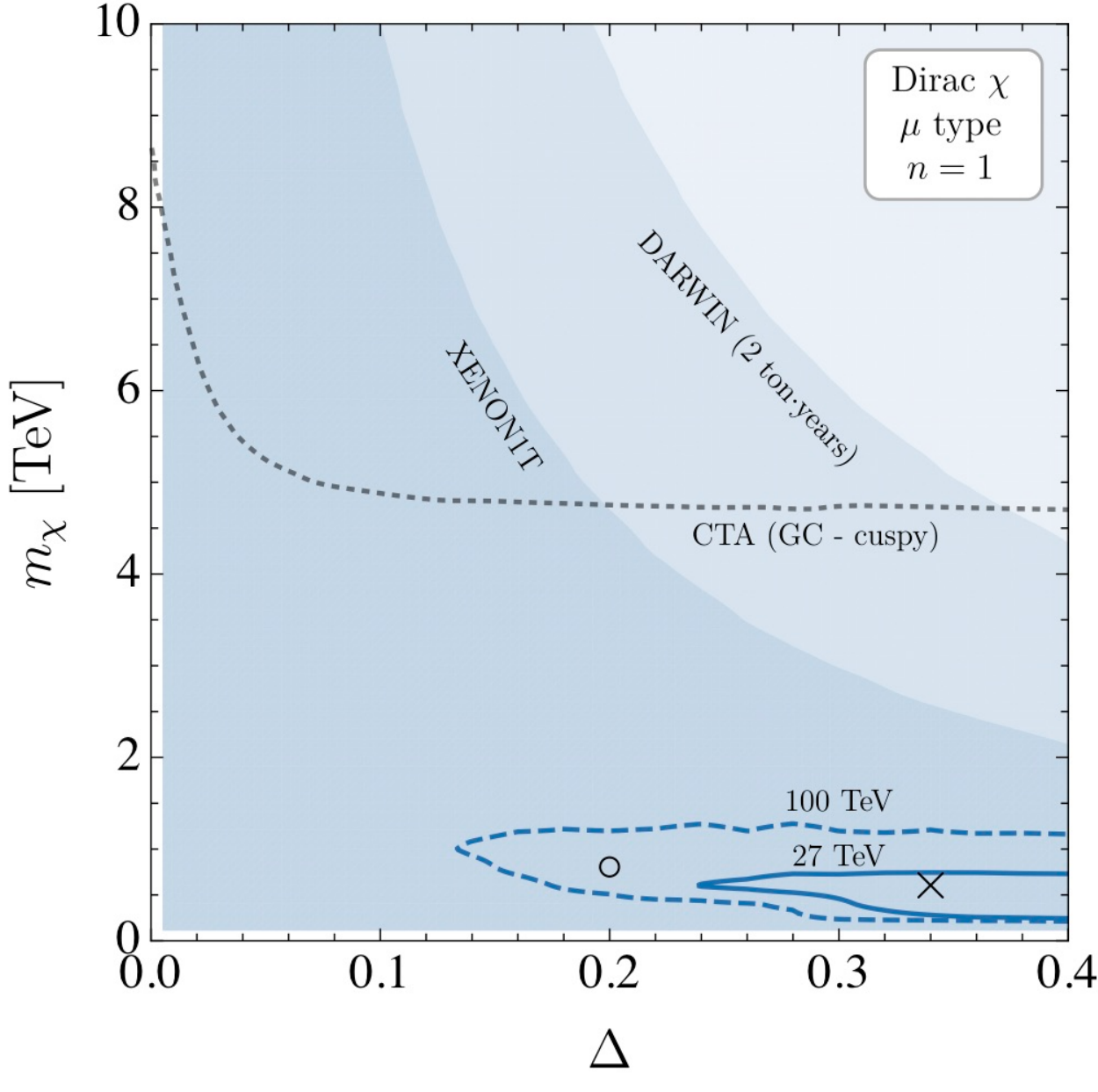}&
    \includegraphics[width=0.42\columnwidth]{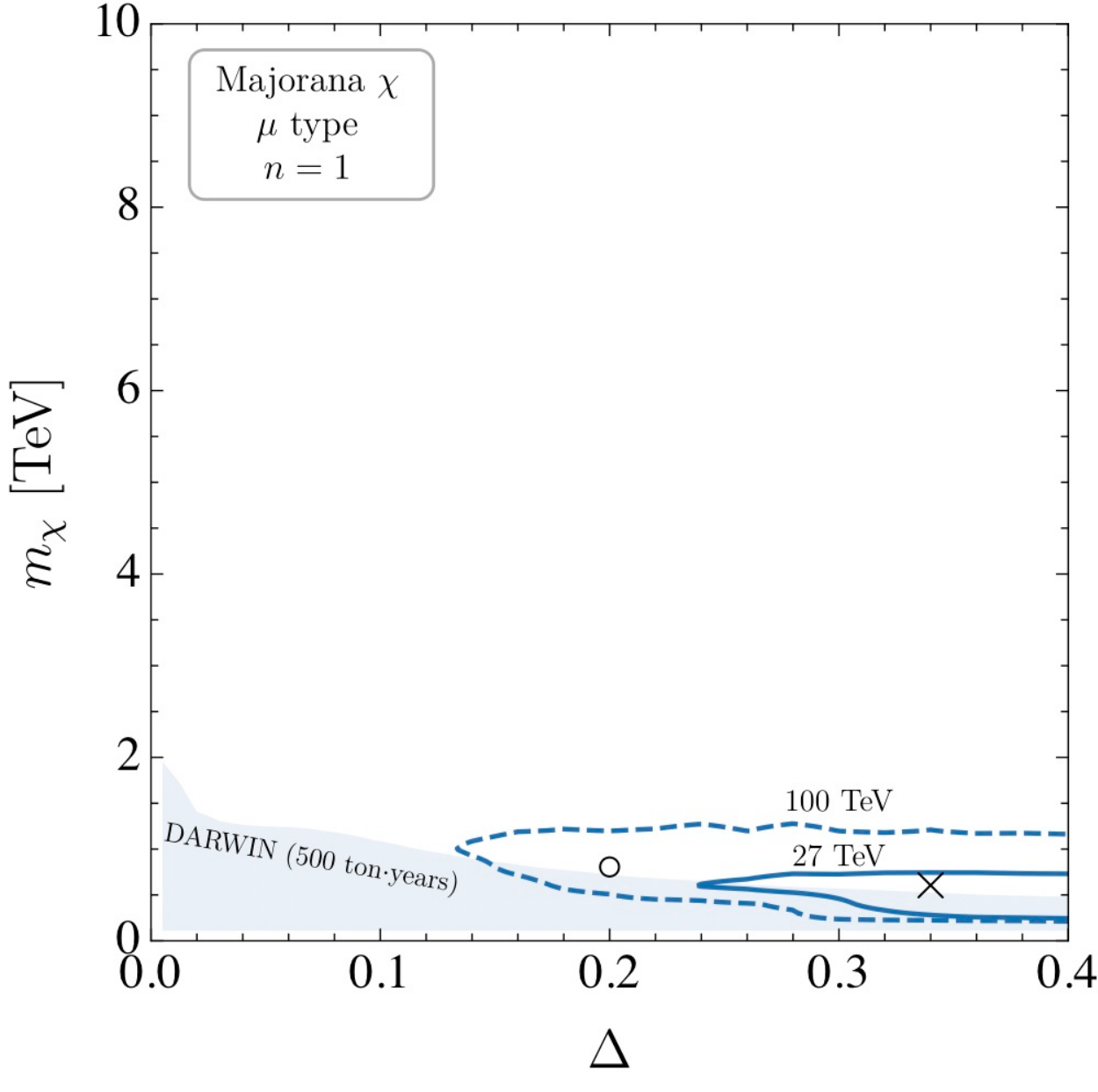}\\
    \includegraphics[width=0.42\columnwidth]{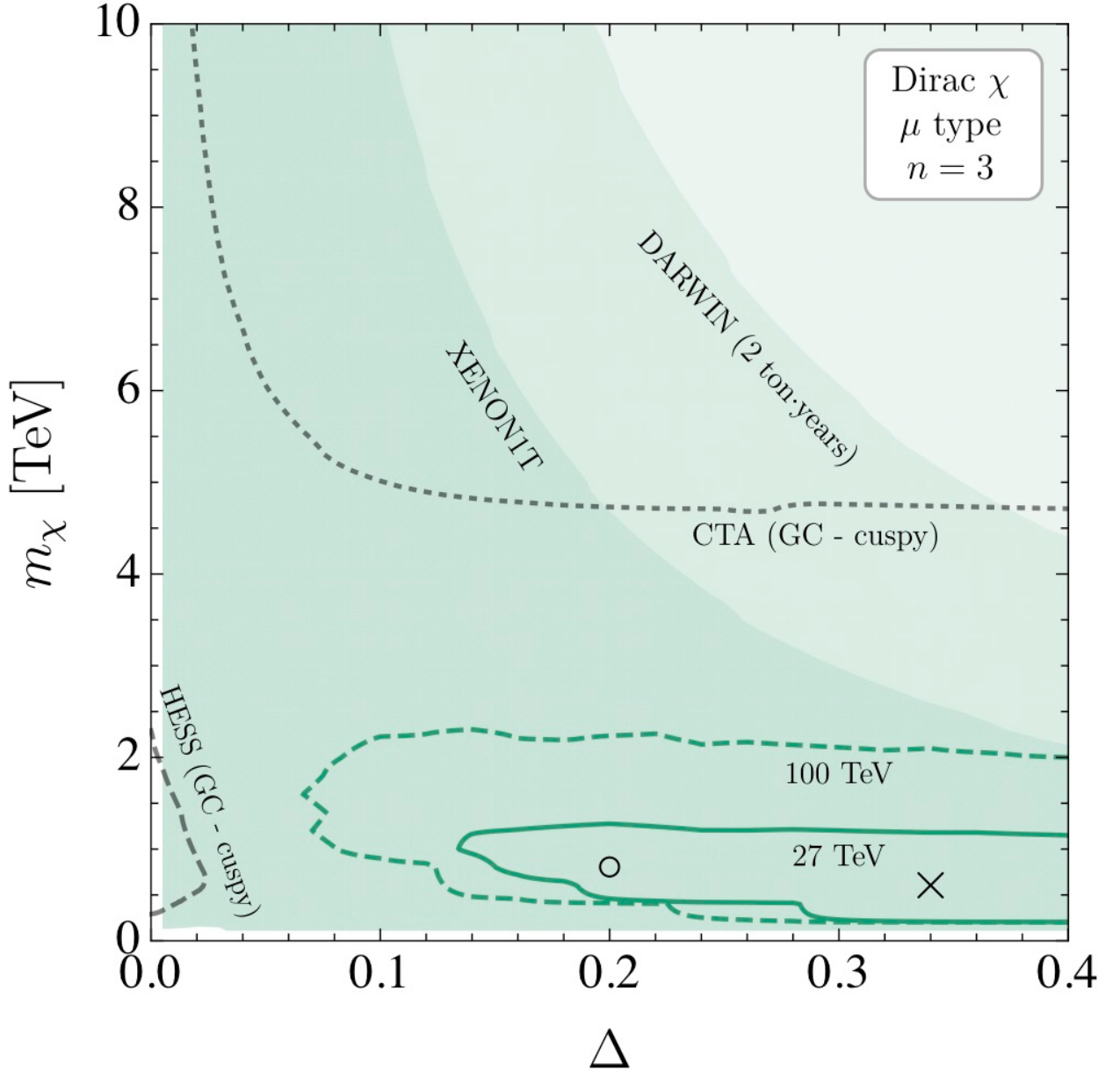}&
    \includegraphics[width=0.42\columnwidth]{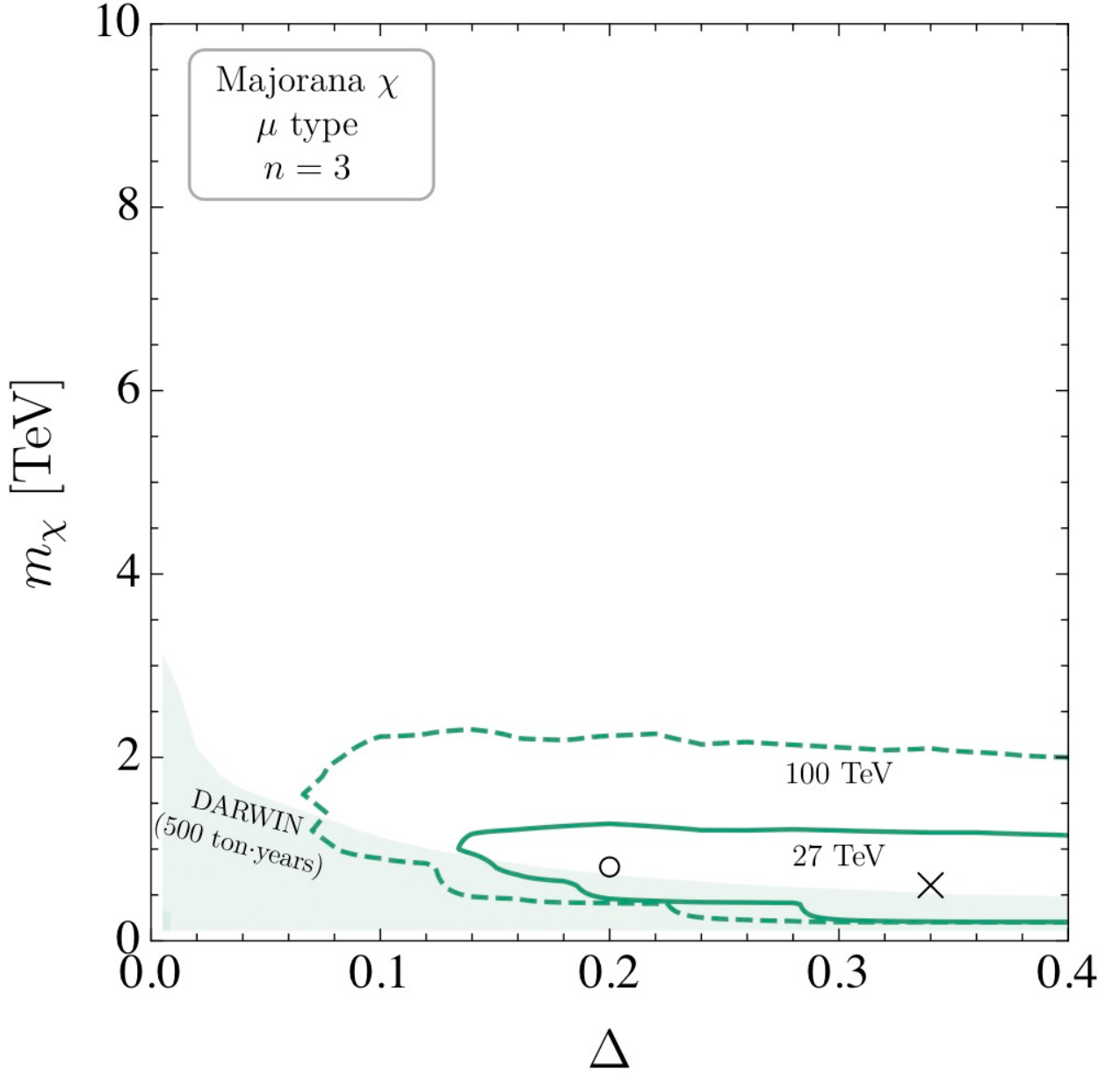}\\
    \includegraphics[width=0.42\columnwidth]{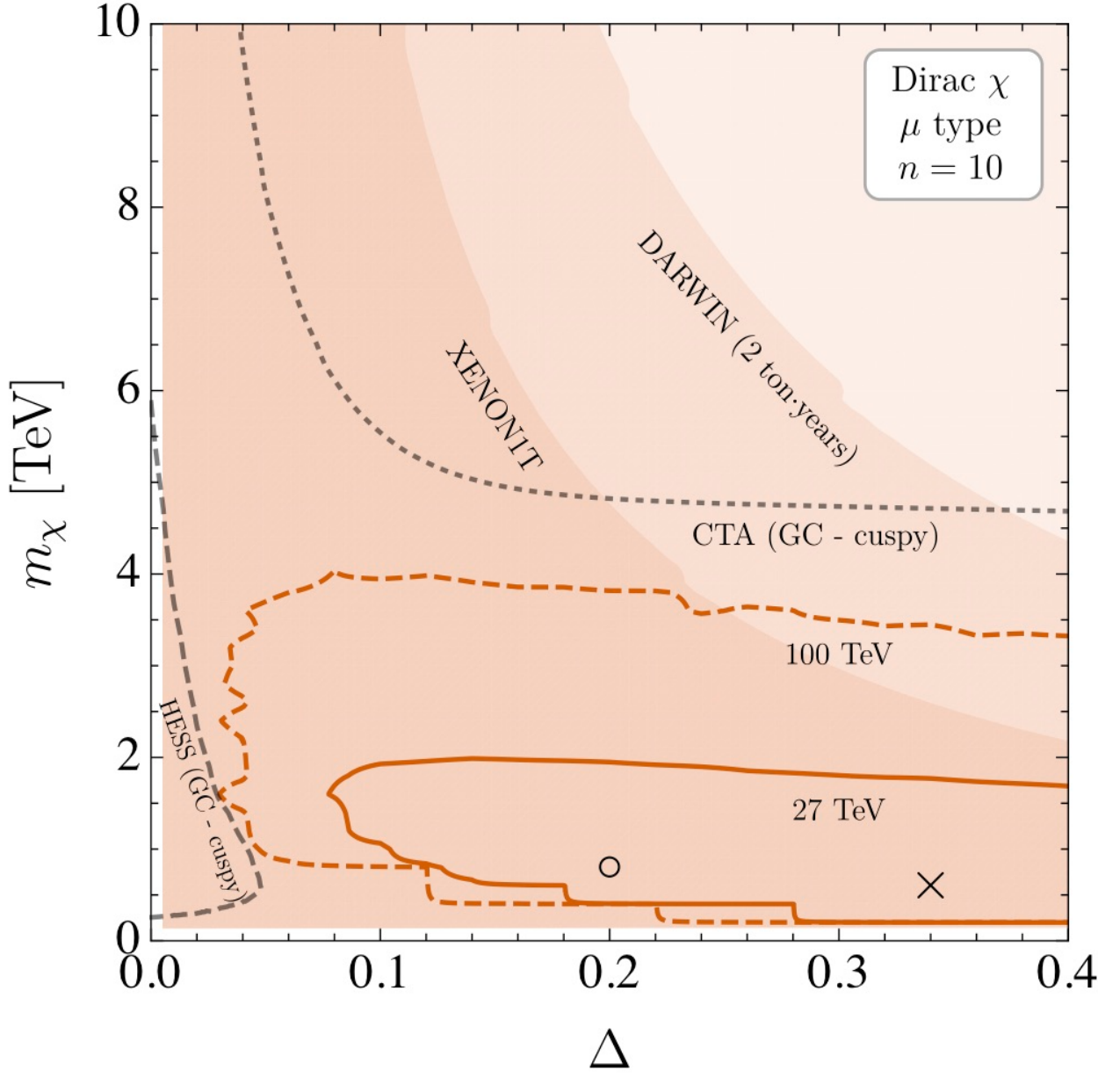}&
    \includegraphics[width=0.42\columnwidth]{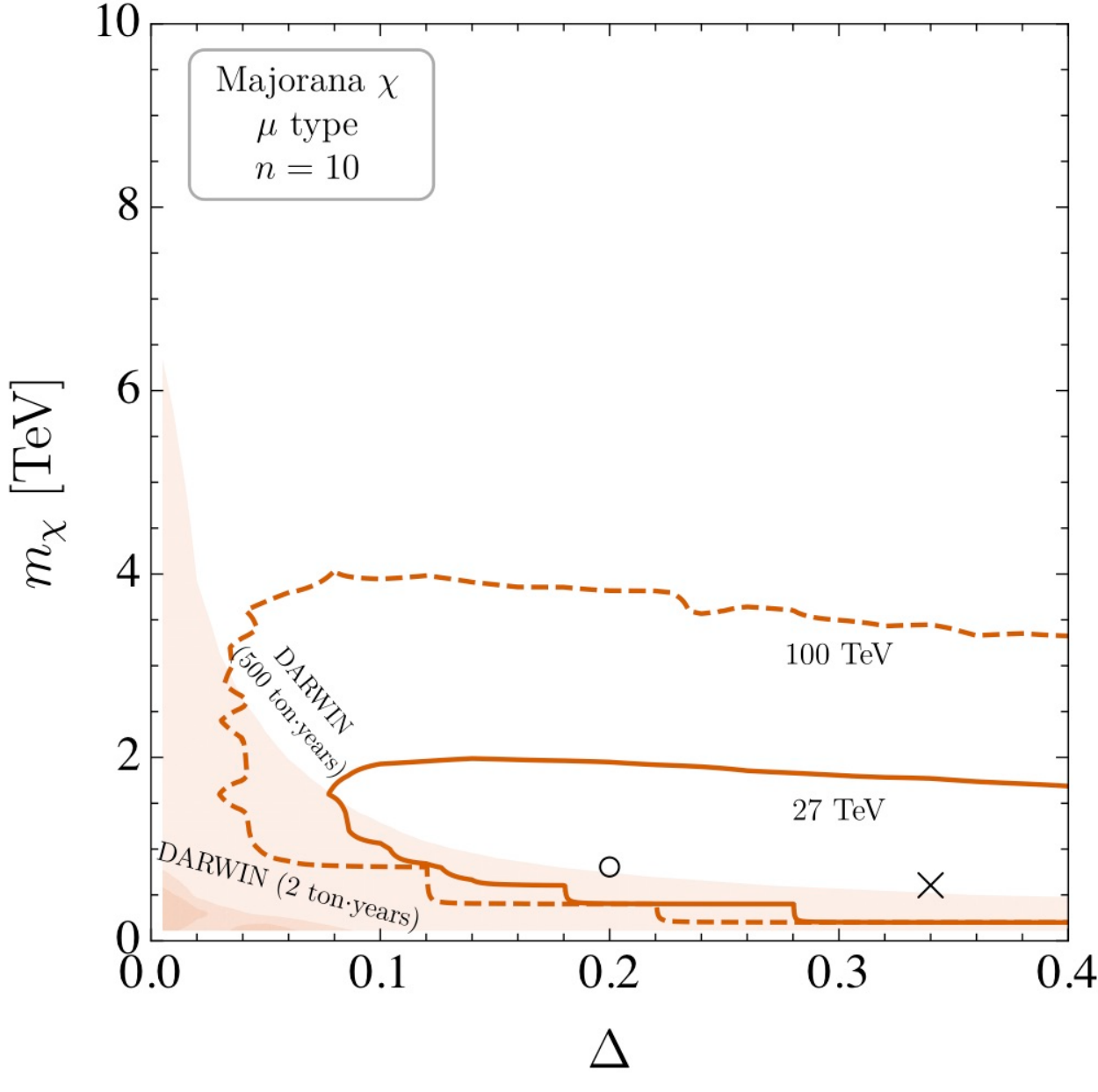}
 \end{tabular}
  \end{center}
  \caption{The reach of future colliders, at $90\%$\,C.L.,   current and future direct detection experiments, at $90\%$\,C.L., 
  and current and future indirect detection experiments,  at $95\%$\,C.L., for Dirac (left) and Majorana (right) DM interacting with a muon and one (top), three (middle) and ten (bottom) coannihilation partners in the $\Delta$-$m_\chi$ plane. The lightly, moderately and strongly shaded regions correspond to the direct detection limits by the future DARWIN experiment with $500\, \text{ton}\cdot\text{years}$, $2\, \text{ton}\cdot\text{years}$ and the XENON1T limits, respectively, which are discussed in detail in \cref{sec:directdetection}. The circle and cross 
  signify our example signals.}
  \label{fig:collider-exclusion}
\end{figure}


In \cref{fig:collider-exclusion} we present the $90\%$\,C.L. sensitivity for 
the muon type models at a $27\TeV$ and a $100\TeV$ proton-proton collider. The reach on electron type models is shown in Appendix~\ref{sec:appColliderE}.
The parameter space probed is where $m_\chi$ is small and $\Delta$ is relatively large.  
The reach is independent of whether dark matter is Majorana or Dirac, 
since it depends on the $\phi$-pair production cross-section 
and the fact that BR$(\phi_i^\pm \to \chi \ell^\pm)=1$.
The large $m_\chi$ region is not probed as $m_\phi$ increases with $m_\chi$, 
and the $\phi$-pair production cross-section decreases rapidly as $m_\phi$ increases, \cref{fig:collider-diagrams} (right).  
We see that in both cases the limits are strongest when there are more coannihilation partners. 
This is because the $pp\to \chi \bar{\chi} \ell^+ \ell^-$ cross-section scales as $n^2$. 
For $n=1$, the $27\TeV$ ($100\TeV$) machine can probe $m_\chi < 0.75\TeV \, (1.2\TeV)$, 
for $n=3$ it can probe $m_\chi < 1.3\TeV \, (2.3\TeV)$ while for $n=10$ the limits are 
$m_\chi < 2.0\TeV \, (4.0\TeV)$.
The small $\Delta$ region is not probed as in this region the 
momentum of the leptons is small and they are not efficiently reconstructed.  
This is a well known problem in the coannihilation region.  The gap for 
lower $\Delta$ can be closed, e.g., by looking for 
ISR~\cite{Gori:2013ala, Schwaller:2013baa} or for disappearing charged 
tracks~\cite{Evans:2016zau,Khoze:2017ixx,Mahbubani:2017gjh}. 
A thorough study of the reach in the challenging small $\Delta$ region is not pursued 
here as in these models there is a nice complementarity with direct detection experiments, which 
can be seen to cover this region. 

We also overlay the direct and indirect detection bounds 
from~\cref{sec:directdetection,sec:indirectdetection}, to give 
a summary of all the relevant current and future experimental constraints.  
We see that the situation is dramatically different for Dirac and Majorana $\chi$.  
For Dirac $\chi$, small masses and mass splittings have already been excluded by XENON1T.  
In the future, DARWIN will probe the full parameter space, while colliders and indirect detection 
will be sensitive for relatively low masses and large or small $\Delta$, respectively.
We see that the challenging small $\Delta$ region at colliders is excluded by the 
existing bound from XENON1T.

For Majorana $\chi$, on the other hand, DARWIN, with the maximum exposure, 
is limited to small masses and small $\Delta$, while indirect constraints do not feature.  
This is due to the velocity suppression of both the DM-nucleus and the annihilation cross-sections.  
The collider bounds are the same as in the Dirac case, since the mass term of $\chi$ 
does not enter into the production and decay of $\phi$-pairs.  In this case, future colliders 
are essential for probing the large $\Delta$ region of the parameter space.

\section{Conclusions}
\label{sec:conclusions}

As nature has not yet provided clues to the mass scale of dark matter, we have to explore many orders of magnitude in mass and coupling strengths. One of the first and best studied regions in parameter space is $10\GeV$ -- $1\TeV$ dark matter, with a large fraction of the experimental progress of the past decade targeting this mass range. Although many of the viable WIMP models in this mass range 
can be exhaustively probed at the next generation of direct and indirect detection and collider experiments, we point out that large regions of viable parameter space are inaccessible when coannihilation is taken into account.

Assuming a minimal and versatile dark matter model, we restricted to the relic surface and studied the reach of current and future  experiments on the dark matter parameter space. We take the dark matter particle to be a gauge-singlet Majorana or Dirac fermion and introduce $n$ charged scalars as coannihilation partners, which together couple to SM right-handed leptons. The relic surfaces of these models demonstrate a viable, perturbative multi-TeV dark matter candidate, whose relic abundance either decreases (in the Majorana case) or \emph{increases} (in the Dirac case) with coannihilation.  

Direct detection experiments are sensitive to our models via loop-induced dipole and anapole interactions.
We compute the interaction cross-section between dark matter and target nuclei, 
taking into account nuclear form factors. 
By explicitly integrating over the DM velocity distribution and including experimental efficiencies 
we calculate the expected event rate, which we compare to bounds from the current XENON1T experiment 
and use to derive projected limits from DARWIN.
We find starkly different result for Dirac and Majorana dark matter.  
Over the parameter space we consider, 
XENON1T excludes the Dirac models at $ m_\chi < 2\TeV$ (for any $\Delta$)
and $\Delta < 0.1$ (for any $m_\chi$).
With 2 $\text{ton}\cdot\text{years}$ of exposure these constraints strengthen to cover 
around 85\% of the parameter space.  DARWIN, with an exposure of $500\,\text{ton}\cdot\text{years}$, can probe 
the entire region for Dirac dark matter. 
The models with Majorana dark matter are, however, currently unconstrained, 
due to velocity suppression in the direct detection cross-section.  
Only DARWIN with an exposure of $500\,\text{ton}\cdot\text{years}$ can make progress, with the ability to exclude
Majorana DM up to $0.5\TeV$ for large $\Delta$ and between
 $2\TeV$ and $6\TeV$ in the region $\Delta < 0.1$.  
The remaining $\sim90\%$ of the parameter space, however, remains unconstrained.
In all cases, increasing coannihilation (both by reducing $\Delta$ and increasing the number 
of partners) tends to make direct detection a better probe of the models.

 In indirect detection experiments, we only find significant limits via continuum photons 
 if the dark matter is Dirac, 
 and if the Milky Way halo profile is cusped.  The models coupled to taus have the best limits, 
 since many continuum photons are produced when the tau decays.  If the Milky Way halo profile 
 is cusped, HESS has excluded the tau models up to $4\TeV$ for all values of $\Delta$, and at least up 
 to $10\TeV$ for $n=10$ and $\Delta < 0.02$.  Again if it is cusped, CTA will be able to probe our 
 whole parameter space for Dirac dark matter which couples to taus via $n\leq10$ 
scalars. 
For the muon (electron) case, even if the Milky Way profile is cusped, CTA can only probe 
the Dirac models to $m_\chi < 5\TeV$ ($m_\chi < 7 \TeV$), unless $\Delta < 0.15$ in 
which case it can probe up to at least $10\TeV$.  
If the Milky Way halo profile is not cusped then HESS and CTA provide no bounds and the 
best limits come from Fermi-LAT -- MAGIC observations of dwarf spheroidals, although they are 
relatively weak, only covering a region with $m_\chi < 2 \TeV$ and $\Delta < 0.05$.
Finally, due to velocity suppression, there are no indirect detection constraints 
on any of the Majorana models.

The collider bounds we find are insensitive to the difference between Majorana and Dirac DM,
 as we consider pair production of the coannihilation partner, which subsequently 
 decays into two same-flavour opposite-sign leptons and missing energy with a 
 branching ratio of 1. 
 We only provide bounds for the models where dark matter couples to electrons and muons 
 since tau reconstruction is difficult to model at a future collider.  In any case, the exclusion reach for 
 the tau models will be worse than for the electron or muon models.
 We simulate signal and background, apply a set of cuts to 
 isolate the signal and derive the expected reach of the HE-LHC with $\sqrt{s} = 27\TeV$ 
 and an integrated luminosity of $15\,$ab$^{-1}$ and a future collider with $\sqrt{s} = 100\TeV$ 
 and $20\,$ab$^{-1}$.  We find that, as the production cross-section is proportional to 
 $n^2$, the bounds strengthen as the number of coannihilation partners increases.  For the muon model, 
 a $27\TeV$ ($100\TeV$) proton-proton collider can exclude the $n=1$ model up to $0.74\TeV$ ($1.2\TeV$), 
 the $n=3$ model up to $1.3\TeV$ ($2.2\TeV$) and the $n=10$ model up to $2\TeV$ ($4\TeV$).  
 Although the analysis does not target the more challenging low $\Delta$ region, 
 small $\Delta$ searches are unlikely to have a larger mass reach and this region is 
 covered by current and future direct detection experiments.
 
We thus see that while viable, perturbative models of Dirac dark matter will be well probed by DARWIN, 
future colliders, and, if the Milky Way Halo is cuspy, by CTA, the future suite of dark matter 
searches will leave around 70\% of the parameter space of our viable models of Majorana dark matter untouched.  
We are limited in one direction by velocity suppressed interactions, and in another direction by 
the large mass scales involved.  
For direct detection, overcoming these limitations could involve the optimisation of current experiments for larger DM masses or a greater emphasis on 
electron recoils (for probing the electron type models at tree level), 
by developing new target materials, e.g,~\cite{Geilhufe:2018gry}, 
the application of novel techniques, e.g.,~\cite{Baum:2018tfw},
or scenarios where the velocity suppression is lifted, e.g., due to infall into neutron stars~\cite{Baryakhtar:2017dbj}.
The situation in indirect detection would be considerably improved with a 
better determination of the dark matter density in the galactic centre, 
and by analyses which do not depend on a cuspy profile, e.g.,~\cite{HESS:2015cda}.  
However, key to probing the velocity/loop/phase-space suppressed Majorana models is 
substantially improving bounds from gamma-ray observations~\cite{Garny:2015wea}. 
We see that future colliders will be able to produce a handful of $10\TeV$ particles in 
optimistic scenarios, but the analysis considered here retains thousands of background events.  
Improved search strategies, 
which choose varying cuts depending on the new physics parameter point under consideration, 
or which optimise the signal against the background using a wider spectrum of information 
(for instance, using neural networks)
could push the sensitivity to larger masses, even though it is intrinsically limited by the steeply falling cross-section with mass. Exploring displaced vertex analyses and using ISR will be useful in the coannihilation region.
We see that, to conclusively test the WIMP paradigm, new experimental 
techniques will need to be developed to surmount these challenges.

\section{Acknowledgments}

We would like to thank Tilman Plehn and Joachim Kopp for collaboration in the early stages of the project, Tim Tait for discussions on coannihilation, Joachim Kopp for advice on direct detection, Alaettin Serhan Mete from ATLAS and Rakhi Mahbubani for useful discussions on the collider analysis, and Paolo Panci and Lucia Rinchiuso for helpful advice on indirect detection.  MJB was supported by the German Research Foundation (DFG) under Grant Nos.~KO 4820/1-1 and
FOR 2239, by the European Research Council (ERC) under the European Union's Horizon 2020
research and innovation programme (grant agreement No. 637506, ``$\nu$Directions''), by
Horizon 2020 INVISIBLESPlus (H2020-MSCA-RISE-2015-690575) and by the Swiss National
Science Foundation (SNF) under contract 200021-175940.
The work of AT was partially supported under the International Cooperative Research and Development
Agreement for Basic Science Cooperation (CRADA No.~FRA-2016-0040) between Fermilab and
Johannes Gutenberg University Mainz, and partially by the Advanced Grant EFT4LHC of the
European Research Council (ERC) and the Cluster of Excellence Precision Physics,
Fundamental Interactions and Structure of Matter (PRISMA -- EXC 1098).

\appendix
\label{sec:app}

\section{Collider Bounds for the Electron Type Models}
\label{sec:appColliderE}

\begin{figure}
  \begin{center}
    \begin{tabular}{cc}
    \includegraphics[width=0.42\columnwidth]{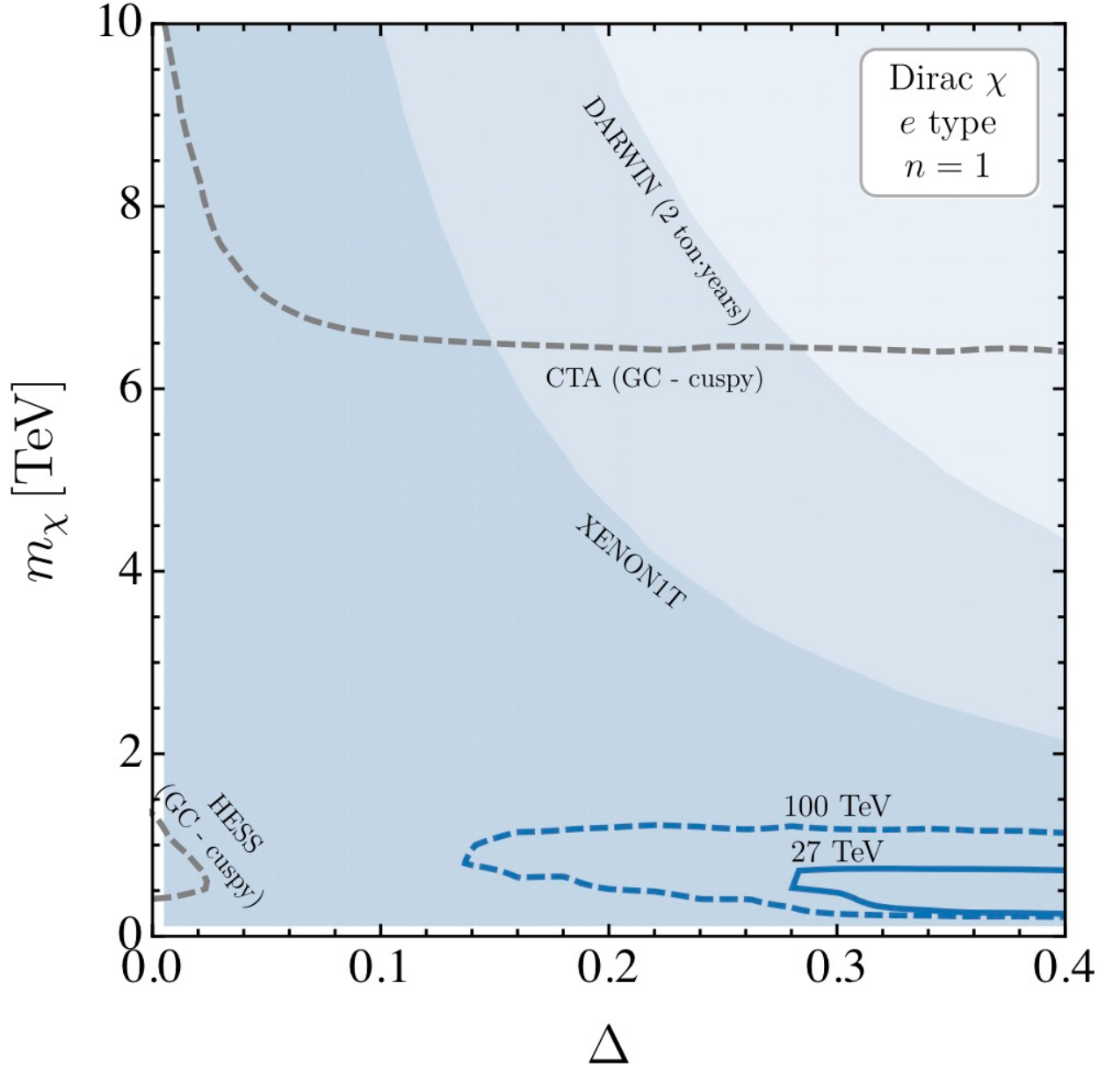}&
    \includegraphics[width=0.42\columnwidth]{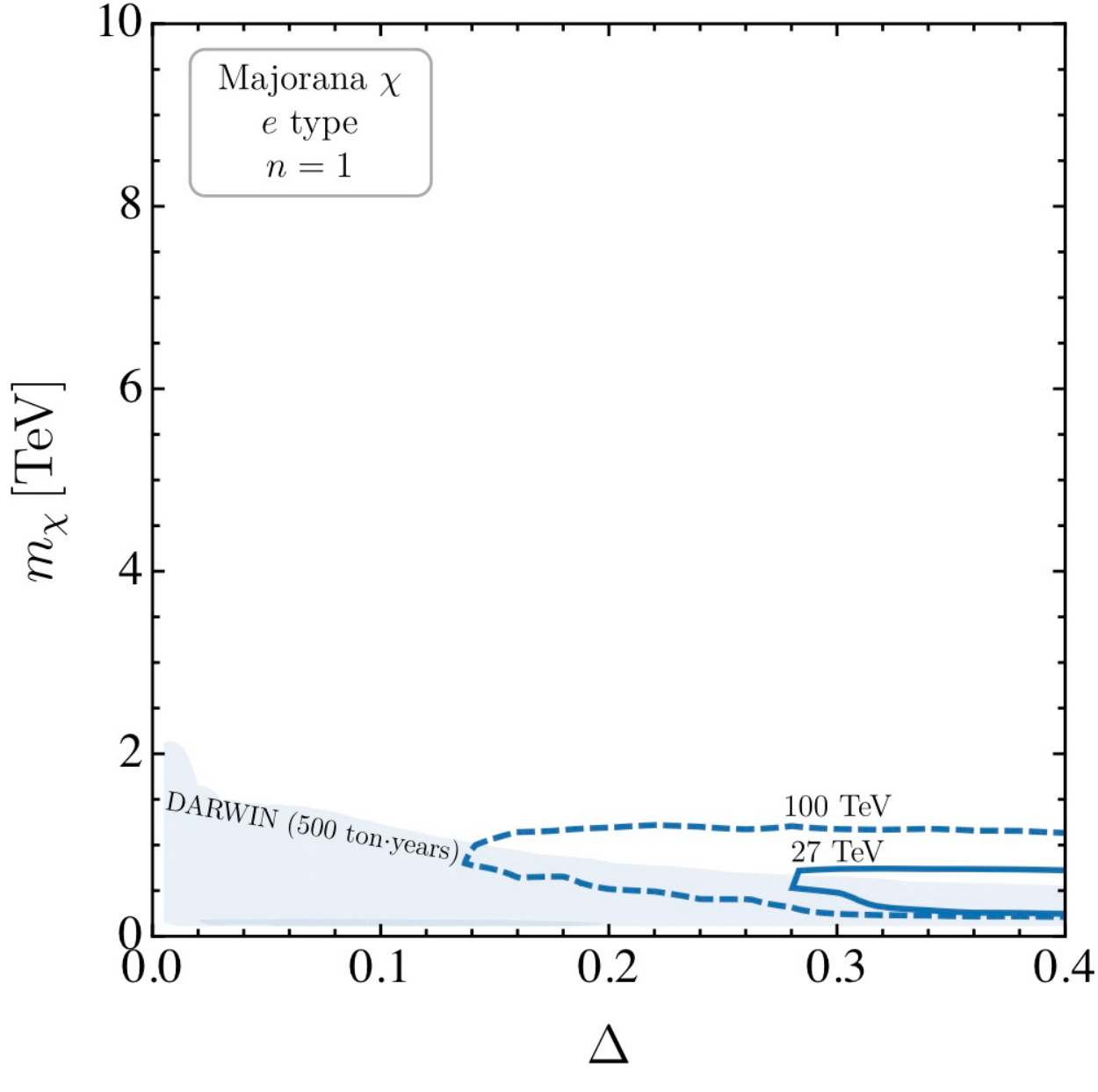}\\
    \includegraphics[width=0.42\columnwidth]{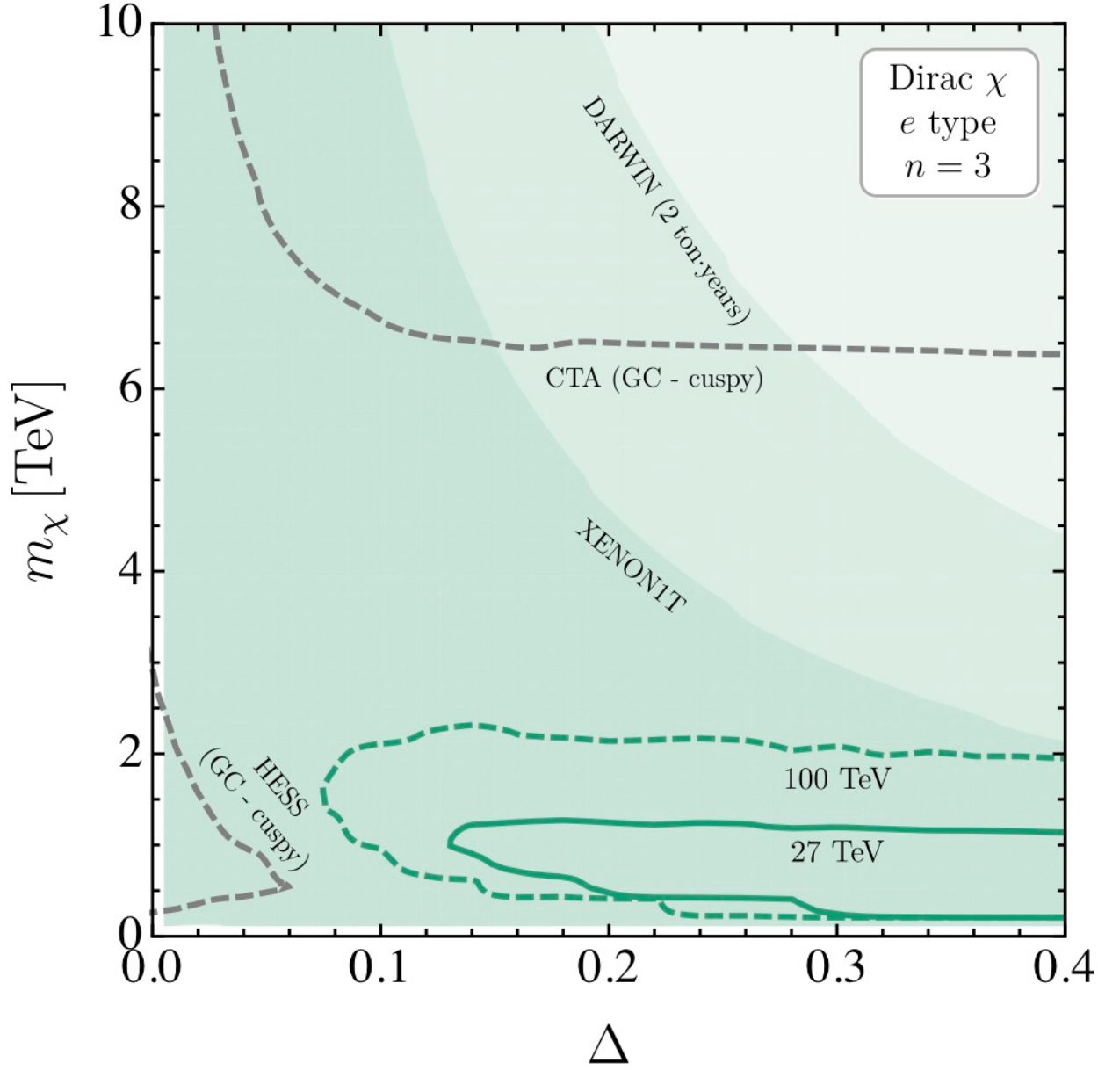}&
    \includegraphics[width=0.42\columnwidth]{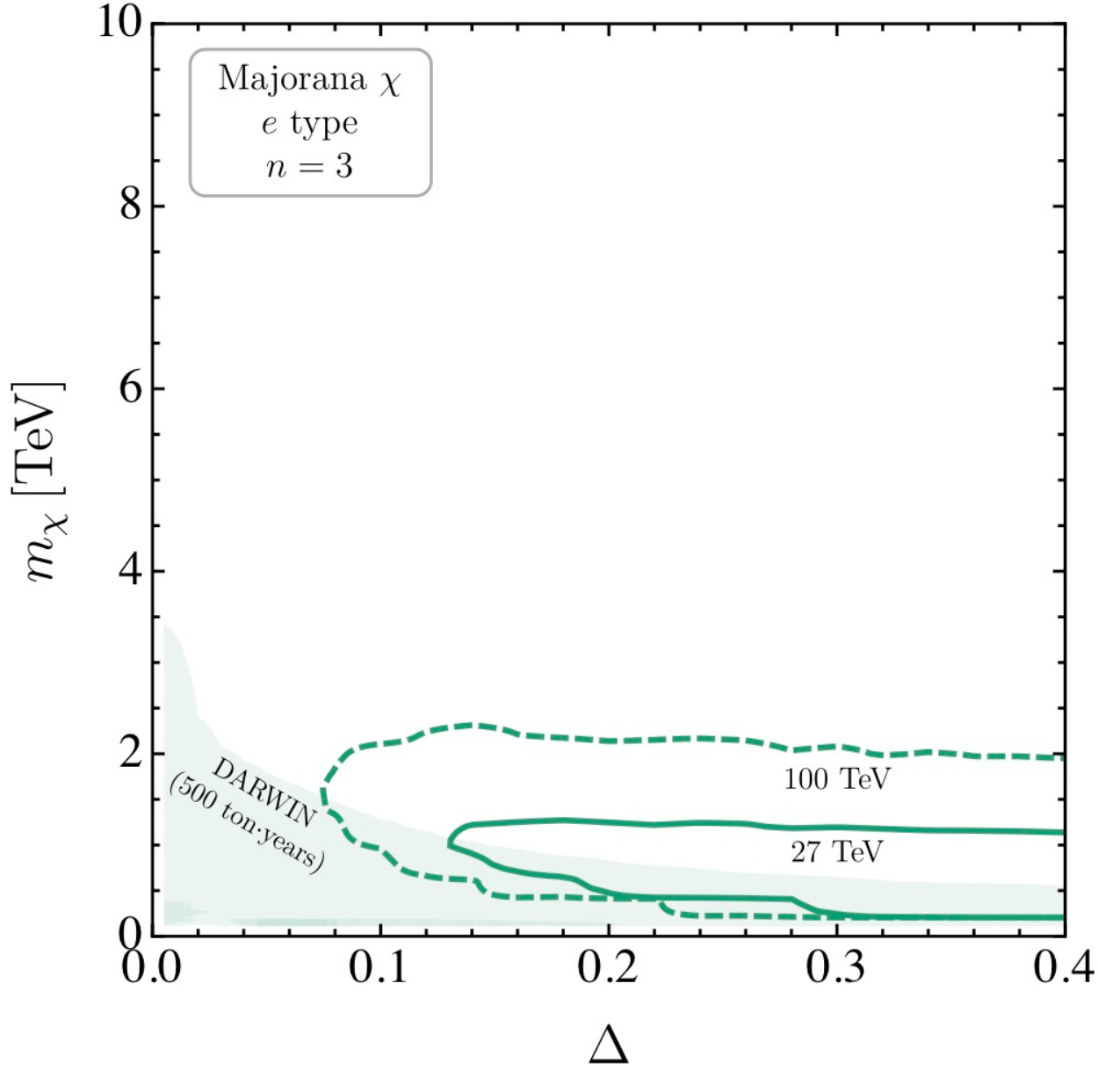}\\
    \includegraphics[width=0.42\columnwidth]{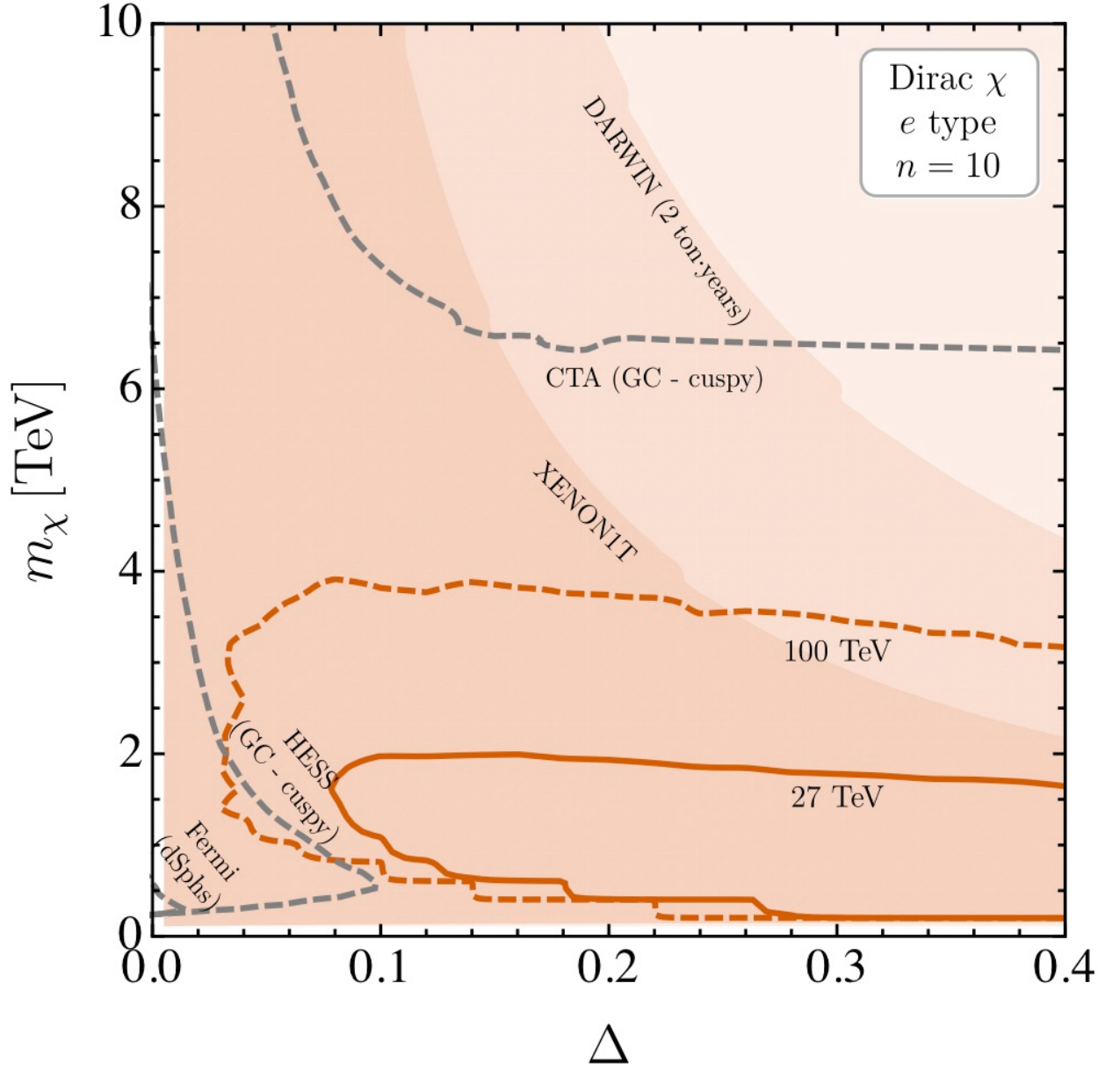}&
    \includegraphics[width=0.42\columnwidth]{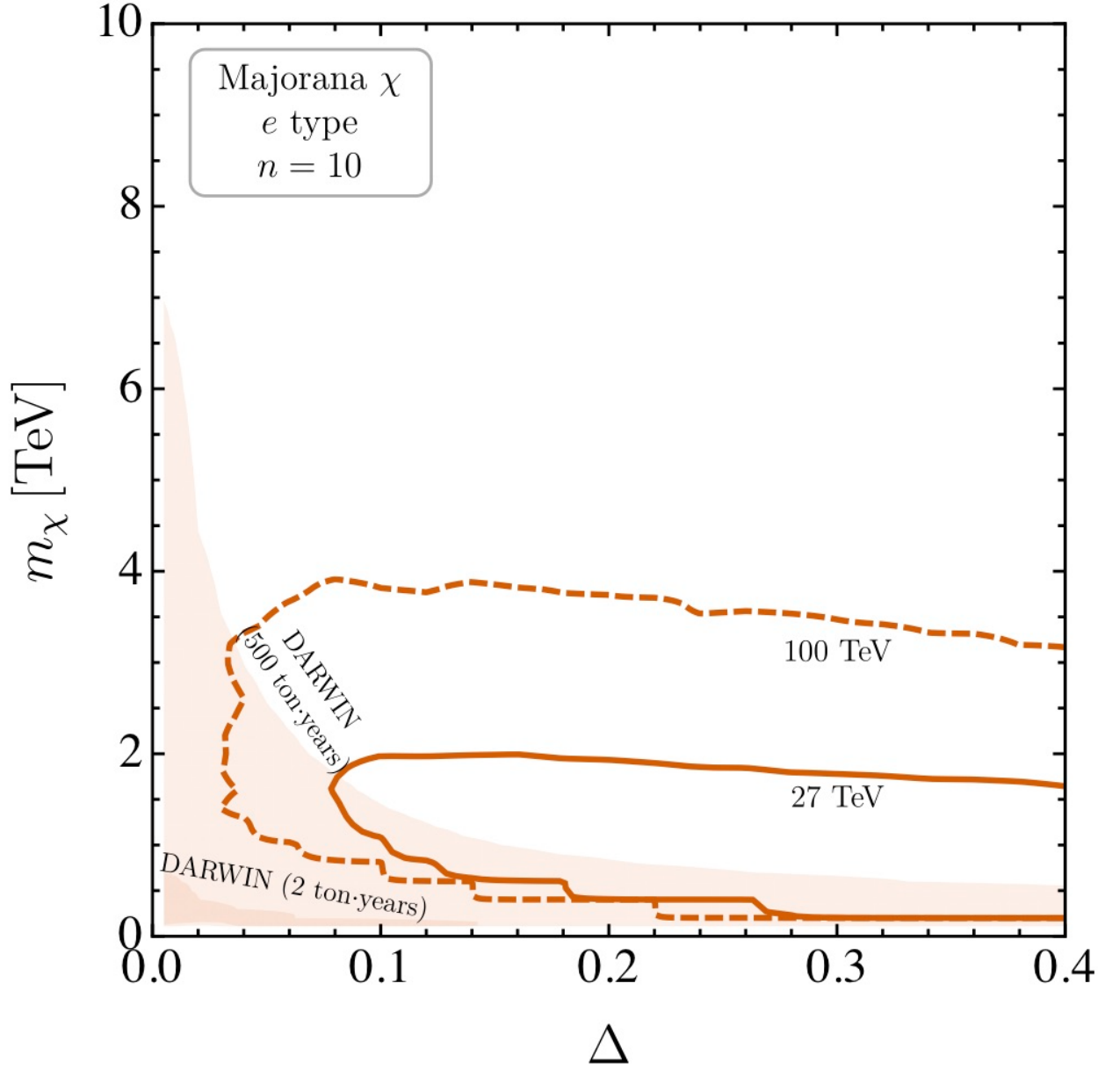}
 \end{tabular}
  \end{center}
    \caption{The reach of future colliders, at $90\%$\,C.L.,   current and future direct detection experiments, at $90\%$\,C.L., 
  and current and future indirect detection experiments,  at $95\%$\,C.L., for Dirac (left) and Majorana (right) DM interacting with an electron and one (top), three (middle) and ten (bottom) coannihilation partners in the $\Delta$-$m_\chi$ plane. The lightly, moderately and strongly shaded regions correspond to the direct detection limits by the future DARWIN experiment with $500\, \text{ton}\cdot\text{years}$, $2\, \text{ton}\cdot\text{years}$ and the XENON1T limits, respectively, which are discussed in detail in \cref{sec:directdetection}. }
  \label{fig:collider-exclusion-electron}
\end{figure}

In this appendix, we present the collider bounds for the electron type models.  
In \cref{fig:collider-exclusion-electron} we show the bounds for Dirac 
dark matter (left) and Majorana dark matter (right), for $n=1,3,10$.
The reach for electron final states is marginally worse than for muon 
final states, \cref{fig:collider-exclusion}, due to the fact that the 
electron reconstruction efficiency is slightly worse than for muons.


\bibliographystyle{JHEP}
\bibliography{./bib}

\providecommand{\href}[2]{#2}\begingroup\raggedright\begin{thebibliography}{10}

\bibitem{Patrignani:2016xqp}
{\bf Particle Data Group} Collaboration, C.~Patrignani et~al., {\it {Review of
  Particle Physics}},  {\em Chin. Phys.} {\bf C40} (2016), no.~10 100001.

\bibitem{Bramante:2014tba}
J.~Bramante, P.~J. Fox, A.~Martin, B.~Ostdiek, T.~Plehn, T.~Schell, and
  M.~Takeuchi, {\it {Relic Neutralino Surface at a 100 TeV Collider}},  {\em
  Phys. Rev.} {\bf D91} (2015) 054015,
  [\href{http://arxiv.org/abs/1412.4789}{{\tt arXiv:1412.4789}}].

\bibitem{Bramante:2015una}
J.~Bramante, N.~Desai, P.~Fox, A.~Martin, B.~Ostdiek, and T.~Plehn, {\it
  {Towards the Final Word on Neutralino Dark Matter}},  {\em Phys. Rev.} {\bf
  D93} (2016), no.~6 063525, [\href{http://arxiv.org/abs/1510.03460}{{\tt
  arXiv:1510.03460}}].

\bibitem{Griest:1990kh}
K.~Griest and D.~Seckel, {\it {Three exceptions in the calculation of relic
  abundances}},  {\em Phys. Rev.} {\bf D43} (1991) 3191--3203.

\bibitem{Edsjo:1997bg}
J.~Edsjo and P.~Gondolo, {\it {Neutralino relic density including
  coannihilations}},  {\em Phys. Rev.} {\bf D56} (1997) 1879--1894,
  [\href{http://arxiv.org/abs/hep-ph/9704361}{{\tt hep-ph/9704361}}].

\bibitem{Profumo:2006bx}
S.~Profumo and A.~Provenza, {\it {Increasing the neutralino relic abundance
  with slepton coannihilations: Consequences for indirect dark matter
  detection}},  {\em JCAP} {\bf 0612} (2006) 019,
  [\href{http://arxiv.org/abs/hep-ph/0609290}{{\tt hep-ph/0609290}}].

\bibitem{Edsjo:2003us}
J.~Edsjo, M.~Schelke, P.~Ullio, and P.~Gondolo, {\it {Accurate relic densities
  with neutralino, chargino and sfermion coannihilations in mSUGRA}},  {\em
  JCAP} {\bf 0304} (2003) 001, [\href{http://arxiv.org/abs/hep-ph/0301106}{{\tt
  hep-ph/0301106}}].

\bibitem{Baker:2015qna}
M.~J. Baker et~al., {\it {The Coannihilation Codex}},  {\em JHEP} {\bf 12}
  (2015) 120, [\href{http://arxiv.org/abs/1510.03434}{{\tt arXiv:1510.03434}}].

\bibitem{Cheng:2018vaj}
H.-C. Cheng, L.~Li, and R.~Zheng, {\it {Coscattering/Coannihilation Dark Matter
  in a Fraternal Twin Higgs Model}},
  \href{http://arxiv.org/abs/1805.12139}{{\tt arXiv:1805.12139}}.

\bibitem{DAgnolo:2018wcn}
R.~T. D'Agnolo, C.~Mondino, J.~T. Ruderman, and P.-J. Wang, {\it {Exponentially
  Light Dark Matter from Coannihilation}},
  \href{http://arxiv.org/abs/1803.02901}{{\tt arXiv:1803.02901}}.

\bibitem{Garny:2014waa}
M.~Garny, A.~Ibarra, S.~Rydbeck, and S.~Vogl, {\it {Majorana Dark Matter with a
  Coloured Mediator: Collider vs Direct and Indirect Searches}},  {\em JHEP}
  {\bf 06} (2014) 169, [\href{http://arxiv.org/abs/1403.4634}{{\tt
  arXiv:1403.4634}}].

\bibitem{Abdughani:2018bhj}
M.~Abdughani, J.~Ren, and J.~Zhao, {\it {TeV SUSY dark matter confronted with
  the current direct and indirect detection data}},
  \href{http://arxiv.org/abs/1805.06206}{{\tt arXiv:1805.06206}}.

\bibitem{Ellis:2018jyl}
J.~Ellis, J.~L. Evans, F.~Luo, K.~A. Olive, and J.~Zheng, {\it {Stop
  Coannihilation in the CMSSM and SubGUT Models}},  {\em Eur. Phys. J.} {\bf
  C78} (2018), no.~5 425, [\href{http://arxiv.org/abs/1801.09855}{{\tt
  arXiv:1801.09855}}].

\bibitem{ElHedri:2017nny}
S.~El~Hedri, A.~Kaminska, M.~de~Vries, and J.~Zurita, {\it {Simplified
  Phenomenology for Colored Dark Sectors}},  {\em JHEP} {\bf 04} (2017) 118,
  [\href{http://arxiv.org/abs/1703.00452}{{\tt arXiv:1703.00452}}].

\bibitem{ElHedri:2018atj}
S.~El~Hedri and M.~de~Vries, {\it {Cornering Colored Coannihilation}},
  \href{http://arxiv.org/abs/1806.03325}{{\tt arXiv:1806.03325}}.

\bibitem{Aboubrahim:2017aen}
A.~Aboubrahim, P.~Nath, and A.~B. Spisak, {\it {Stau coannihilation, compressed
  spectrum, and SUSY discovery potential at the LHC}},  {\em Phys. Rev.} {\bf
  D95} (2017), no.~11 115030, [\href{http://arxiv.org/abs/1704.04669}{{\tt
  arXiv:1704.04669}}].

\bibitem{Duan:2018rls}
G.~H. Duan, K.-I. Hikasa, J.~Ren, L.~Wu, and J.~M. Yang, {\it {Probing
  bino-wino coannihilation dark matter under the neutrino floor at the LHC}},
  \href{http://arxiv.org/abs/1804.05238}{{\tt arXiv:1804.05238}}.

\bibitem{Dutta:2017nqv}
B.~Dutta, K.~Fantahun, A.~Fernando, T.~Ghosh, J.~Kumar, P.~Sandick, P.~Stengel,
  and J.~W. Walker, {\it {Probing Squeezed Bino-Slepton Spectra with the Large
  Hadron Collider}},  {\em Phys. Rev.} {\bf D96} (2017), no.~7 075037,
  [\href{http://arxiv.org/abs/1706.05339}{{\tt arXiv:1706.05339}}].

\bibitem{Aaboud:2018jiw}
{\bf ATLAS} Collaboration, M.~Aaboud et~al., {\it {Search for electroweak
  production of supersymmetric particles in final states with two or three
  leptons at $\sqrt{s}=13\,$TeV with the ATLAS detector}},
  \href{http://arxiv.org/abs/1803.02762}{{\tt arXiv:1803.02762}}.

\bibitem{HELHC}
{\it {HL/HE-LHC Physics Workshop}}, .

\bibitem{Hinchliffe:2015qma}
I.~Hinchliffe, A.~Kotwal, M.~L. Mangano, C.~Quigg, and L.-T. Wang, {\it
  {Luminosity goals for a $100\,$TeV pp collider}},  {\em Int. J. Mod. Phys.}
  {\bf A30} (2015), no.~23 1544002,
  [\href{http://arxiv.org/abs/1504.06108}{{\tt arXiv:1504.06108}}].

\bibitem{Fujii:2017vwa}
K.~Fujii et~al., {\it {Physics Case for the 250 GeV Stage of the International
  Linear Collider}},  \href{http://arxiv.org/abs/1710.07621}{{\tt
  arXiv:1710.07621}}.

\bibitem{CLIC:2016zwp}
{\bf CLICdp, CLIC} Collaboration, M.~J. Boland et~al., {\it {Updated baseline
  for a staged Compact Linear Collider}},
  \href{http://arxiv.org/abs/1608.07537}{{\tt arXiv:1608.07537}}.

\bibitem{FCCee}
J.~Wenninger, M.~Benedikt, K.~Oide, and F.~Zimmermann, {\it {Future Circular
  Collider Study Lepton Collider Parameters}}, .

\bibitem{Akerib:2016vxi}
{\bf LUX} Collaboration, D.~S. Akerib et~al., {\it {Results from a search for
  dark matter in the complete LUX exposure}},  {\em Phys. Rev. Lett.} {\bf 118}
  (2017), no.~2 021303, [\href{http://arxiv.org/abs/1608.07648}{{\tt
  arXiv:1608.07648}}].

\bibitem{Cui:2017nnn}
{\bf PandaX-II} Collaboration, X.~Cui et~al., {\it {Dark Matter Results From
  54-Ton-Day Exposure of PandaX-II Experiment}},  {\em Phys. Rev. Lett.} {\bf
  119} (2017), no.~18 181302, [\href{http://arxiv.org/abs/1708.06917}{{\tt
  arXiv:1708.06917}}].

\bibitem{Aprile:2018dbl}
{\bf XENON} Collaboration, E.~Aprile et~al., {\it {Dark Matter Search Results
  from a One Tonne$\times$Year Exposure of XENON1T}},  {\em Phys. Rev. Lett.}
  {\bf 121} (2018), no.~11 111302, [\href{http://arxiv.org/abs/1805.12562}{{\tt
  arXiv:1805.12562}}].

\bibitem{Aalbers:2016jon}
{\bf DARWIN} Collaboration, J.~Aalbers et~al., {\it {DARWIN: towards the
  ultimate dark matter detector}},  {\em JCAP} {\bf 1611} (2016) 017,
  [\href{http://arxiv.org/abs/1606.07001}{{\tt arXiv:1606.07001}}].

\bibitem{Ahnen:2016qkx}
{\bf Fermi-LAT, MAGIC} Collaboration, M.~L. Ahnen et~al., {\it {Limits to dark
  matter annihilation cross-section from a combined analysis of MAGIC and
  Fermi-LAT observations of dwarf satellite galaxies}},  {\em JCAP} {\bf 1602}
  (2016), no.~02 039, [\href{http://arxiv.org/abs/1601.06590}{{\tt
  arXiv:1601.06590}}].

\bibitem{HESS:2015cda}
{\bf H.E.S.S.} Collaboration, A.~Abramowski et~al., {\it {Constraints on an
  Annihilation Signal from a Core of Constant Dark Matter Density around the
  Milky Way Center with H.E.S.S.}},  {\em Phys. Rev. Lett.} {\bf 114} (2015),
  no.~8 081301, [\href{http://arxiv.org/abs/1502.03244}{{\tt
  arXiv:1502.03244}}].

\bibitem{Carr:2015hta}
{\bf CTA} Collaboration, J.~Carr et~al., {\it {Prospects for Indirect Dark
  Matter Searches with the Cherenkov Telescope Array (CTA)}},  {\em PoS} {\bf
  ICRC2015} (2016) 1203, [\href{http://arxiv.org/abs/1508.06128}{{\tt
  arXiv:1508.06128}}]. [34,1203(2015)].

\bibitem{Staub:2008uz}
F.~Staub, {\it {SARAH}},  \href{http://arxiv.org/abs/0806.0538}{{\tt
  arXiv:0806.0538}}.

\bibitem{Belanger:2014vza}
G.~Bélanger, F.~Boudjema, A.~Pukhov, and A.~Semenov, {\it {micrOMEGAs4.1: two
  dark matter candidates}},  {\em Comput. Phys. Commun.} {\bf 192} (2015)
  322--329, [\href{http://arxiv.org/abs/1407.6129}{{\tt arXiv:1407.6129}}].

\bibitem{Ade:2015xua}
{\bf Planck} Collaboration, P.~A.~R. Ade et~al., {\it {Planck 2015 results.
  XIII. Cosmological parameters}},  {\em Astron. Astrophys.} {\bf 594} (2016)
  A13, [\href{http://arxiv.org/abs/1502.01589}{{\tt arXiv:1502.01589}}].

\bibitem{Gori:2013ala}
S.~Gori, S.~Jung, and L.-T. Wang, {\it {Cornering electroweakinos at the LHC}},
   {\em JHEP} {\bf 10} (2013) 191, [\href{http://arxiv.org/abs/1307.5952}{{\tt
  arXiv:1307.5952}}].

\bibitem{Schwaller:2013baa}
P.~Schwaller and J.~Zurita, {\it {Compressed electroweakino spectra at the
  LHC}},  {\em JHEP} {\bf 03} (2014) 060,
  [\href{http://arxiv.org/abs/1312.7350}{{\tt arXiv:1312.7350}}].

\bibitem{Evans:2016zau}
J.~A. Evans and J.~Shelton, {\it {Long-Lived Staus and Displaced Leptons at the
  LHC}},  {\em JHEP} {\bf 04} (2016) 056,
  [\href{http://arxiv.org/abs/1601.01326}{{\tt arXiv:1601.01326}}].

\bibitem{Khoze:2017ixx}
V.~V. Khoze, A.~D. Plascencia, and K.~Sakurai, {\it {Simplified models of dark
  matter with a long-lived co-annihilation partner}},  {\em JHEP} {\bf 06}
  (2017) 041, [\href{http://arxiv.org/abs/1702.00750}{{\tt arXiv:1702.00750}}].

\bibitem{Mahbubani:2017gjh}
R.~Mahbubani, P.~Schwaller, and J.~Zurita, {\it {Closing the window for
  compressed Dark Sectors with disappearing charged tracks}},  {\em JHEP} {\bf
  06} (2017) 119, [\href{http://arxiv.org/abs/1703.05327}{{\tt
  arXiv:1703.05327}}]. [Erratum: JHEP10,061(2017)].

\bibitem{Hochberg:2017wce}
Y.~Hochberg, Y.~Kahn, M.~Lisanti, K.~M. Zurek, A.~G. Grushin, R.~Ilan, S.~M.
  Griffin, Z.-F. Liu, S.~F. Weber, and J.~B. Neaton, {\it {Detection of sub-MeV
  Dark Matter with Three-Dimensional Dirac Materials}},  {\em Phys. Rev.} {\bf
  D97} (2018), no.~1 015004, [\href{http://arxiv.org/abs/1708.08929}{{\tt
  arXiv:1708.08929}}].

\bibitem{Knapen:2017xzo}
S.~Knapen, T.~Lin, and K.~M. Zurek, {\it {Light Dark Matter: Models and
  Constraints}},  {\em Phys. Rev.} {\bf D96} (2017), no.~11 115021,
  [\href{http://arxiv.org/abs/1709.07882}{{\tt arXiv:1709.07882}}].

\bibitem{Knapen:2017ekk}
S.~Knapen, T.~Lin, M.~Pyle, and K.~M. Zurek, {\it {Detection of Light Dark
  Matter With Optical Phonons in Polar Materials}},
  \href{http://arxiv.org/abs/1712.06598}{{\tt arXiv:1712.06598}}.

\bibitem{Kopp:2014tsa}
J.~Kopp, L.~Michaels, and J.~Smirnov, {\it {Loopy Constraints on Leptophilic
  Dark Matter and Internal Bremsstrahlung}},  {\em JCAP} {\bf 1404} (2014) 022,
  [\href{http://arxiv.org/abs/1401.6457}{{\tt arXiv:1401.6457}}].

\bibitem{Agrawal:2011ze}
P.~Agrawal, S.~Blanchet, Z.~Chacko, and C.~Kilic, {\it {Flavored Dark Matter,
  and Its Implications for Direct Detection and Colliders}},  {\em Phys. Rev.}
  {\bf D86} (2012) 055002, [\href{http://arxiv.org/abs/1109.3516}{{\tt
  arXiv:1109.3516}}].

\bibitem{Bai:2014osa}
Y.~Bai and J.~Berger, {\it {Lepton Portal Dark Matter}},  {\em JHEP} {\bf 08}
  (2014) 153, [\href{http://arxiv.org/abs/1402.6696}{{\tt arXiv:1402.6696}}].

\bibitem{Ibarra:2015fqa}
A.~Ibarra and S.~Wild, {\it {Dirac dark matter with a charged mediator: a
  comprehensive one-loop analysis of the direct detection phenomenology}},
  {\em JCAP} {\bf 1505} (2015), no.~05 047,
  [\href{http://arxiv.org/abs/1503.03382}{{\tt arXiv:1503.03382}}].

\bibitem{Banks:2010eh}
T.~Banks, J.-F. Fortin, and S.~Thomas, {\it {Direct Detection of Dark Matter
  Electromagnetic Dipole Moments}},  \href{http://arxiv.org/abs/1007.5515}{{\tt
  arXiv:1007.5515}}.

\bibitem{8ad88931fb144c15b303a77d05f9ab6e}
T.~Alford, L.~Feldman, and J.~Mayer, {\em Fundamentals of nanoscale film
  analysis}.
\newblock Springer US, 2007.

\bibitem{McCabe:2010zh}
C.~McCabe, {\it {The Astrophysical Uncertainties Of Dark Matter Direct
  Detection Experiments}},  {\em Phys. Rev.} {\bf D82} (2010) 023530,
  [\href{http://arxiv.org/abs/1005.0579}{{\tt arXiv:1005.0579}}].

\bibitem{Kahlhoefer:2016eds}
F.~Kahlhoefer and S.~Wild, {\it {Studying generalised dark matter interactions
  with extended halo-independent methods}},  {\em JCAP} {\bf 1610} (2016),
  no.~10 032, [\href{http://arxiv.org/abs/1607.04418}{{\tt arXiv:1607.04418}}].

\bibitem{Savage:2006qr}
C.~Savage, K.~Freese, and P.~Gondolo, {\it {Annual Modulation of Dark Matter in
  the Presence of Streams}},  {\em Phys. Rev.} {\bf D74} (2006) 043531,
  [\href{http://arxiv.org/abs/astro-ph/0607121}{{\tt astro-ph/0607121}}].

\bibitem{Fairbairn:2008gz}
M.~Fairbairn and T.~Schwetz, {\it {Spin-independent elastic WIMP scattering and
  the DAMA annual modulation signal}},  {\em JCAP} {\bf 0901} (2009) 037,
  [\href{http://arxiv.org/abs/0808.0704}{{\tt arXiv:0808.0704}}].

\bibitem{Fukushima:2014yia}
K.~Fukushima, C.~Kelso, J.~Kumar, P.~Sandick, and T.~Yamamoto, {\it {MSSM dark
  matter and a light slepton sector: The incredible bulk}},  {\em Phys. Rev.}
  {\bf D90} (2014), no.~9 095007, [\href{http://arxiv.org/abs/1406.4903}{{\tt
  arXiv:1406.4903}}].

\bibitem{Abdallah:2016ygi}
{\bf H.E.S.S.} Collaboration, H.~Abdallah et~al., {\it {Search for dark matter
  annihilations towards the inner Galactic halo from 10 years of observations
  with H.E.S.S}},  {\em Phys. Rev. Lett.} {\bf 117} (2016), no.~11 111301,
  [\href{http://arxiv.org/abs/1607.08142}{{\tt arXiv:1607.08142}}].

\bibitem{Rinchiuso:2017pcx}
{\bf HESS} Collaboration, L.~Rinchiuso and E.~Moulin, {\it {Dark matter
  searches toward the Galactic Centre halo with H.E.S.S}},  in {\em
  {Proceedings, 52nd Rencontres de Moriond on Very High Energy Phenomena in the
  Universe: La Thuile, Italy, March 18-25, 2017}}, 2017.
\newblock \href{http://arxiv.org/abs/1711.08634}{{\tt arXiv:1711.08634}}.

\bibitem{Lefranc:2015pza}
V.~Lefranc, E.~Moulin, P.~Panci, and J.~Silk, {\it {Prospects for Annihilating
  Dark Matter in the inner Galactic halo by the Cherenkov Telescope Array}},
  {\em Phys. Rev.} {\bf D91} (2015), no.~12 122003,
  [\href{http://arxiv.org/abs/1502.05064}{{\tt arXiv:1502.05064}}].

\bibitem{Carpenter:2016thc}
L.~M. Carpenter, R.~Colburn, J.~Goodman, and T.~Linden, {\it {Indirect
  Detection Constraints on s and t Channel Simplified Models of Dark Matter}},
  {\em Phys. Rev.} {\bf D94} (2016), no.~5 055027,
  [\href{http://arxiv.org/abs/1606.04138}{{\tt arXiv:1606.04138}}].

\bibitem{Bai:2013iqa}
Y.~Bai and J.~Berger, {\it {Fermion Portal Dark Matter}},  {\em JHEP} {\bf 11}
  (2013) 171, [\href{http://arxiv.org/abs/1308.0612}{{\tt arXiv:1308.0612}}].

\bibitem{Garny:2015wea}
M.~Garny, A.~Ibarra, and S.~Vogl, {\it {Signatures of Majorana dark matter with
  t-channel mediators}},  {\em Int. J. Mod. Phys.} {\bf D24} (2015), no.~07
  1530019, [\href{http://arxiv.org/abs/1503.01500}{{\tt arXiv:1503.01500}}].

\bibitem{TheFermi-LAT:2017vmf}
{\bf Fermi-LAT} Collaboration, M.~Ackermann et~al., {\it {The Fermi Galactic
  Center GeV Excess and Implications for Dark Matter}},  {\em Astrophys. J.}
  {\bf 840} (2017), no.~1 43, [\href{http://arxiv.org/abs/1704.03910}{{\tt
  arXiv:1704.03910}}].

\bibitem{Cirelli:2010xx}
M.~Cirelli, G.~Corcella, A.~Hektor, G.~Hutsi, M.~Kadastik, P.~Panci, M.~Raidal,
  F.~Sala, and A.~Strumia, {\it {PPPC 4 DM ID: A Poor Particle Physicist
  Cookbook for Dark Matter Indirect Detection}},  {\em JCAP} {\bf 1103} (2011)
  051, [\href{http://arxiv.org/abs/1012.4515}{{\tt arXiv:1012.4515}}].
  [Erratum: JCAP1210,E01(2012)].

\bibitem{Abramowski:2014tra}
{\bf H.E.S.S.} Collaboration, A.~Abramowski et~al., {\it {Search for dark
  matter annihilation signatures in H.E.S.S. observations of Dwarf Spheroidal
  Galaxies}},  {\em Phys. Rev.} {\bf D90} (2014) 112012,
  [\href{http://arxiv.org/abs/1410.2589}{{\tt arXiv:1410.2589}}].

\bibitem{Lefranc:2016dgx}
V.~Lefranc, G.~A. Mamon, and P.~Panci, {\it {Prospects for annihilating Dark
  Matter towards Milky Way's dwarf galaxies by the Cherenkov Telescope Array}},
   {\em JCAP} {\bf 1609} (2016), no.~09 021,
  [\href{http://arxiv.org/abs/1605.02793}{{\tt arXiv:1605.02793}}].

\bibitem{Accardo:2014lma}
{\bf AMS} Collaboration, L.~Accardo et~al., {\it {High Statistics Measurement
  of the Positron Fraction in Primary Cosmic Rays of 0.5–500 GeV with the
  Alpha Magnetic Spectrometer on the International Space Station}},  {\em Phys.
  Rev. Lett.} {\bf 113} (2014) 121101.

\bibitem{Aguilar:2014mma}
{\bf AMS} Collaboration, M.~Aguilar et~al., {\it {Electron and Positron Fluxes
  in Primary Cosmic Rays Measured with the Alpha Magnetic Spectrometer on the
  International Space Station}},  {\em Phys. Rev. Lett.} {\bf 113} (2014)
  121102.

\bibitem{Aguilar:2016kjl}
{\bf AMS} Collaboration, M.~Aguilar et~al., {\it {Antiproton Flux,
  Antiproton-to-Proton Flux Ratio, and Properties of Elementary Particle Fluxes
  in Primary Cosmic Rays Measured with the Alpha Magnetic Spectrometer on the
  International Space Station}},  {\em Phys. Rev. Lett.} {\bf 117} (2016),
  no.~9 091103.

\bibitem{Lefranc:2016fgn}
V.~Lefranc, E.~Moulin, P.~Panci, F.~Sala, and J.~Silk, {\it {Dark Matter in
  $\gamma$ lines: Galactic Center vs dwarf galaxies}},  {\em JCAP} {\bf 1609}
  (2016), no.~09 043, [\href{http://arxiv.org/abs/1608.00786}{{\tt
  arXiv:1608.00786}}].

\bibitem{Aad:2014vma}
{\bf ATLAS} Collaboration, G.~Aad et~al., {\it {Search for direct production of
  charginos, neutralinos and sleptons in final states with two leptons and
  missing transverse momentum in $pp$ collisions at $\sqrt{s} =$ 8 TeV with the
  ATLAS detector}},  {\em JHEP} {\bf 05} (2014) 071,
  [\href{http://arxiv.org/abs/1403.5294}{{\tt arXiv:1403.5294}}].

\bibitem{Alwall:2014hca}
J.~Alwall, R.~Frederix, S.~Frixione, V.~Hirschi, F.~Maltoni, O.~Mattelaer,
  H.~S. Shao, T.~Stelzer, P.~Torrielli, and M.~Zaro, {\it {The automated
  computation of tree-level and next-to-leading order differential cross
  sections, and their matching to parton shower simulations}},  {\em JHEP} {\bf
  07} (2014) 079, [\href{http://arxiv.org/abs/1405.0301}{{\tt
  arXiv:1405.0301}}].

\bibitem{Sjostrand:2006za}
T.~Sjostrand, S.~Mrenna, and P.~Z. Skands, {\it {PYTHIA 6.4 Physics and
  Manual}},  {\em JHEP} {\bf 05} (2006) 026,
  [\href{http://arxiv.org/abs/hep-ph/0603175}{{\tt hep-ph/0603175}}].

\bibitem{deFavereau:2013fsa}
{\bf DELPHES 3} Collaboration, J.~de~Favereau, C.~Delaere, P.~Demin,
  A.~Giammanco, V.~Lemaître, A.~Mertens, and M.~Selvaggi, {\it {DELPHES 3, A
  modular framework for fast simulation of a generic collider experiment}},
  {\em JHEP} {\bf 02} (2014) 057, [\href{http://arxiv.org/abs/1307.6346}{{\tt
  arXiv:1307.6346}}].

\bibitem{FCCDelphes}
{\bf FCC Working Group} Collaboration, Z.~Drasal, {\it {FCCSW Main page for
  documentation and resources.}}, .

\bibitem{Aad:2016wpd}
{\bf ATLAS} Collaboration, G.~Aad et~al., {\it {Measurement of total and
  differential $W^+W^-$ production cross sections in proton-proton collisions
  at $\sqrt{s}=$ 8 TeV with the ATLAS detector and limits on anomalous
  triple-gauge-boson couplings}},  {\em JHEP} {\bf 09} (2016) 029,
  [\href{http://arxiv.org/abs/1603.01702}{{\tt arXiv:1603.01702}}].

\bibitem{Gehrmann:2014fva}
T.~Gehrmann, M.~Grazzini, S.~Kallweit, P.~Maierhöfer, A.~von Manteuffel,
  S.~Pozzorini, D.~Rathlev, and L.~Tancredi, {\it {$W^+W^-$ Production at
  Hadron Colliders in Next to Next to Leading Order QCD}},  {\em Phys. Rev.
  Lett.} {\bf 113} (2014), no.~21 212001,
  [\href{http://arxiv.org/abs/1408.5243}{{\tt arXiv:1408.5243}}].

\bibitem{Heinemeyer:2013tqa}
{\bf LHC Higgs Cross Section Working Group} Collaboration, J.~R. Andersen
  et~al., {\it {Handbook of LHC Higgs Cross Sections: 3. Higgs Properties}},
  \href{http://arxiv.org/abs/1307.1347}{{\tt arXiv:1307.1347}}.

\bibitem{Grazzini:2016swo}
M.~Grazzini, S.~Kallweit, D.~Rathlev, and M.~Wiesemann, {\it {$W^{\pm}Z$
  production at hadron colliders in NNLO QCD}},  {\em Phys. Lett.} {\bf B761}
  (2016) 179--183, [\href{http://arxiv.org/abs/1604.08576}{{\tt
  arXiv:1604.08576}}].

\bibitem{Cascioli:2014yka}
F.~Cascioli, T.~Gehrmann, M.~Grazzini, S.~Kallweit, P.~Maierhöfer, A.~von
  Manteuffel, S.~Pozzorini, D.~Rathlev, L.~Tancredi, and E.~Weihs, {\it {ZZ
  production at hadron colliders in NNLO QCD}},  {\em Phys. Lett.} {\bf B735}
  (2014) 311--313, [\href{http://arxiv.org/abs/1405.2219}{{\tt
  arXiv:1405.2219}}].

\bibitem{Khachatryan:2016yzq}
{\bf CMS} Collaboration, V.~Khachatryan et~al., {\it {Measurements of the
  $\mathrm{t}\overline{\mathrm{t}}$ production cross section in lepton+jets
  final states in pp collisions at 8 $\,\text {TeV}$ and ratio of 8 to 7
  $\,\text {TeV}$ cross sections}},  {\em Eur. Phys. J.} {\bf C77} (2017),
  no.~1 15, [\href{http://arxiv.org/abs/1602.09024}{{\tt arXiv:1602.09024}}].

\bibitem{Gabaldon:2125723}
C.~Gabaldon, {\it {Single top measurements at the LHC: $s$-channel and $Wt$
  production}},  Tech. Rep. ATL-PHYS-PROC-2016-012, CERN, Geneva, Jan, 2016.

\bibitem{Lester:1999tx}
C.~G. Lester and D.~J. Summers, {\it {Measuring masses of semiinvisibly
  decaying particles pair produced at hadron colliders}},  {\em Phys. Lett.}
  {\bf B463} (1999) 99--103, [\href{http://arxiv.org/abs/hep-ph/9906349}{{\tt
  hep-ph/9906349}}].

\bibitem{Barr:2003rg}
A.~Barr, C.~Lester, and P.~Stephens, {\it {m(T2): The Truth behind the
  glamour}},  {\em J. Phys.} {\bf G29} (2003) 2343--2363,
  [\href{http://arxiv.org/abs/hep-ph/0304226}{{\tt hep-ph/0304226}}].

\bibitem{Barr:2009jv}
A.~J. Barr, B.~Gripaios, and C.~G. Lester, {\it {Transverse masses and
  kinematic constraints: from the boundary to the crease}},  {\em JHEP} {\bf
  11} (2009) 096, [\href{http://arxiv.org/abs/0908.3779}{{\tt
  arXiv:0908.3779}}].

\bibitem{Cowan:2010js}
G.~Cowan, K.~Cranmer, E.~Gross, and O.~Vitells, {\it {Asymptotic formulae for
  likelihood-based tests of new physics}},  {\em Eur. Phys. J.} {\bf C71}
  (2011) 1554, [\href{http://arxiv.org/abs/1007.1727}{{\tt arXiv:1007.1727}}].
  [Erratum: Eur. Phys. J.C73,2501(2013)].

\bibitem{Geilhufe:2018gry}
R.~M. Geilhufe, B.~Olsthoorn, A.~Ferella, T.~Koski, F.~Kahlhoefer, J.~Conrad,
  and A.~V. Balatsky, {\it {Materials Informatics for Dark Matter Detection}},
  \href{http://arxiv.org/abs/1806.06040}{{\tt arXiv:1806.06040}}.

\bibitem{Baum:2018tfw}
S.~Baum, A.~K. Drukier, K.~Freese, M.~G\'orski, and P.~Stengel, {\it {Searching
  for Dark Matter with Paleo-Detectors}},
  \href{http://arxiv.org/abs/1806.05991}{{\tt arXiv:1806.05991}}.

\bibitem{Baryakhtar:2017dbj}
M.~Baryakhtar, J.~Bramante, S.~W. Li, T.~Linden, and N.~Raj, {\it {Dark Kinetic
  Heating of Neutron Stars and An Infrared Window On WIMPs, SIMPs, and Pure
  Higgsinos}},  {\em Phys. Rev. Lett.} {\bf 119} (2017), no.~13 131801,
  [\href{http://arxiv.org/abs/1704.01577}{{\tt arXiv:1704.01577}}].

\end{thebibliography}\endgroup

\end{document}